\documentclass[a4paper, 12pt]{article}

\usepackage{amssymb,amsmath,amsfonts,bbm,pifont,upgreek,bbold,accents}  %Luigi
\usepackage[colorlinks=true]{hyperref}
\hypersetup{urlcolor=blue, citecolor=red}

\usepackage[dvips]{graphicx}
\font\teneufm=eufm10
\font\seveneufm=eufm7
\font\fiveeufm=eufm5
\newfam\eufmfam
\textfont\eufmfam=\teneufm
\scriptfont\eufmfam=\seveneufm
\scriptscriptfont\eufmfam=\fiveeufm
\def\eufm#1{{\fam\eufmfam\relax#1}}

\newcommand\beq[1]{\begin{equation}\label{#1} }
\newcommand{\eeq}{\end{equation} }

\newcommand\beqa[1]{\begin{eqnarray} \label{#1}}
\newcommand{\eeqa}{\end{eqnarray} }
\newcommand{\beqano}{\begin{eqnarray*} }
\newcommand{\eeqano}{\end{eqnarray*} }
\newcommand\arr[1]{\left\{\begin{array}{l}#1\end{array}\right.}
\renewcommand{\theequation}{\arabic{section}.\arabic{equation}}

\newtheorem{theorem}{Theorem}[section]
\newtheorem{definition}{Definition}[section]
\newtheorem{proposition}{Proposition}[section]
\newtheorem{lemma}{Lemma}[section]

\newtheorem{sublemma}{Sublemma}[section]
\newtheorem{remark}{Remark}[section]
\newtheorem{notationalremark}{Notational Remark}[section]
\newtheorem{corollary}{Corollary}[section]
\newtheorem{assumption}{Assumption}[section]
\newtheorem{claim}{Claim}[section]

\newtheorem{tools}{$\negsp\negsp$}[subsection]

\newcommand\thm[1]{\begin{theorem}\label{#1}}
\newcommand\thmtwo[2]{\begin{theorem}[#1]\label{#2}}
\newcommand\ethm{\end{theorem} }
\newcommand\dfn[1]{\begin{definition}\label{#1} \rm}
\newcommand\dfntwo[2]{\begin{definition}[#1]\label{#2} \rm}
\newcommand\edfn{\end{definition} }
\newcommand\pro[1]{\begin{proposition}\label{#1}}
\newcommand\protwo[2]{\begin{proposition}[#1]\label{#2}}
\newcommand\epro{\end{proposition} }
\newcommand\lem[1]{\begin{lemma}\label{#1}}
\newcommand\lemtwo[2]{\begin{lemma}[#1]\label{#2}}
\newcommand\elem{\end{lemma} }
\newcommand\sublem[1]{\begin{sublemma}\label{#1}}
\newcommand\sublemtwo[2]{\begin{sublemma}[#1]\label{#2}}
\newcommand\esublem{\end{sublemma} }
\newcommand\rem[1]{\begin{remark}\label{#1} \rm}
\newcommand\erem{\end{remark} }
\newcommand\notrem[1]{\begin{notationalremark}\label{#1} \rm}
\newcommand\enotrem{\end{notationalremark} }
\newcommand\cor[1]{\begin{corollary}\label{#1}}
\newcommand\cortwo[2]{\begin{corollary}[#1]\label{#2}}
\newcommand\ecor{\end{corollary} }
\newcommand\asmp[1]{\begin{assumption}\label{#1}}
\newcommand\asmptwo[2]{\begin{assumption}[#1]\label{#2}}
\newcommand\easmp{\end{assumption} }
\newcommand\clm[1]{\begin{claim}\label{#1}}
\newcommand\eclm{\end{claim} }
\newcommand{\proof}{\par\medskip\noindent{\bf Proof\ }}
\newcommand\equ[1]{{\rm (\ref{#1})}}

\expandafter\chardef\csname pre amssym.def
at\endcsname=\the\catcode`\@
\catcode`\@=11
\def\undefine#1{\let#1\undefined}
\def\newsymbol#1#2#3#4#5{\let\next@\relax
 \ifnum#2=\@ne\let\next@\msafam@\else
 \ifnum#2=\tw@\let\next@\msbfam@\fi\fi
 \mathchardef#1="#3\next@#4#5}
\def\mathhexbox@#1#2#3{\relax
 \ifmmode\mathpalette{}{\m@th\mathchar"#1#2#3}%
 \else\leavevmode\hbox{$\m@th\mathchar"#1#2#3$}\fi}
\def\hexnumber@#1{\ifcase#1 0\or 1\or 2\or 3\or 4\or 5\or 6\or 7\or
8\or
 9\or A\or B\or C\or D\or E\or F\fi}
\ifcase\@ptsize
 \font\tenmsb=msbm10
 \font\sevenmsb=msbm7
 \font\fivemsb=msbm5
\or
 \font\tenmsb=msbm10 scaled \magstephalf
 \font\sevenmsb=msbm7 scaled \magstephalf
 \font\fivemsb=msbm5  scaled \magstephalf
\or
 \font\tenmsb=msbm10 scaled \magstep1
 \font\sevenmsb=msbm7 scaled \magstep1
 \font\fivemsb=msbm5 scaled \magstep1
\fi
\newfam\msbfam
\textfont\msbfam=\tenmsb
\scriptfont\msbfam=\sevenmsb
\scriptscriptfont\msbfam=\fivemsb
\edef\msbfam@{\hexnumber@\msbfam}
\def\Bbb#1{\fam\msbfam\relax#1}
\def\widehat#1{\setboxz@h{$\m@th#1$}%
 \ifdim\wdz@>\tw@ em\mathaccent"0\msbfam@5B{#1}%
 \else\mathaccent"0362{#1}\fi}
\def\widetilde#1{\setboxz@h{$\m@th#1$}%
 \ifdim\wdz@>\tw@ em\mathaccent"0\msbfam@5D{#1}%
 \else\mathaccent"0365{#1}\fi}

\def\RIfM@{\relax\ifmmode}
\def\nonmatherr@#1{\errmessage{\string#1\space allowed only in math mode}}
\def\Bbb{\RIfM@\expandafter\Bbb@\else
 \expandafter\nonmatherr@\expandafter\Bbb\fi}
\def\Bbb@#1{{\Bbb@@{#1}}}
\def\Bbb@@#1{\fam\msbfam\relax#1}
\def\setboxz@h{\setbox\z@\hbox}
\def\wdz@{\wd\z@}
\catcode`\@=\csname pre amssym.def at\endcsname
\newcommand{\ii}{{\rm i}  }

\newcommand{\eg}{{\tt e.g.\,}}

\newcommand{\nl}{{\smallskip\noindent}}
\newcommand{\noi}{{\noindent}}

\newcommand{\negsp}{\hspace{-.09truecm}}  %%% equivalente a \!

\newcommand{\dst}{\displaystyle}
\newcommand\ovl[1]{\overline {#1} }

\newcommand\su[1]{\frac{1}{{#1}} }

\newcommand\sign{{\, \rm sign\, }}

\newcommand{\integer}{{\Bbb Z}   }
\newcommand{\complex}{{\Bbb C}   }

\renewcommand{\a }{{\alpha}   }
\renewcommand{\b}{{\beta}   }
\newcommand{\g}{{\gamma}   }

\renewcommand{\d}{{\delta}   }
\newcommand{\D}{{\Delta}   }

\renewcommand{\l}{{\lambda}   }
\renewcommand{\L}{{\Lambda}   }
\newcommand{\m}{{\mu}   }
\newcommand{\n}{{\nu}   }

\newcommand{\p}{{\pi}   }
\renewcommand{\P}{{\Pi}   }
\renewcommand{\r}{{\rho}   }
\newcommand{\s}{{\sigma}   }

\renewcommand{\t}{{\tau}   }
\newcommand{\f}{{\varphi}   }

\renewcommand{\o}{{\omega}   }

\newcommand{\cA}{{\cal A} }
\newcommand{\cB}{{\cal B} }
\newcommand{{\cE}}{{\cal  E} }

\newcommand{\cR}{{\cal R} }
\newcommand{{\cH}}{{\cal H} }
\newcommand{\cK}{{\cal K} }
\newcommand{\cC}{{\cal C} }
\newcommand{\cD}{{\cal D} }
\newcommand{\cF }{{\cal F} }
\newcommand{\cG}{{\cal G} }
\newcommand{{\cJ}}{{\cal J}}
\newcommand{\cL}{{\cal L} }
\newcommand{\cM}{{\cal M} }
\newcommand{\cP}{{\cal P} }
\newcommand{\cI}{{\cal I} }
\newcommand{\cN}{{\cal N} }

\newcommand{\cU}{{\cal U} }

\newcommand\ppu{{(1) }}
\newcommand\ppd{{(2) }}
\newcommand\ppt{{(3) }}

\newcommand\ppk{{(k) }}

\newcommand\ppj{{(j) }}

\newcommand\ppi{{(i) }}

\newcommand\ppo{{(0) }}

\newcommand{\CC}{{\rm C}}

\newcommand{\EE}{{\rm E}}
\newcommand\FF{{\tt F}}
\newcommand\GG{{\rm G}}
\newcommand\HH{{\rm H}}
\newcommand\II{{\rm I}}
\newcommand\JJ{{\rm J}}

\newcommand\MM{{\rm M}}

\newcommand\OO{{\rm O}}

\newcommand\RR{{\rm R}}

\newcommand\UU{{\rm U}}

\newcommand\XX{{\rm X}}
\newcommand\YY{{\rm Y}}
\newcommand\ZZ{{\rm Z}}
\newcommand\dd{{\rm d}}
\newcommand\ee{{\rm e}}
\newcommand\hh{{\rm h}}

\newcommand\rr{{\rm r}}
\newcommand\mm{{\rm m}}

\newcommand\cO{{\cal O}}

\begin{document}

\title{\bf 
Exponential stability of Euler integral in the three--body problem\footnote{{\bf MSC2000 numbers:}
primary:
34C20, 70F10,  37J10, 37J15, 37J40;
secondary: 
34D10,  70F07, 70F15, 37J25, 37J35. {\bf Keywords:} Three--body problem;  Normal form theory; Two--centre problem; Euler Integral; Prediction of collisions.}}

\author{ 
Gabriella Pinzari\thanks{
This research is funded by the ERC project 677793 Stable and Chaotic Motions in the Planetary Problem.
}\\
\vspace{-.2truecm}
{\footnotesize Dipartimento di Matematica}
\\{\footnotesize University of Padova}
%\vspace{-.2truecm}
%\\{\footnotesize Monte Sant'Angelo -- Via Cinthia I-80126 Napoli (Italy)}
%\vspace{-.2truecm}
\\{\scriptsize gabriella.pinzari@math.unipd.it}
}\date{August,  2018}
\maketitle

\begin{abstract}\footnotesize{
The first integral characteristic of the two--centres problem  is proven to be an approximate integral (in the sense of N.N.Nekhorossev) to the three--body problem, at least if the masses are very different  and the particles are constrained on the same plane. The proof uses a new normal form result, carefully designed around the degeneracies of the problem, and a new study of the phase portrait of the unperturbed problem. Applications to the prediction of collisions between the two minor bodies are shown.
}
\end{abstract}

\maketitle
\newpage
\tableofcontents

%\appendix

\renewcommand{\theequation}{\arabic{equation}}
\setcounter{equation}{0}
\newpage
\section{Description of the result}

 \nl
A relevant problem in dynamics of $N$--particle systems is related to the occurrence of {\it collisions}, i.e.,  equalities of the kind  $${\tt x}_i(t_0)={\tt x}_j(t_0)$$ for some  $i\ne j$, $t_0>0$,  where ${\tt x}_i\in {\mathbb R}^3$ represents the set of Cartesian coordinates  of the $i^{\rm th}$ particle of the system, $t\to\underline{\tt x}(t)=({\tt x}_1(t), \cdots, {\tt x}_N(t))$ is a given time law for such coordinates. 
 The theoretical interest in the study of collisions relies in the fact that  often these are associated to singularities of the vectorfield and hence to loss of meaning of the equations of motion.  An outstanding example is the one of gravitational systems, where  the occurrence or the absence of collisions for a given motion has a central interest by itself, also on the practical side: improving techniques to  {\it predict}, within a prefixed error, the occurrence of a collision is a daily job of astronomers  (see~\cite{milaniG10} and references therein).
For this problem, an important   part of the mathematical literature has been devoted to develop {\it regularizing techniques}  (see~\cite{KustaanheimoS65, waldvogel76} and quoted references), consisting of changes of coordinates and time $(t, \underline{\tt x})\to (\t, \underline{\tt z})$,  such that, in the new coordinates, the law $\t\to\underline{\tt z}(\t)=({\tt z}_1(\t), \cdots, {\tt z}_N(\t))$ has a meaning even if a  collision occurs. An important (often hard) part of the work consists  then in proving that a given solution of interest for the system is  collision--free~\cite{chencinerM00, ferrarioT04} or eventually collisional~\cite{guardiaKZ18}.

 \nl
In this paper we address the question  in  the case of the {\it three--body problem}. This is the system composed of  three  point--wise  masses, undergoing gravitational attraction. We assume that the  masses  are of three different and well separated sizes, and are constrained on the same plane. Gravitating systems attracted the attention of eminent mathematicians since the beginning of the rational thought, mainly because of
 their physical interpretation. The one considered in the paper emulates a Sun--Earth--Asteroid system.   We aim to show that, for this problem, it is possible to predict wether a collision between the two minor bodies will occur within a given time according to the initial value of a certain function -- an approximate integral for the system -- that we shall call {\it Euler integral}. 
  The main ingredient of proof will be given by a  connection  with the so--called {\it two--centre problem}, the integrable problem solved by Euler~\cite[p. 247]{jacobi09}, that we shall recall below.

\nl
Let us now describe the mathematical setting, trying to keep technicalities to a minimum.  
Let $m_0$, 
$m'= \m m_0$, $m=\varepsilon\m m_0$ be the masses of  three particles interacting through gravity, where
 $\m$, $\varepsilon$ are very small numbers. 
After the reduction of translation invariance according to the heliocentric method (see~\cite{siegelM71} for notices), the motions of the system are governed by the Hamilton equations 
 of

\beqa{H3B initial}
\HH({ \underline{\tt y}}, { \underline{\tt x}})=\frac{\|{{\tt y}'}\|^2}{2\m \mm'}-\frac{\m \mm'{\cal M}'}{\|{{\tt x}'}\|}+\frac{\|{{\tt y}}\|^2}{2\varepsilon\m \mm}-\frac{\varepsilon\m \mm{\cal M}}{\|{{\tt x}}\|}+\frac{{{\tt y}'}\cdot {{\tt y}}}{m_0}-\frac{\varepsilon\m^2 \mm{\cal M}}{\|{{\tt x}}-{{\tt x}'}\|}
\eeqa
where $({ \underline{\tt y}}, { \underline{\tt x}}):=({\tt y}', {\tt y}, {\tt x}', {\tt x})\in {\mathbb R}^2\times {\mathbb R}^2\setminus\{{\tt x}'=0,\ {\tt x}=0,\ {\tt x}'={\tt x}\}$ and

\beqa{masses}\mm'=\frac{m_0}{1+\m}\ ,\quad \mm=\frac{m_0}{1+\varepsilon\m}\ ,\quad {\cal M}'= m_0(1+\m)\ ,\quad {\cal M}=m_0(1+\varepsilon\m) \eeqa
are the {\it reduced masses}.
We rescale impulses and time, switching to the Hamiltonian %Rescaling time and impulses as
%$$ t\to \frac{t}{\varepsilon}\ ,\quad { {\tt y}}\to\b\; { {\tt y}}\quad {\rm with}\quad \b=\varepsilon\m$$
%the $\frac{t}{\varepsilon}$--time evolution of $({ \underline{\tt y}}, { \underline{\tt x}})$ is governed by the Hamilton equations of 
$$\widehat\HH({\underline{\tt y}}, {\underline{\tt x}}):=\frac{{1}}{\m}\HH(\varepsilon\m {\underline{\tt y}}, {\underline{\tt x}})\ .$$
We obtain (neglecting the ``hat'')
\beqa{3BP}
{\tt H}(  {\underline{{\tt y}}}, {\underline{{\tt x}}})&=&-\frac{\mm'{\cal M}'}{\|{{{\tt x}}'}\|}+\varepsilon\left(
\frac{\|  {{{\tt y}}}\|^2}{2 \mm}-\frac{\mm{\cal M}}{\|{{{\tt x}}}\|}-\frac{\m \mm{\cal M}}{\|{{{\tt x}}}-{{{\tt x}}'}\|}
\right)\nonumber\\
&+&\varepsilon^2\left(\frac{\| {{{\tt y}}'}\|^2}{2\mm'}+\frac{\m}{m_0} {{{\tt y}}'}\cdot  {{{\tt y}}}\right)\ .
%\nonumber\\
%&=&\JJ_0+\varepsilon\JJ_1+\varepsilon^2 f
\eeqa

\nl
Setting the terms weighted by $\varepsilon^2$ to ${\tt 0}$,  the Hamiltonian reduces to
\beqa{truncated}
{\tt H}_{0}({{{\tt y}}}, {{\tt x}}; {\tt x}')&=&-\frac{\mm'{\cal M}'}{\|{{{\tt x}}'}\|}+\varepsilon\left(
\frac{\|  {{{\tt y}}}\|^2}{2 \mm}-\frac{\mm{\cal M}}{\|{{{\tt x}}}\|}-\frac{\m \mm{\cal M}}{\|{{{\tt x}}}-{{{\tt x}}'}\|}
\right)\ .%\nonumber\\
%\nonumber\\
%&=&\JJ_0+\varepsilon\JJ_1+\varepsilon^2 f
\eeqa
The  motions of $ \HH_{0}$  are immediate: (i) ${\tt x}'$ remains constant; (ii) the motion of $({\tt y}$, ${\tt x})$ are ruled, apart for an inessential scaling factor $\varepsilon$, by
 \beqa{newH2C}\JJ=
\frac{\|  {{\tt y}}\|^2}{2 \mm}-\frac{\mm{\cal M}}{\|{{\tt x}}\|}-\frac{\m \mm{\cal M}}{\|{{\tt x}}-{{\tt x}'}\|}\ ;\eeqa
(iii) the motion of ${\tt y}'$ are found by an elementary quadrature.

\noi
%The Hamiltonians $\HH_{0}$ and $\JJ$ differ
%by a first integral of them both, so, they are dynamically equivalent. In particular, they have the same first integrals.
$\JJ$ is %well known: it is 
the 3 ($2$ in the planar  problem)--degrees of freedom Hamiltonian governing the motion of a moving particle
with mass $\mm$, attracted by two fixed masses  $\cM$, $\m\cM$, located at the origin ${\tt 0}$  and at ${\tt x}'$, respectively. It the Hamiltonian of the {\it two--centre problem}. Euler in the XVIII century showed that 
it admits an independent integral of motion, which, through out this paper, we shall refer to as {\it Euler integral} and denote as $\EE$.
The expression of $\EE$ in terms of  initial   coordinates $(\underline{\tt y}, \underline{\tt x})$  -- actually not easy to be found in the literature -- is

\beqa{EEE}\EE=\|{\tt M}\|^2-{\tt x}'\cdot {\tt L}+ \m\mm^2{\cal M}\frac{({\tt x}'-{\tt x})\cdot {\tt x}'}{\|{\tt x}'-{\tt x}\|}\eeqa
where 
\beqa{CL}
{\tt M}:={\tt x}\times {\tt y}\ ,\qquad {\tt L}:={{\tt y}}\times{\tt M}-\mm^2{\cal M}\frac{{{\tt x}}}{\|{{\tt x}}\|}\eeqa 
are the  {\it angular momentum} and the {\it eccentricity vector} associated to $({\tt y}, {\tt x})$.
Observe that, when $\m=0$, $\JJ$ reduces to the {\it Kepler Hamiltonian}
\beqa{kepler}\JJ_0({\tt y}, {\tt x})=
\frac{\|  {{\tt y}}\|^2}{2 \mm}-\frac{\mm{\cal M}}{\|{{\tt x}}\|}\eeqa
and $\EE$ reduces to
\beqa{EEE0}\EE_0({\tt y}, {\tt x})=\|{\tt M}\|^2-{\tt x}'\cdot {\tt L}\ .\eeqa
It is not surprising that $\EE_0$ is function of
${\tt M}$ and
${\tt L}$, well known first integrals to $\JJ_0$.

\nl
Now we turn  to describe the result of the paper.
The formula in~\equ{3BP} seems to suggest that the the motions of $\HH$ and of $\HH_{0}$ are ``close'' one to the other. On the other hand, the Euler integral $\EE$ in~\equ{EEE} remains constant during the motions of $\HH_{0}$, so it seems reasonable to conjecture that $\EE$ varies a little even under the dynamics of $\HH$. We shall prove that, at least in the planar case, this is true.

\nl
{\bf Theorem~A} {\it
Let $\dd=2$. Under suitable assumptions on the initial data, $\EE$  affords little changes along the trajectories of the Hamiltonian $\HH$, over exponentially  long times.
}

\nl
A more precise statement of Theorem~A will be given  in the course of the paper (see Theorem~\ref{theor: Euler}). In particular,  the expression ``exponentially  long times'' will be quantified in terms of small quantities, characteristic of the problem. Here, we describe how Theorem~A is related to the prediction of  collisions between the two minor bodies  in the Hamiltonian~\equ{3BP}.

\nl
In Section~\ref{integration 2CP} -- elaborating previous work~\cite{pinzari13, pinzari18} -- we shall introduce a system of canonical coordinates similar in some respect to the coordinates of the rigid body, but with six degree of freedom instead of three -- that we denote  as $$\cK=(\ZZ, \CC, \Theta, \GG, \L, \RR', \zeta, \g, \vartheta, {\rm g}, \ell, \rr')$$ 
defined in a region of phase space where $\JJ_0({\tt y}, {\tt x})<0$,
such that
$\rr'=\|{\tt x}'\|$, $\GG=\|{\tt M}\|$ and, if $\cE({\tt y}, {\tt x})$ denotes the  instantaneous ellipse\footnote{The instantaneous ellipse $\cE({\tt y}_0, {\tt x}_0)$ through $({\tt y}_0, {\tt x}_0)$ is defined as the solution of $\JJ_0({\tt y}, {\tt x})$ with initial datum a given $({\tt y}_0, {\tt x}_0)\in {\mathbb R}^\dd\times {\mathbb R}^\dd\setminus\{0\}$ such that 
$\JJ_0({\tt y}_0, {\tt x}_0)<0$.
} through $({\tt y}, {\tt x})$, $a$ its semi--major axis, $\ee$ its eccentricity, then $\L=\mm\sqrt{\cM a}$, $\ee=\sqrt{1-\frac{\GG^2}{\L^2}}$ and, finally, in the case of the planar problem,  ${\tt x}'$ and ${\tt P}$ form a convex angle equal to $|\p-{\rm g}|$ (see Section~\ref{details}).

\nl
Using the well--known relation
$${\tt L}=\mm^2\cM \ee {\tt P}$$
we find that
$\EE$ in~\equ{EEE} takes the intriguing aspect
\beqa{E0}\EE=\GG^2+\mm^2\cM\rr'\sqrt{1-\frac{\GG^2}{\L^2}}\cos{\rm g}+\m \mm^2\cM\rr'\hat\EE_1\ ,\eeqa
with $\hat\EE_1$ being a  function of $(\L, \GG,\rr',  \ell, {\rm g})$ defined as
$$\hat\EE_1(\L, \GG,\rr',  \ell, {\rm g}):=\left(\frac{{\tt x}'\cdot ({\tt x}'-{\tt x})}{\|{\tt x}'\|\|{\tt x}'-{\tt x}\|}\right)\circ\cK$$
and hence  verifying
$$|\hat\EE_1|\le 1\ .$$
Now, a collision between ${\tt x}$ and ${\tt x}'$ occurs when ${\tt x}'$ belongs to  $\cE({\tt y}, {\tt x})$, or, in other words,  
 the {\it focal equation}
\beqa{coll cond}\rr'=\frac{p}{1+\ee\cos(\p-{\rm g})}=\frac{\GG^2}{\mm^2\cM\left(1-\sqrt{1-\frac{\GG^2}{\L^2}}\cos{\rm g}\right)}\eeqa
is satisfied, where
 $$p=(1-\ee^2)a=\frac{\GG^2}{\mm^2\cM}$$ is the {\it parameter} of ${\cE}({\tt y}, {\tt x})$, is satisfied. 
Combining~\equ{E0}
and~\equ{coll cond}, we find \beqa{E*}\EE=\mm^2\cM\rr'+\m \mm^2\cM\rr'\hat\EE_1\ ,\qquad |\hat\EE_1|\le 1\ .\eeqa
Therefore, Theorem~A carries the following consequence. It was conjectured by the author in~\cite{pinzari17}.
 \vskip.1in
\noi
{\bf Corollary A}
{\it Under the same assumptions as in  {\rm Theorem~A}, if  $|\EE-\mm^2\cM\rr'|$ is sufficiently grater than $\m \mm^2\cM\rr'$, in the planar three--body problem, collisions between the two minor bodies are excluded over exponentially long times. }

 \vskip.1in
\noi
The proof Theorem~A includes a {\it geometric part} and a {\it analytic part}.
The {\it geometric part}
(developed in Sections~\ref{integration 2CP} and~\ref{2centres3}) aims to find a system a canonical coordinates   such that the function $\HH_{0}$ in~\equ{truncated} depends only on $\rr'$ and two {\it action--coordinates} $\II=(\cL, \cG)$ of a suitable action--angle coordinates set. The  {\it analytic part} (developed in Section~\ref{A weak normal form theory}) is finalized to determine  which extent of time it is true that the actions  $\II=(\cL, \cG)$ remain confined closely to their initial values.

\nl
 The {\it geometric part} starts with the study of the phase portrait of the {\it integral  map} $(\JJ$, $\EE)$. We look  at zones, in the phase space, where the energy level satisfy the assumptions of Liouville--Arnold Theorem. The case $\m=0$ is completely explicit: as $\JJ_0$ depends only on $\L$, while $\EE_0$ depends only on $(\L, \GG, \rr', {\rm g})$, the  full phase portrait, i.e., the
 manifold in the  space of $(\L, \GG, \ell, {\rm g})$ defined by the solutions of
\beqa{level sets***}\JJ_0(\L)={\tt J}\ ,\qquad \EE_0(\L, \GG, {\rm g};\rr')={\tt E}\eeqa
splits as the direct product of two uncoupled portraits: the one in the variables $(\L, \ell)$  being the flat torus ${\mathbb R}\times {\mathbb T}$; the one in the variables $(\GG, {\rm g})$, depending parametrically on $\rr'$ and $\L$. The latter is studied in the case that the ratio $\d:=\frac{\rr'}{a}$ takes values in the interval $(0, 2)$. They  are represented in Figures~\ref{figure12},~\ref{figure5} and~\ref{figure34}, with ${\rm g}$ on  the abscissas'; $\GG$ on  the ordinate's axis. They include one saddle ${\tt P}_0$ and two centres,  ${\tt P}_\pm$, 
given by
$${\tt P}_0=(0,0)\ ,\qquad  {\tt P}_{-}=(\p,0)\ ,\qquad {\tt P}_+=\left(0,\L\sqrt{1-\frac{\d^2}{4}}\right)\ .$$
 Librations around the centers and rotational motions  are delimitated by {\it two separatrices}.
Librations (visible in  Figure~\ref{figure12}, left) actually exist only for $\d\in (0, 1)$ and ${\tt E}_0<{\tt E}<{\tt E}_{\rm max}$, where ${\tt E}_0$, is the value of ${\tt E}$ at the saddle; 
${\tt E}_{\rm Max}$ is the maximum value of $\EE_0$. %A remarkable aspect is that action--angle coordinates $\ \ \hat{\!\!\!\cA}_0=(\cL_0, \cG_0, \l_0, \g_0)$ can be defined in any
%connected component of the phase space excluding the two separatrices, with $\cL_0=\L$ and $\cG_0$ {\it continuous also across a separatrix}.
It is to be remarked, however, that, as $\JJ_0$ is independent of $\GG$ and ${\rm g}$, every point of any  level set in such figures
  is a {\it fixed point} for the dynamics of $\JJ_0$. 
However, when $\m$ is positive this is no longer true and, for sufficiently small $\m$,  it is possible to continue all the level sets~\equ{level sets***}, apart for the ones ``too much close to the separatrices'',  to smooth and compact level sets for $(\JJ, \EE)$. An application of the Liouville--Arnold theorem allows then to define a set of ``mixed'' canonical coordinates, that we denote as  ${\cA}=(\hat\RR', \cL, \cG, \hat\rr', \l, \g)$, with $\hat\rr'=\rr'$,  such that $(\cL, \cG,  \l, \g)$ are ``action--angle'' coordinates to $\JJ(\cdot, \rr')$ (and $\EE(\cdot, \rr')$), for fixed any $\rr'$, while $(\hat\RR', \hat\rr')$ are ``rectangular coordinates''.

\nl 
The {\it analytic part} consists of a ``weak'' (see the comment (iv) below for the meaning we give to such word) normal form result (Theorem~\ref{NFC}) for Hamiltonians of the form  \beqa{HH}&&\HH(\II, \f, y, x)=\hh(\II)+\frac{\o_0(\II)}{2}(x^2+y^2)+f(\II, \f, y, x)\eeqa
where $(\II, \f)\in {\mathbb R}^n\times {\mathbb T}^n$ are $2n$--dimensional ``action--angle coordinates'', while $(y, x)$ are ``rectangular coordinates''. For definiteness, we restrict to the case, of interest in the economy of the 
paper, that the dimension of such rectangular coordinates is 2. Clearly, a more general setting might be explored. To clarify the motivations that led us to study such kind of Hamiltonians,  we add some technical comment on the nature of the problem.

\nl
\begin{itemize}
\item[(i)]  In terms of the coordinates ${\cA}$,  the Hamiltonian ${\tt H}$ in~\equ{3BP} takes the form

\beqa{pert funct}
\HH(\RR', \II,  \rr', \f; \varepsilon, \m)&=&\HH_0(\II,  \rr'; \varepsilon, \m)+ {\rm f}(\RR', \II,  \rr', \f; \varepsilon, \m)\eeqa
where
\beqa{pert funct1}\HH_0(\II,  \rr'; \varepsilon,  \m)&=&-\frac{\mm\cM}{\rr'}+ \hh(\II, \rr'; \varepsilon, \m)\eeqa
corresponds to the term in~\equ{truncated}, while ${\rm f}(\RR', \II,  \rr', \f; \varepsilon, \m)$, corresponds to the $\varepsilon^2$ part of~\equ{3BP} (the exact definition of $f$ is given in Equation~\equ{model} below). Observe that the ``perturbing term'' ${\rm f}$ in~\equ{pert funct} is not periodic with respect to the coordinate $\RR'$, hence standard perturbative techniques (see item (iii) below)  do not apply. 
\\
In Section~\ref{A normal form lemma without small divisors} we prove that it is still possible to discuss normal for theories to Hamiltonians of the form
\beqa{aperiodic}\HH(y, \II, x, \f)=\HH_0(\II, y)+{\rm f}(\II, y, \f, x)\eeqa
where $(\II, \f)$ are ``action--angle'', while $(y, x)$ are ``rectangular'' coordinates, with ${\rm f}$ not periodic with respect to $x$. The assumptions that are needed in order that the theories work look even nicer with respect to the standard case where the couple $(y, x)$ does not appear. As an example, the problem of {\it small divisors} does not exist for such problems -- the quantity $\partial_\II \HH_0(\II, y)$ might also vanish identically. Basically, the only request is that some smallness of $f$ with respect to $\HH_0$ holds. The difficulty in the application of such kind of theories is that, in general, such smallness condition is not ensured for long times, so  the normal form that one obtains risks to be useless. 
 As an example, consider the $\II$--independent case
\beqa{ovl f}
\ovl\HH_0=-\frac{\mm\cM}{\rr'}\ ,\quad \ovl{\rm f}=\frac{\varepsilon^2{\RR'}^2}{2\mm'}+\frac{\varepsilon^2{\Phi'_0}^2}{2\mm'{\rr'}^2}\ .
\eeqa
The Hamiltonian
$$\ovl\HH:=\ovl\HH_0+\ovl{\rm f}=
\frac{\varepsilon^2{\RR'}^2}{2\mm'}+\frac{\varepsilon^2{\Phi_0'}^2}{2\mm'{\rr'}^2}-\frac{\mm\cM}{\rr'}$$
is exactly soluble, since it corresponds to be the two--body problem Hamiltonian, with masses $\mm'\varepsilon^{-2}$, $\cM'\varepsilon^{2}$.
For negative values of the energy $\ovl\cH=\ovl\HH$, the motions of $\ovl\HH$
 are evolve on Keplerian ellipses, with  period  $T=T_0\varepsilon^2$, where $T_0=2\p \frac{{\L'}^3}{{\mm'}^3{\cM'}^2}$ and eccentricity $\ee'=\sqrt{1-\frac{{\Phi_0'}^2}{{\L'}^2}}$, with
 $\ovl\cH=-\frac{{\mm'}^3{\cM'}^2}{2\varepsilon^2{\L'}^2}$  the  energy.
Assume that $\L'=\OO(\varepsilon^{-1})$, so $\ee'=1-\OO(\varepsilon^2)$. Let $t=0$ be the time of aphelion crossing. So, at $t=0$,
\beqa{t=0}\ovl\HH_0=-\frac{{\mm'}^3\cM^2}{\varepsilon^2{\L'}^2(1+\ee')}=\OO(1)\ ,\quad \ovl{\rm f}=\frac{{\mm'}^3\cM^2(1-\ee')}{2\varepsilon^2{\L'}^2(1+\ee')}=\OO(\varepsilon^{2})\ .\eeqa
During each period, at the time when $\RR'$ reaches its maximum, given by $\frac{\mm^2\cM}{{\L'}\varepsilon^{2}}\ee'$,  $\rr'$ takes the value $\frac{\varepsilon^2{\L'}^2}{\mm^2\cM}$. At that  time, 
$\ovl\HH_0$, $\ovl{\rm f}$ are of the same order:
\beqa{comparable}
\ovl\HH_0=-\frac{{\mm'}^3\cM^2}{\varepsilon^2{\L'}^2}\ ,\qquad \ovl{\rm f}=\frac{{\mm'}^3\cM^2}{2\varepsilon^2{\L'}^2}\ .
\eeqa
As a matter of fact, from the exact solution, we know that $\ovl\HH_0$ and $\ovl{\rm f}$ remain bounded as in~\equ{t=0} only for a fraction of the period $T=T_0\varepsilon^2$ (corresponding to an interval around the aphelion crossing), and hence the amount of time that~\equ{t=0}  remain true cannot exceed  $\OO(\varepsilon^2)$. Note than on a circular orbit, i.e., for $\Phi_0=\L'$, relations in \equ{comparable} hold for all $t$.

\item[(ii)] 
The example in the item above above is not so ``exotic'' in the economy of the paper, because it is possible (see Section~\ref{setup} for the details) to split further the function ${\rm f}$  in~\equ{pert funct} as ${\rm f}=\hh'+\tilde{\rm f}$, and hence rewrite $\HH$ as
\beqa{step1}\HH(\hat\RR', \cL, \cG, \hat\rr', \l, \g; \CC, \varepsilon, \m)&=&\hh(\cL, \cG, \hat\rr'; \varepsilon, \m)+\hh'(\hat\RR', \hat\rr', \cG; \CC, \varepsilon, \m)\nonumber\\
&+&\tilde{\rm f}(\hat\RR', \cL, \cG, \hat\rr', \l, \g; \CC, \varepsilon, \m)\eeqa
where $\hh$ is as in \equ{pert funct1},  $\hh'$ is precisely as $\ovl\HH$, with a certain ${\Phi_0'}$, depending on $\cG$ and $\CC$ only, and $\tilde{\rm f}$ is a suitably small term. 
\\ 
By the considerations in (i), we give up any attempt of applying {\it directly} the above mentioned Lemma~\ref{NFL} to the Hamiltonian \equ{pert funct}. Rather, we start from the system written in the form \equ{step1} and look at the expansion of $\hh'$  with respect to the coordinates $(\hat\RR', \hat\rr')$ centered  around its minimum. We recall that the minimum point for $\hh'$ corresponds to circular motions for ${\tt x}'$. The Hamiltonian $\HH$ is carried  to the form~\equ{HH} (see Section~\ref{setup} for the details). 

\item[(iii)] In Section~\ref{A weak normal form theory} we present a normal form result (Theorem~\ref{NFC}) designed around the Hamiltonian $\HH$ in \equ{HH}. The novelty of this theorem with respect to previous similar results is that it holds {\it without  assumptions on $\hh$}. 
We recall, at this respect, the celebrated Nekhorossev's result~\cite{nehorosev77}, remarkably refined by J. P\"oschel~\cite{poschel93} and Guzzo et al. \cite{guzzoCB16}. It  
states that, for  close to be integrable systems of the form
 $${\tt H}(\II, \f)={\tt h}({\tt I})+{\tt f}({\tt I}, \f)\qquad ({\tt I}, \f)\in V\times {\mathbb T}^n\ ,\quad V\subset {\mathbb R}^n$$
the actions ${\tt I}$ remain confined closely to their initial values over exponentially long times 
provided that  the ``unperturbed part'' ${\tt h}({\tt I})$ 
satisfies a transversal condition known as
{\it steepness}. This condition allows, thanks to a  analysis of the {\it geometry of resonances}, to overcome the problem of the so--called {\it small divisors}.  A sufficient condition for steepness -- which is also necessary for systems with 2 degrees of freedom --  is {\it quasi--convexity}.  According to~\cite{poschel93},  ${\tt h}$ is said to be  {\it $l$, $m$ quasi--convex} if, at each point ${\tt I}$ of a neighborhood of $V$, at least one of inequalities
\beqa{quasi convex}| \xi\cdot \big(\partial_{\tt I}{\tt h}({\tt I}) \xi\big)|> l \|\xi\| \quad\quad |\xi\cdot \big(\partial^2_{\tt I}{\tt h}({\tt I}) \xi\big)|\ge m \|\xi\|^2 \eeqa
holds for all $\xi\in {\mathbb R}^n$.
Condition~\equ{quasi convex} has an extension, called {\it three--jet} condition, to
systems with three--degrees of freedom, which one might hope to apply to the Hamiltonian~\equ{HH}.
\\
The main obstacle to the application of Nekhorossev theory to the Hamiltonian~\equ{HH} relies not so much in the linearity (implying not steepness) with respect to $(x^2+y^2)$ (which could, with some work, be overcome) but, rather, in the fact that the the function $\hh(\II)$ in~\equ{HH} that arises from the application verifies~\equ{quasi convex}, with $m$ of order $\varepsilon^2$, too small compared to $f$, which cannot be smaller than $\varepsilon^2$.

\item[(iv)] The proof of Theorem~\ref{NFC} uses  the Lemma~\ref{NFL}, mentioned in (i), where the absence of  {\it small denominators} allows to avoid the {\it geometry of resonances}. The thesis of Theorem~\ref{NFC} is  ``weaker'' compared to standard results in~\cite{nehorosev77, poschel93, guzzoCB16}, because the domain in the coordinates $(y, x)$ in~\equ{HH} where the normal form is achieved is an {\it annulus} around the origin, rather than a neighborhood of it. The physical meaning of this assumption, in the use we do of Theorem~\ref{NFC} in the paper,  is that the eccentricity of the orbits of ${\tt x}'$ has to be disclosed from $0$ -- compare the comment in (i) at this respect. 
\end{itemize}

\vskip.1in
\noi
We conclude this introduction with a brief  overview of papers addressing problems related to the  paper.\\
 As mentioned, Euler solved the two--centre problem. He
showed that, adopting a well--suited system of canonical coordinates  usually referred to {\it elliptic}  or {\it ellipsoidal} (see~\cite{bekovO78} for a review, or Appendix~\ref{2centres} for a brief account), the Hamilton--Jacobi equations of the two--centre problem separates in two independent equations, each depending on one degree of freedom only. This separation gives rise to the Euler integral, showing only {\it integrability by quadratures}. The  two--centre problem received a renewed attention only recently. In the early 2000's, Waalkens, Richter and Dullin~\cite{waalkensDR06} studied monodromy properties of the problem and raised for the first time the question of the existence of action--angle coordinates. Their starting point was the Hamiltonian written in Cartesian coordinates, combined with a Levi--Civita regularization, made possible by the separability of the Hamiltonian. Their point of view is quite different from the one used in the paper, due mainly to the fact that the regularization in~\cite{waalkensDR06} carries to fix a energy level at time. Ten years later, Dullin and Montgomery faced the study of syzygies in the two--centres problem. Very recently, Biscani and Izzo produced an explicit solution for the spatial problem~\cite{biscaniI16}. 
On the side of normal form theory with small divisors problem, much has been written. We refer to
\cite{llave01, giorgilliLS09, poschel93, guzzoCB16} and references therein for notices. The attention, in Hamiltonian mechanics, to normal forms to systems where also non--periodic coordinates appear is pretty recent. 
Fortunati and Wiggins~\cite{fortunatiW16} proved  a normal form result for an Hamiltonian with a--periodic coordinates,  under the 
assumption that the perturbing term has an exponential decay with respect to the coordinate $x$. Such assumption allows to overcome the difficulties mentioned in (i). The theory in~\cite{fortunatiW16} is clearly not applicable to our setting (where $f$ increases quadratically with $x$), so  Lemma~\ref{NFL} below may be regarded as a  variation of their result, without such decay assumption.

\newpage\section{A weak normal form theory}\label{A weak normal form theory}
In this section, we present a normal form theory for the Hamiltonian $\HH$ in~\equ{HH}. 
To motivate  the result, we begin with some quantitative considerations.

\nl
Let $\cI\subset {\mathbb R}^n$ open and connected, $0<\d<\D$; let $$\cA_{\d, \D}:=\big\{(x, y)\in {\mathbb R}^2:\ \d^2<\frac{x^2+y^2}{2}<\D^2\big\}$$ and put 
 \beqa{MdD}\cM:=\cI\times {\mathbb T}^n\times \cA_{\d, \D}%=\bigcup_{j=1}^p {\cal U}_j
  \eeqa
Let $0<\epsilon_0<1$ be so small a number, compared to the diameter of $\cI$, $\d$ and $\D$, that the sets $\cI_1\subset\cI$, $\cA_1\subset \cA_{\d, \D}$ defined as
\beqano
&&\cI_1:=\{\II\in \cI:\ B^n_{\varepsilon_0\r}\subset \cI\}\nonumber\\
&& \cA_1:=\left\{(y, x):\ \d^2(1+\epsilon_0)<\frac{x^2+y^2}{2}<\D^2(1-\epsilon_0)\right\}
\eeqano
are  not empty. Consider the sub--manifold of $\cM$
\beqa{M1}\cM_1:=\cI_1\times {\mathbb T}^n\times \cA_1\ .\eeqa
The question we aim to give an answer is
 which is the amount of time such that forward or backward orbits generated by the Hamiltonian $\HH$ in~\equ{HH} with initial data in $\cM_1$   do not leave $\cM$ for all $0\le t\le T$. This amounts to ask which is the maximum $T>0$ such that  \beqa{thesis***}|\II(\pm T)-\II(0)|\le \epsilon_0\r\ ,\qquad |\JJ(\pm T)-\JJ(0)|\le \epsilon_0\d^2\ .\eeqa 
  
 \nl
Let us look, to fix ideas, to forward orbits. Cauchy inequalities  show that, if \beqa{T00}T\le \epsilon_0\frac{\r s}{\EE}\eeqa then, \beqa{DeltaI***}|{\tt D}\II| \le \frac{\EE T}{s}\le\epsilon_0\r\ .\eeqa To evaluate $|{\tt D}\JJ|$, we use an energy conservation argument. From

\beqa{dJ0}0={\tt D}\HH={\tt D}\hh+\frac{\o_0(\II(0))+{\tt D}\o_0}{2}{\tt D}\JJ+\frac{{\tt D}\o_0}{2}\JJ(0)+{\tt D} f\eeqa
and
$$|{\tt D}\hh|\le M|{\tt D}\II|\ ,\quad |\o_0(\II(0))|\ge a\ ,\quad  |\JJ(0)|\le \D^2\ ,\quad |{\tt D} f|\le 2\EE$$
and, as soon as
\beqa{T1}T\le \frac{a s}{2\EE M_0'}\eeqa
we have
$$|{\tt D}\o_0|\le M_0' |{\tt D}\II |\le M_0'  \frac{\EE T}{s}\le \frac{a}{2}\ .$$
We find, using also the bound for $|{\tt D}\II|$ in~\equ{DeltaI***},
\beqa{dJ}\frac{a}{4}|{\tt D}\JJ|&\le& \left|
\frac{\o_0(\II(0))+{\tt D}\o_0}{2}
\right||{\tt D}\JJ|\le M|{\tt D}\II|+\frac{M_0'\D^2}{2}|{\tt D}\II|+2\EE\nonumber\\
&=&\left(M+\frac{M_0'\D^2}{2}\right)|{\tt D}\II|+2\EE\nonumber\\
&\le&\left(M+\frac{M_0'\D^2}{2}\right)
 \frac{\EE T}{s}
+2\EE
\eeqa
We obtain $|{\tt D}\JJ|\le \epsilon_0\d^2$ provided that \beqa{4E}\frac{16\EE}{a\d^2}\le \epsilon_0\quad {\rm and}\quad T\le s \frac{\epsilon_0}{8\EE}\left(\frac{M}{a\d^2}+\frac{M_0'\D^2}{2a{\d^2}}\right)^{-1} \eeqa
Collecting the previous bounds, the a--priori stability time can be taken to be
\beqa{T0}T=T_0:= s \min\left\{\frac{\r\epsilon_0}{\EE},\ \frac{\epsilon_0}{8\EE}\left(\frac{M}{a\d^2}+\frac{M_0'\D^2}{2a{\d^2}}\right)^{-1},\  \frac{a}{2\EE M_0'}\right\}\eeqa
provided that also the first condition in~\equ{T0} is met.
Theorem~\ref{NFC} below is, in a sense, an improvement of the ``a--priori bound'' in~\equ{T0}.

\nl
In order to state it, we need to fix the following notation, after~\cite{poschel93}. For given a holomorphic function $$f:\ (\II, \f, y, x)\in {\cal I}_\r\times {\mathbb T}_s^n\times B^2_\d\to \complex$$
where ${\cal I}\subset {\mathbb R}^n$, is  open and connected, while $B^2_\d$ is the complex two--dimensional ball with radius $\d$ centered at the origin, and, as usual, for a given set $A$ in a metric space, we denote $A_\theta:=\cup_{x_0\in A}\{B_\theta(x_0)\}$, while ${\mathbb T}_s:={\mathbb T}+\ii [-s,s]$, with ${\mathbb T}:={\mathbb R}/(2\p\integer)$ the standard torus,
we define
$$\|f\|_{r, s, \d%,\d
}:=\sum_{k%,h,j
}\|f_{k%hj
}\|_{\r, \d}e^{s|k|}%\d^{h+j}
$$
where $f_{k%hj
}(y,\II, x)$ are the coefficients of the Taylor--Fourier expansion
$$f=\sum_{k}f_{k}(\II, y, x)e^{\ii k \cdot\f}\ ,$$
while
$\|f_{k%hj
}\|_{\r, \d}:=\sup_{(\II, y, x)\in\cI_\r\times B^2_\d}|f_k(\II, y, x)|$.

\begin{theorem}\label{NFC} For some positive number $p_*$ the following holds. Let $\cI\subset {\mathbb R}^n$ open and connected, $0<\d<\D$, $\epsilon_0>0$ small; $\cM_1$ as in~\equ{M1}.
Let \beqa{domain}(\II, \f, y, x)\in \cI_\r\times {\mathbb T}_s^n\times B^2_{\D+\d}\to \HH(\II, \f, y, x)\eeqa be a holomorphic function of the form~\equ{HH}.
Let $0<c<\d+\D$; $\EE:=\|f\|_{\r, s, \d+\D}$. Let
$$\o_1:=\partial_\II\left(\hh+\frac{\o_0(\II)}{2}y^2\right)$$
and put

\beqa{IN}
&&M_0:=\sup\|\o_0\|\ ,\quad M_1:=\sup\left\|\o_1\right\|\ ,\quad a:=\inf |\o_0|\nonumber\\
&&M:=\sup |\partial_\II\hh|\ ,\quad M_0':=\sup|\partial_\II\o_0|\ ,\quad c=\frac{4\r s}{\d}\nonumber\\
&&\epsilon=32 p_* \max \left\{\frac{16 \r sM_0}{a\d^2}\frac{\D}{\d}\ ,\ \frac{2\EE}{a{\r s}}\frac{\D}{\d}\ ,\  
 \frac{2\r  M_1}{p_*  a}\frac{1}{\d^2}
 \right\}\nonumber\\
&&\epsilon':=\left(\frac{4M}{a\d^2}+\frac{2M_0'}{a}\frac{\D^2}{\d^2}\right)\epsilon\r+\frac{8\EE}{a\d^2}\ .
  \eeqa
Assume that the following inequalities are satisfied:
\beqa{assump3}
2\frac{M_0'}{a}\epsilon_0\r\le 1
\eeqa
 \beqa{simplify}\frac{4\r s}{\d( \D+\d)}\le1\ ,\quad \frac{{\tt c}_n\r s}{2p_*\D\d}\le 1\eeqa
 and
 \beqa{eps}\epsilon\le\frac{\epsilon_0}{2}\ ,\quad \epsilon'\le\frac{\epsilon_0}{2}\ ,\quad T\le T_1 2^{\left[
 \frac{1}{\epsilon}
 \right]}\eeqa
where
\beqa{T1}T_1:=\frac{s\epsilon_0}{2}\min\left\{\r,\ \left(\frac{4M}{a\d^2}+\frac{2M_0'\D^2}{a\d^2}\right)^{-1} \right\}\left(
\frac{M_0}{2}c^2+\EE
\right)^{-1}\eeqa
Then any solution
$t\in [-T,T]\to \g(t)=(x(t), y(t), \II(t), \f(t))$  of $\HH$ such that $\g(0)\in \cM_1$ verifies~\equ{thesis***}.
\end{theorem}

\nl
The proof of Theorem~\ref{NFC} is based on the following result, where we use the (standard) notation
   $$\ovl f(\II, y, x):=f_0(\II, y, x)=\frac{1}{(2\p)^n}\int_{{\mathbb T}^n}f(\II, \f, y, x)d\f\ . $$

\begin{proposition}\label{NFP}
Let $N\in {\mathbb N}$, $\r$, $s$, $\D>\d>0$, $\cI\subset {\mathbb R}^n$ open and connected. Let $\HH$, $a$, $M_0$, $M_1$ be as in Theorem~\ref{NFC}. There exists a pure number $p_*$  and ${\tt c}_n$ depending only on $n$ such that, for all  ${\tt c}_*\ge {\tt c}_n$, all $0<c<\d+\D$ such that
\beqa{first cond}8Nc\frac{M_1}{a\d}<s\eeqa
and
\beqa{nonstandardsmallness} 
&&{\tt c}_*\left(% \|\frac{1}{\o_y}\|_{r,\r} 
 {\frac{c^3M_0}{a\d\r s} }+\frac{2{c}\EE}{a{\d\r s}} \right)N<1\quad {\rm and}\quad {\tt c}_*\left(% \|\frac{1}{\o_y}\|_{r,\r} 
 {4\frac{c^2M_0}{a\d^2} }+\frac{8\EE}{a{\d^2}} \right)N<1 \eeqa
 where
$\EE:=\|f\|_{\r, s, \D+\d}$, the following holds. %for any positive number $c$ satisfying
%$$c<4\frac{\r s}{\d}\ ,\quad \frac{|\o_0|}{2}c^2\le \EE\ .$$
Let
\beqano
&&\cR^\ppu_{\d/2, c/2}:=\big\{y\in {\mathbb R}:\quad \frac{\d}{2}<  |y|< \D+\frac{\d}{2}\big\}_{\d/2}\times \big\{x\in {\mathbb R}:\quad |x|< \frac{c}{2}\big\}_{c/2}\nonumber\\ 
&&\cR^\ppj_{\d/2, c/2}:=R_{(j-1)\theta_0}\cR^\ppu_{\d/2, c/2}\ ,%\quad (\cU_j)_{\d/2, c/2, \r, s}:=\cR^\ppj_{\d/2, c/2}\times \cI_\r\times {\mathbb T}^n_s\nonumber\\
\eeqano
where $R_{\theta}\in SO(2)$ denotes the the $2\times 2$ matrix corresponding to a rotation by $\theta$ in the plane. Then for each $p\in {\mathbb N}$, any $\theta_0\in {\mathbb T}$, any
$j=1,\ \cdots, p$, 
it is possible to find a real--analytic canonical transformation
\beqano
\phi_j:\quad\quad (\cU_j)_{\r/3, s/3, \d/6, c/6}&\to&\quad (\cU_j)_{\r, s, \d/2, c/2}\nonumber\\
(\tilde \II_j, \tilde \f_j, \tilde y_j,\tilde  x_j)&\to& (\II, \f, y, x)
\eeqano
where $(\cU_j)_{\r, s, \d, c}:=\cI_\r\times {\mathbb T}^n_s\times \cR^\ppj_{\d, c}$,
which carries $\HH$ to a function of the form
\beqa{Hi*}
\HH_j&:=&\HH\circ\phi_j=\hh(\tilde\II_j)+\frac{\o_0(\tilde\II_j)}{2}(\tilde x_j^2+\tilde y_j^2)+\ovl f(\tilde \II_j, \tilde y_j, \tilde x_j)+g_{j}(\tilde \II_j, \tilde y_j, \tilde x_j)\nonumber\\
&+&f_{j, *}(\tilde \II_j, \tilde \f_j, \tilde y_j, \tilde x_j)\eeqa
where
\beqa{Hi**}
\|g_j\|_{\r/3, s/3, \d/6, c/6}&\le&\max\left\{
{\tt c}_n\left(% \|\frac{1}{\o_y}\|_{r,\r} 
 {\frac{c^3M_0}{a\d\r s} }+\frac{2{c}\EE}{a{\d\r s}} \right)\ ,\
 {\tt c}_n\left(% \|\frac{1}{\o_y}\|_{r,\r} 
 {4\frac{c^2M_0}{a\d^2} }+\frac{8\EE}{a{\d^2}} \right)
 \right\}\nonumber\\
&&{\textrm{\tiny$\times$}}\left(\frac{M_0}{2} c^2+\EE\right) \nonumber\\
 \|f_{j,*}\|_{\r/3, s/3, \d/6, c/6}&\le& \left(\frac{M_0}{2} c^2+\EE\right)2^{-N}\ .
\eeqa
and such that the following bounds hold, for all $j=1$, $\cdots$, $p$:
\beqa{transformation estimates}
&&\max\Big\{\frac{|\tilde\II_j- \II|}{\r},\ \frac{|\tilde\varphi_j-\varphi|}{s},\ 
\frac{2 |\tilde y_j- y|}{\d},\ \frac{2 |\tilde x_j- x|}{c} \Big\} \le %\nonumber\\
%&\leq& 
%4\max\left\{\left(% \|\frac{1}{\o_y}\|_{r,\r} 
% {\frac{c^3M_0}{a\d\r s} }+\frac{2{c}\EE}{a{\d\r s}} \right),\ \left(% \|\frac{1}{\o_y}\|_{r,\r} 
% {4\frac{c^2M_0}{a\d^2} }+\frac{8\EE}{a{\d^2}} \right)\right\}\nonumber\\
% &\le& 
\frac{4}{{\tt c}_*N}
\ .
\eeqa
In particular, for 
\beqa{p*}p:=\left[\frac{2\p}{\theta_0}\right]+1\ ,\quad \theta_0:=\tan^{-1}\frac{c}{2\D}\eeqa
 the collection of  $\{\cU_j, \phi_j\}_{j=1, \cdots, p}$ is an atlantis for the manifold $\cM$ in~\equ{MdD}.
%\beqa{second smallness}\frac{{\tt c}_{n}}{|\o_0|\d^2 c} N% \|\frac{1}{\o_y}\|_{r,\r} 
%\| f\|_{\r, s, \D+\d} <1\ .\eeqa
%Then the same conclusions hold, with $\ovl f$ replaced by $\ovl{\ovl f}$, $g_i$ depending on $(\tilde y_i, \tilde x_i)$ only via $\frac{\tilde y_i^2+\tilde x_i^2}{2}$, $\d^2$ replaced by $\d^2 c$ in~\equ{Hi**} and~\equ{transformation estimates}.
\end{proposition}

\nl
The proof of Proposition~\ref{NFP} is deferred to  the next Section~\ref{proof of NFP}. Here we prove how Theorem~\ref{NFC} follows from it.

\proof  {\bf of Theorem~\ref{NFC}} 
To fix ideas, we prove the theorem for forward orbits, since the backward case is specular.
We prove that, for any $0<c<\d+\D$, ${\tt c}_*\ge {\tt c}_n$ %such that condition
%\beqa{assump3}2\frac{M_0'}{a}\left(\ovl\epsilon\r+\frac{T}{s} \left(\frac{M_0}{2} c^2+\EE\right)2^{-\ovl N}\right)\le 1
%\eeqa
% holds, then 
the following inequality hold
\beqa{DeltaI}
&&|\II(T)-\II(0)|\le \ovl\epsilon\r+\frac{T}{s} \left(\frac{M_0}{2} c^2+\EE\right)2^{-\ovl N}%\nonumber\\
%&&|\JJ(T)-\JJ(0)|\le \ovl \epsilon'\d^2+\frac{T}{s} \left(\frac{4M}{a}+\frac{2M_0'\D^2}{a}\right)
%\left(
%\frac{M_0}{2}c^2+\EE
%\right)2^{-\ovl N}
\eeqa
with 
\beqa{NN}
&&\ovl\epsilon:=32 p_* \max \left\{
 \frac{c^2M_0}{a\r s}\frac{\D}{\d}\ ,\ \frac{4cM_0}{a\d}\frac{\D}{\d}\ ,\ \frac{2\EE}{a{\r s}}\frac{\D}{\d}\ ,\ \frac{8\EE}{a{\d c}}\frac{\D}{\d}\ ,\  \frac{4  M_1}{{\tt c}_*  s a}\frac{\D}{\d}
 \right\}\nonumber\\
%&& \ovl\epsilon':=\left(\frac{4M}{a\d^2}+\frac{2M_0'}{a}\frac{\D^2}{\d^2}\right)\ovl\epsilon\r+\frac{8\EE}{a\d^2}\nonumber\\
  &&\ovl N:=%\frac{8\r p_*}{{\tt c}_*\partial\II}\frac{\D}{c}=
\left[\min\left\{
 \frac{1}{4{\tt c}_*} \frac{a\d\r s}{M_0c^3}\ ,\  \frac{1}{16{\tt c}_*}\frac{a\d^2}{M_0c^2}\ ,\  \frac{1}{4{\tt c}_*}\frac{a{\d\r s}}{2{c}\EE}\ ,\  \frac{1}{32{\tt c}_*}\frac{a{\d^2}}{\EE}\ ,\  \frac{\d s}{16 c}\frac{a}{M_1} \right\}\right]\ .\nonumber\\
 \eeqa

 \nl
 The proof of%the former inequality in
~\equ{DeltaI} is based on a patchwork application of  Proposition~\ref{NFP}, made possible by the annular symmetry of the domain. %; the latter will be obtained via energy conservation argument. We start with the former.

 \nl
Let $\ovl N$ be as in~\equ{NN}.
 Then we find
 $$\frac{8\ovl Nc}{\d}\frac{M_1}{a}\le \frac{s}{2}$$
 \beqano 
{\tt c}_*\left(% \|\frac{1}{\o_y}\|_{r,\r} 
 {\frac{c^3M_0}{a\d\r s} }+\frac{2{c}\EE}{a{\d\r s}} \right)\ovl N\le\su4+\su4=\su2<1\eeqano

\beqano 
{\tt c}_*\left(% \|\frac{1}{\o_y}\|_{r,\r} 
 {4\frac{c^2M_0}{a\d^2} }+\frac{8\EE}{a{\d^2}} \right)\ovl N\le \su4+\su4=\su2<1 \eeqano
 Then the assumptions of Proposition~\ref{NFP} are verified, and we find number $p\le p_*\frac{\D}{c}$, a finite collection of open sets  $\{\cU_j\}_{j=1, \cdots, p}\subset \cM$, with $\bigcup_{j=1}^p\cU_j=\cM$ and real--analytic, symplectic maps
 $$\phi_j:\quad \cU_j\to \cU_j$$
 such that
\beqa{estimates1}
\HH_j:=\HH\circ\phi_j=\frac{\o_0}{2}(\tilde x_j^2+\tilde y_j^2)+\ovl f(\tilde y_j, \tilde x_j, \tilde \II_j)+g_{j}(\tilde y_j, \tilde x_j, \tilde \II_j)+f_{j, *}(\tilde y_j, \tilde x_j, \tilde \II_j, \tilde \f_j)\eeqa
where
\beqa{estimates2}
%\sup_{{\cal U}_j}|g_j|\le {\tt c}_*\frac{\EE^2 }{a\d^2}\ ,\qquad
 \sup_{{\cal U}_j}|f_{j, *}|\le \left(\frac{M_0}{2} c^2+\EE\right)2^{-\ovl N}\ .
\eeqa
with
\beqa{estimates}
\max&&\Big\{\frac{|\tilde\II_j- \II|}{\r},\ \frac{|\tilde\varphi_j-\varphi|}{s},\ 
\frac{2 |\tilde y_j- y|}{\d},\ \frac{2 |\tilde x_j- x|}{c} \Big\}\le \frac{4}{{\tt c}_*\ovl N}\nonumber\\
&=& 16 \max \left\{
 \frac{c^3M_0}{a\d\r s}\ ,\ \frac{4c^2M_0}{a\d^2}\ ,\ \frac{2{c}\EE}{a{\d\r s}}\ ,\ \frac{8\EE}{a{\d^2}} \ ,\  \frac{4 c M_1}{{\tt c}_* \d s a}
 \right\}
\ .
\eeqa

 \nl
Let now $t\in [0,T]\to \g(t)=(x(t), y(t), \II(t), \f(t))$ be a curve in $\cM$. Fix times $t_0:=0$ $\le t_1\le $ $\cdots$ $\le t_{k+1}:=T$
and $\cU^\ppo$, $\cdots$, $\cU^\ppk$, with $\cU^\ppi\in \{\cU_1, \cdots, \cU_p\}$
such in a way that $\forall\ i=0$, $\cdots$, $k$, $\g(t)\in \cU^\ppi$ for all $t\in [t_i, t_{i+1}]$. \\
For $i=1$, $\cdots$, $k$, denote as $\tilde\g_i(t):=(\tilde x_i(t), \tilde y_i(t), \tilde\II_i(t), \tilde\f_i(t))$ the curve 
$$\tilde\g_i(t):\ [t_i, t_{i+1}]\to \cU^\ppi$$  defined as
$$ \tilde\g_i(t):=\tilde\phi_j^{-1}\circ(\g|_{[t_i, t_{i+1}]})(t)\quad {\rm if}\quad \cU^\ppi=\cU_j\ .$$

\nl
Define, inductively, two finite sequences\footnote{The estimate of the difference $|\II(T)-\II(0)|$ unavoidably passes through the estimate of the terms $$\left|\sum_{i=1}^{k+1}\II(t_i)-\tilde\II_{j(i)}(t_i)\right|$$
where $j(i)$ is defined so that $\cU_{j(i)}=\cU^\ppi$. A inaccurate evaluation of this summand, based  on~\equ{transformation estimates} and the triangular inequality would lead to $(k+2)\frac{4\r}{{\tt c}_*\ovl N}$. This bound would be, however, of no help, since,
during the time $T$, the curve $t\to\g(t)$ might visit each $\cU_j$ many times, so nothing excludes $k\to \infty$ very fast. This justifies the construction below, where we sample the sets $\{\cU_{j_m}\}_{m=0, \cdots, q+1}$,  according to the times of last visit, rather than according
to all their visits $\{\cU_{j(i)}\}_{i=0, \cdots, k+1}$. This gives a much better evaluation, because now   $q\sim p\ll k$.}
\beqano
&&{i_0}\ ,\quad {i_1}\ ,\quad \cdots\ ,\quad {i_{q}}\ ,\quad {i_{q+1}}\in  \{0, \cdots, k+1\}\qquad \nonumber\\
&&{j_0}\ ,\quad {j_1}\ ,\quad \cdots\ ,\quad {j_{q}}\in \{1\ ,\cdots\ , p\}
\eeqano
 via the following relations:
$$i_0=0\ ,\qquad  \cU^\ppo=\cU_{j_0}$$
and, given
  $${i_m}\ ,\qquad {j_m}\ ,$$
if $0\le i_m<k$,  define $i_{m+1}$, $j_{m+1}$ via the relations
$${i_{m+1}}:=\max\{ i\in \{1, \cdots, k\}:\ \cU^\ppi=\cU_{j_m}\}+1 \qquad \cU^{(i_{m+1})}=\cU_{j_{m+1}}\ .$$
If $i_m=k$, put
$$m=q\ ,\quad i_{q+1}:=i_q+1=k+1$$
  By construction, $q+1\in \{1, \cdots, p\}$ and
  $$0={i_0}< {i_1}< \cdots< {i_q}<i_{q+1}={k+1}\quad \Longrightarrow\quad t_{i_{q+1}}=t_{{k+1}}=T$$
Then we have, by the triangular inequality,
\beqa{telescopic}
|\II(T)-\II(0)|&=&|\II(t_{i_{q+1}})-\II(t_{i_0})|\le \sum_{m=0}^{q}|\II(t_{i_{m+1}})-\II(t_{i_m})|\nonumber\\
&\le& \sum_{m=0}^{q}\big(|\II(t_{i_{m+1}})-\tilde\II_{i_{m+1}}(t_{i_{m+1}})| +|\tilde\II_{i_{m+1}}(t_{i_{m+1}})-\tilde \II_{i_m}(t_{i_m})| \nonumber\\
&&+|\tilde \II_{i_m}(t_{i_m})-\II(t_{i_m})|\big)
\eeqa

\nl
But, by Equations~\equ{estimates1}--\equ{estimates2} and Hamilton equations,
 
\beqa{tele1}
\sum_{m=0}^q
|\tilde\II_{i_{m+1}}(t_{i_{m+1}})-\tilde \II_{i_m}(t_{i_m})|&\le& \sum_{m=0}^q\sum_{i=i_m}^{i_{m+1}-1}|\tilde\II_{i}(t_{i})-\tilde\II_{{i+1}}(t_{i+1})|\nonumber\\
&\le& \sum_{m=0}^q(t_{i_{m+1}}-t_{i_{m}})\frac{\EE 2^{-\ovl N}}{s}=T\frac{\EE 2^{-\ovl N}}{s}\eeqa
and, by Equation~\equ{estimates},

 \beqa{tele2}&&\sum_{m=0}^q\big( |\II(t_{i_{m+1}})-\tilde\II_{i_{m+1}}(t_{i_{m+1}})|+|\tilde \II_{i_m}(t_{i_m})-\II(t_{i_m})|\big)\nonumber\\
 &&\le 32(q+1)\r  \max \left\{
 \frac{c^3M_0}{a\d\r s}\ ,\ \frac{4c^2}{\d^2}\ ,\ \frac{2{c}\EE}{a{\d\r s}}\ ,\ \frac{8\EE}{a{\d^2}}\ ,\  \frac{4 c M_1}{{\tt c}_* \d s a}
 \right\}\nonumber\\
 &&\le 32\r p_* \max \left\{
 \frac{c^3M_0}{a\d\r s}\ ,\ \frac{4c^2M_0}{a\d^2}\ ,\ \frac{2{c}\EE}{a{\d\r s}}\ ,\ \frac{8\EE}{a{\d^2}}\ ,\  \frac{4 c M_1}{{\tt c}_* \d s a}
 \right\}\frac{\D}{c}\nonumber\\
 &&= 32\r p_* \max \left\{
 \frac{c^2M_0}{a\r s}\frac{\D}{\d}\ ,\ \frac{4cM_0}{a\d}\frac{\D}{\d}\ ,\ \frac{2\EE}{a{\r s}}\frac{\D}{\d}\ ,\ \frac{8\EE}{a{\d c}}\frac{\D}{\d}\ ,\  \frac{4  M_1}{{\tt c}_*  s a}\frac{\D}{\d}
 \right\}\nonumber\\
 &&=\ovl\epsilon\r
 \eeqa

\nl
 having used
 $$q+1\le p\le p_*\frac{\D}{c}\ .$$
 
 \nl
Collecting~\equ{tele1} and~\equ{tele2}  into~\equ{telescopic},  we have proved the former inequality in~\equ{DeltaI}.

\nl
We now conclude the proof of the theorem. Due to~\equ{simplify}, we can choose
 $$c=\frac{4\r s}{\d}\le \D+\d\ ,\quad {\tt c}_*= 2 p_*\frac{\D\d}{\r s}\ge {\tt c}_n\ .$$
With these values, we have
$$\ovl\epsilon=\epsilon\ ,\quad %\ovl\epsilon'=\epsilon'\ ,\quad 
\ovl N=N:=\left[
 \frac{1}{\epsilon}
 \right]$$
where $\epsilon$, $\epsilon'$ and $N$ are as in~\equ{IN}. So, by~\equ{DeltaI} and~\equ{eps},
\beqa{DeltaInew}
&&|\II(T)-\II(0)|\le \epsilon\r+\frac{T}{s} \left(\frac{M_0}{2} c^2+\EE\right)2^{- N}\le \epsilon_0\r%\nonumber\\
%&&|\JJ(T)-\JJ(0)|\le  \epsilon'\d^2+\frac{T}{s} \left(\frac{4M}{a}+\frac{2M_0'\D^2}{a}\right)
%\left(
%\frac{M_0}{2}c^2+\EE
%\right)2^{- N}
\eeqa
To bound $|{\tt D}\JJ|$, we use an energy conservation argument analogue to in~\equ{dJ0}--\equ{dJ}, but replacing~\equ{4E} with (by
\equ{DeltaInew} and~\equ{assump3})
$$|{\tt D}\o_0|\le M_0' |{\tt D}\II |\le M_0'\r\epsilon_0 \le \frac{a}{2}\ ,$$
we find, using also the bound for $|{\tt D}\II|$ in~\equ{DeltaInew},
\beqano\frac{a}{4}|{\tt D}\JJ|&\le& \left|
\frac{\o_0(\II(0))+{\tt D}\o_0}{2}
\right||{\tt D}\JJ|\le M|{\tt D}\II|+\frac{M_0'\D^2}{2}|{\tt D}\II|+2\EE\nonumber\\
&=&\left(M+\frac{M_0'\D^2}{2}\right)|{\tt D}\II|+2\EE\nonumber\\
&\le&\left(M+\frac{M_0'\D^2}{2}\right)
\left(\epsilon\r+\frac{T}{s} \left(\frac{M_0}{2} c^2+\EE\right)2^{- N}\right)
+2\EE
\eeqano
which we rewrite as
$$|{\tt D}\JJ|\le  \epsilon'\d^2+\frac{T}{s} \left(\frac{4M}{a}+\frac{2M_0'\D^2}{a}\right)
\left(
\frac{M_0}{2}c^2+\EE
\right)2^{- N}$$
where $\epsilon'$ is as in~\equ{IN}. Under conditions~\equ{eps}, the second inequality in~\equ{thesis***} immediately follows.
$\quad\square$

 \subsection{A normal form lemma without small divisors}\label{A normal form lemma without small divisors}

   The proof of Proposition~\ref{NFP} is based on a normal form lemma with a--periodic coordinates, which here we aim to state.
  
  \nl
 We consider an abstract Hamiltonian of the form
 \beq{full system}\HH(y,\II,p, x,\f,q)=\hh_0(y,\II,\JJ(p,q))+f_0(y,\II,p, x,\f,q)\eeq
where
%$$\hh_0(y,\II,\JJ)=\JJ(y,\II)+{\tt P}(y,\II,\JJ)\ ,\qquad {\tt P}(y,\II,0)\equiv0$$
%and
$$\JJ(p,q)=(p_1q_1,\cdots, p_mq_m)\ .$$
and, if
$$\cP_{r,\r,\xi,s,\d}=\Upsilon_r\times \cI_\r \times \Xi_\xi\times {\mathbb T}^n_s\times B^{2m}_{\d}\ ,$$
of $\cP$ where, as usual, we assume that  $\HH$ is holomorphic in $\cP_{r,\r,\xi,s,\d}$.

%\nl
\nl
We denote as ${\cal O}_{r,\r,\xi,s,\d}$ the set of complex holomorphic functions $\phi:\ \cP_{\hat r,\hat \r,\hat \xi,\hat s,\hat d}\to \complex$ for some $\hat r>r$, $\hat\r>\r$, $\hat\xi>\xi$, $\hat s>s$, $\hat\d>\d$, equipped with 
 the norm
$$\|\phi\|_{r,\r,\xi,s,\d}:=\sum_{k,h,j}\|\phi_{khj}\|_{r,\r,\xi}e^{s|k|}\d^{h+j}$$
where $\phi_{khj}(y,\II, x)$ are the coefficients of the Taylor--Fourier expansion
$$\phi=\sum_{k,h,j}\phi_{khj}(y,\II,x)e^{\ii k s}p^h q^j\ .$$
and $\|\phi_{khj}\|_{r,\r,\xi}:=\sup_{\Upsilon_r\times \cI_\r\times \Xi_\xi}|\phi_{khj}|$. Observe that  $\|g_{khj}\|_{r,\r,\xi}$ is well defined because of the boundedness of $\Upsilon$, $\cI$ and $\Xi$, while $\|\phi\|_{r,\r,\xi,s,\d}$ is well defined by the usual properties of  holomorphic functions.

%\nl
%For a given vector--valued  function $\underline\phi=(\phi_1,\cdots, \phi_k)\in {\cal O}_{r,\r,\xi,s,\d}^k$, we let
%$$\|\underline\phi\|_{r,\r,\xi,s,\d}:=\sum_{i=1}^k \|\phi_i\|_{r,\r,\xi,s,\d}\ .$$

\nl
If $\phi\in {\cal O}_{r,\r,\xi,s,\d}$, we define its ``off--average''  and   ``average''  as
$$\widetilde\phi:=\sum_{k,h,j:\atop (k,h-j)\ne (0,0)}g_{khj}(y,\II, x)e^{\ii k s}p^h q^j\ ,\qquad
\ovl\phi:=\phi-\widetilde\phi
\ .$$
We decompose
$${\cal O}_{r,\r,\xi,s,\d}={\cal Z}_{r,\r,\xi,s,\d}\oplus {\cal N}_{r,\r,\xi,s,\d}\ .$$
where ${\cal Z}_{r,\r,\xi,s,\d}$, ${\cal N}_{r,\r,\xi,s,\d}$
are the ``zero--average'' and the the ``normal'' classes
\beqa{zero average}&&{\cal Z}_{r,\r,\xi,s,\d}:=\{\phi\in {\cal O}_{r,\r,\xi,s,\d}:\quad \phi=\widetilde\phi\}=\{\phi\in {\cal O}_{r,\r,\xi,s,\d}:\quad \ovl\phi=0\}\\
\label{phi independent}&&{\cal N}_{r,\r,\xi,s,\d}:=\{\phi\in {\cal O}_{r,\r,\xi,s,\d}:\quad\phi=\ovl\phi\}=\{\phi\in {\cal O}_{r,\r,\xi,s,\d}:\quad \widetilde\phi=0\}\ .\eeqa
respectively. %By definition, in the Hamiltonian
%~\equ{full system}, 
%we have $\hh_0\in {\cal N}_{r,\r,\xi,s,\d}$, is $x$--independent and $f\in {\cal O}_{r,\r,\xi,s,\d}$.

\nl
We shall prove the following result.

\begin{lemma}\label{NFL}
For any $n$, $m$, there exists a number ${\tt c}_{n,m}\ge 1$ such that, for any $N\in {\mathbb N}$ such that the following inequalities are satisfied
\beq{normal form assumptions} 4N{\cal X}\|\frac{\o_\II}{\o_y}\|_{r,\r}<
s
\ ,\quad
4N{\cal X}\|\frac{\o_\JJ}{\o_y}\|_{r,\r}<
1\ ,\quad {\tt c}_{n,m}N\frac{{\cal X}}{{\tt d}}  { \|\frac{1}{\o_y}\|_{r,\r} 
\|% \frac
{f_0}%{\o_y}
\|_{r,\r,\xi,s,\d}} <1 \eeq
with ${\tt d}:=\min\big\{\r\s, r\xi, {\d}^2\big\}$, ${\cal X}:=\sup\big\{|x|:\ x\in \Xi_\xi\big\}$ and $\omega_{y,\II,\JJ}:=\partial_{y,\II,\JJ} \hh_0$, one can find an operator $$\Psi_N:\quad {\cal O}_{r,\r,\xi,s,\d}\to \cO_{1/3 (r, \r,\xi, s, \d)}$$
which carries $\HH$ to
$$\HH_N:=\Psi_N[\HH]=\hh_0+\ovl f_0+g_N+f_N$$
where
$g_N\in \cN_{1/3 (r, \r,\xi, s, \d)}$, $f_N\in \cO_{1/3 (r, \r,\xi, s, \d)}$ and, moreover, the following inequalities hold 
\beqa{thesis}
&&\|g_N\|_{1/3 (r, \r,\xi, s, \d)}\le {\tt c}_{n,m} \frac{{\cal X}}{\tt d}\|%\frac{1}{\o_y}\|_{r,\r} \|
{\frac{\widetilde f_0}{\o_y}}\|_{r,\r,\xi,s,\d}\| f_0\|_{r,\r,\xi,s,\d}\nonumber\\
&&    \|f_N\|_{1/3 (r, \r,\xi, s, \d)}\le \frac{1}{2^{N+1}} \|f_0\|_{r,\r,\xi,s,\d}\ .\eeqa
Furthermore, 
if
$$(I, \f, p, q, y, x):=\Psi_N(I_N, \f_N, p_N, q_N, y_N, x_N)$$
the following uniform bounds hold:
\beqa{phi close to id***}
&&{\tt d}\max\Big\{\frac{|I-I_N|}{\r},\ \frac{|\varphi-\varphi_N|}{s},\ 
\frac{|p-p_N|}{\d},\ \frac{ |q-q_N|}{\d},\ \frac{|y-y_N|}{r},\frac{ |x-x_N|}{\xi} \Big\}\nonumber\\
&&\le\max\Big\{s|I-I_N|,\ \r|\varphi-\varphi_N|,\ 
\d |p-p_N|,\ \d |q-q_N|,\ \xi |y-y_N|,\ r |x-x_N| \Big\}\nonumber\\
&&\leq4\, {\cal X}   % \|\frac{1}{\o_y}\|_{r,\r}
 \| \frac{f_0}{\o_y}\|_{r,\r,\xi,s,\d}\ .
\eeqa
\end{lemma}

\nl
{\bf Ideas of proof} The proof of Lemma~\ref{NFL} is based on  the well--settled framework acknowledged to J\"urgen P\"oschel~\cite{poschel93}. As in~\cite{poschel93}, we shall obtain the   Normal Form Lemma  via iterate applications of one--step transformations (Iterative Lemma, see below) where the dependence of $\f$ and $(p,q)$ other than the combinations  $\JJ(p,q)$ is eliminated at higher and higher orders. It  goes as follows.

\nl
We assume that, at a certain step, we have a system of the form
\beq{step i}\HH=\hh_0(y,\II, \JJ(p,q))+g(y,\II, \JJ(p,q), x)+f(y,\II,  x, \f, p,q)\eeq
where  $f\in {\cal O}_{r,\r,\xi,s,\d}$, while $\hh_0$, $g\in {\cal N}_{r,\r,\xi,s,\d}$, with $\hh_0$ is independent of $x$ (the first step corresponds to take $g\equiv0$). 

\nl
After splitting $f$ on its Taylor--Fourier basis 
$$f=\sum_{k,h,j} f_{khj}(y,\II,x)e^{\ii k \f}p^h  q^j\ .$$
one looks for a time--1 map 
$$\Phi=e^{\cL_\phi}$$
generated by a small Hamiltonian 
$\phi$
which will be taken  in the  class ${\cal Z}_{r,\r,\xi,s,\d}$ in~\equ{zero average}.
One lets
\beq{exp}\phi=\sum_{(k,h,j):\atop{(k,h-j)\ne (0,0)}} \phi_{khj}(y,\II,x)e^{\ii k \f}p^h q^j\ .\eeq

\nl
The operation
$$\phi\to \{\phi,\hh_0\}$$
acts diagonally  on the monomials  in the expansion~\equ{exp}, carrying
\beq{diagonal}\phi_{khj}\to -\big(\o_y\partial_x \phi_{khj}+\l_{khj} \phi_{khj}\big)\ ,\quad {\rm with}\quad \l_{khj}:=(h-j)\cdot\o_\JJ+\ii k\cdot\o_\II\ .\eeq
Therefore, one defines
$$\{\phi,\hh_0\}=:-D_\omega\phi\ .$$
The formal application of $\Phi=e^{\cL_\phi}$ yields:
\beqa{f1}
e^{\cL_\phi} \HH&=&e^{\cL_\phi} (\hh_0+g+f)=\hh_0+g-D_\omega \phi+f+\Phi_2(\hh_0)+\Phi_1(g)+\Phi_1(f)
\eeqa
where the $\Phi_h$'s are the queues of $e^{\cL_\phi}$, defined in Section~\ref{Time--one flows}.

\nl
Next, one requires that   the residual term $-D_\omega \phi+f$ lies in the class ${\cal N}_{r,\r,\xi,s,\d}$ in~\equ{phi independent}. This
amounts  to solve 
the ``homological'' equation
\beq{homological equation}\widetilde{(-D_\omega \phi+f\big)}=0\eeq
for $\phi$.

\nl
Since we have chosen $\phi\in{\cal Z}_{r,\r,\xi,s,\d}$, by~\equ{diagonal}, we have that also $D_\omega \phi\in{\cal Z}_{r,\r,\xi,s,\d}$. So, Equation~\equ{homological equation} becomes
\beq{homol}-D_\omega \phi+\widetilde f=0\ .\eeq
In terms of  the Taylor--Fourier modes, the equation becomes
\beq{homol eq} \o_y\partial_x \phi_{khj}+\l_{khj} \phi_{khj}=f_{khj}\qquad \forall\ (k,h,j):\ (k,h-j)\ne (0,0)\ .\eeq

\nl In the standard situation, one typically proceeds to solve such equation via Fourier series:
\beq{periodic case}f_{khj}(y,\II,x)=\sum_{\ell}f_{khj\ell}(y,\II)e^{\ii \ell x}\ ,\qquad \phi_{khj}(y,\II,x)=\sum_{\ell}\phi_{khj\ell}(y,\II)e^{\ii \ell x}\eeq
so as to find
$\dst\phi_{khj\ell}=\frac{f_{khj\ell}}{\m_{khj\ell}}$
with the usual  denominators $\m_{khj\ell}:=\l_{khj}+\ii\ell \omega_y$ which one requires not to vanish %(equivalently, $\frac{\l_{khj}}{\o_y}\notin \ii\integer$)
via, \eg, a ``diophantine inequality'' to be held for all $(k,h,j,\ell)$ with $(k,h-j)\ne (0,0)$. In this standard case, there is not much freedom in the choice of $\phi$. In fact, such solution is determined up to solutions  of the homogenous equation
\beq{homogeneous}D_\omega\phi_0=0\eeq
which, in view of the Diophantine condition, has the only trivial solution $\phi_0\equiv0$. {\it The situation is different if $f$ is not periodic in $x$, or $\phi$ is not needed so}. In such a case, it is possible to find a solution of~\equ{homol eq}, corresponding to a non--trivial solution of~\equ{homogeneous},  where small divisors do not appear.

\nl
This is
\beq{solution}\phi_{khj}(y,\II, x)=%B_{khj}(y,\II,x) e^{-\frac{\l_{khj}}{\o_y} x}
\frac{1}{\o_y}\int_0^xf_{khj}(y,\II,\t)e^{\frac{\l_{khj}}{\o_y}(\t-x)}d\t\qquad \forall\ (k,h,j):\ (k,h-j)\ne (0,0)
\eeq
and $\phi_{0hh}(y,\II, x)\equiv0$. %Note that in the particular case that $f$ is periodic in $x$, and hence it affords an expansion like~\equ{periodic case}, the solution~\equ{solution} may be  written as
%$$\phi_{khj}=e^{-\frac{\l_{khj}}{\o_y}x}\sum_{\ell}f_{khj\ell}(y,\II) \frac{e^{\frac{\m_{khj\ell}}{\o_y}x}-1}{\m_{khj\ell}}=:e^{-\frac{\l_{khj}}{\o_y}x}\widehat\phi_{khj}(y,\II,x, p,q)\ . $$
Complete details are in the following section.

 \subsection{Proof of Lemma~\ref{NFL}}\label{Time--one flows}

\begin{definition}[Time--one flows and their queues]\rm
Let $\cL_\phi(\cdot):=\big\{\phi, \cdot\big\}$, where $\{f, g\}:=\sum_{i=1^k}(\partial_{p_i}f\partial_{q_i}g-\partial_{p_i}g\partial_{q_i}f)$, where $\Omega=\sum_{i=1}^k dp_i\wedge dq_i$ is the standard two--form, denotes Poisson parentheses.

\nl
For a given $\phi\in {\cal O}_{r,\r,\xi,s,\d}$, we denote as $\Phi_h$, $\Phi$ the formal series

\beq{queue}\Phi_h:=\sum_{j\ge h}\frac{{\cal L}_\phi^j}{j!}\ \qquad \Phi:=\Phi_0\ .\eeq
It is customary to let, also $\Phi:=e^{\cL_\phi}$.
\end{definition}

\begin{lemma}[\cite{poschel93}]\label{base lemma}
There exists an integer number $\ovl{\tt c}_{n,m}$ such that, for any $\phi\in {\cal O}_{r,\r,\xi,s,\d}$ and any $r'<r$, $s'<s$, $\r'<\r$, $\xi'<\xi$, $\d'<\d$ such that
$$\frac{\ovl{\tt c}_{n,m}\|\phi\|_{r,\r,\xi,s,\d}}{d}<1\qquad d:=\min\big\{\r'\s', r'\xi', {\d'}^2\big\}$$
then  the series in~\equ{queue} converge uniformly so as to define 
the family $\{\Phi_h\}_{h=0,1,\cdots}$ of operators 
 $$\Phi_h:\quad  {\cal O}_{r,\r,\xi,s,\d}\to \cO_{r-r',\r-\r',\xi-\xi',s-s',\d-\d'}\ .$$ 
 Moreover, the following bound holds (showing, in particular, uniform  convergence):

\beq{geometric series}\|\cL^j_\phi[g]\|_{r-r',\r-\r',\xi-\xi',s-s',\d-\d'}\le j!\big(\frac{\ovl{\tt c}_{n,m}\|\phi\|_{r,\r,\xi,s,\d}}{d}\big)^j\|g\|_{r,\r,\xi,s,\d}\ .\eeq
for all $g\in{\cal O}_{r,\r,\xi,s,\d}$.
\end{lemma}

\begin{remark}[\cite{poschel93}]\rm The bound~\equ{geometric series} immediately implies
\beq{h power}\|\Phi_hg\|_{r-r',\r-\r',\xi-\xi',s-s',\d-\d'}\le \frac{\big(\frac{\ovl{\tt c}_{n,m}\|\phi\|_{r,\r,\xi,s,\d}}{d}\big)^h}{1-\frac{\ovl{\tt c}_{n,m}\|\phi\|_{r,\r,\xi,s,\d}}{d}}\|g\|_{r,\r,\xi,s,\d}\qquad \forall g\in {\cal O}_{r,\r,\xi,s,\d}\ .\eeq
\end{remark}

\begin{lemma}[Iterative Lemma]\label{iterative lemma}
There exists a number $\widetilde{\tt c}_{n,m}>1$ such that the following holds. For any choice  of positive numbers  $r'$, $\r'$, $s'$, $\xi'$. $\d'$ satisfying
%\beqano
%&&r'<r\ ,\quad \r'<\r\ ,\quad \xi'<\xi\\
%&& s'<s\ ,\quad \d'<\d\ ,\quad {\cal X}\|\frac{\o_\II}{\o_y}\|_{r,\r}<s-s'\ ,\quad
%{\cal X}\|\frac{\o_\JJ}{\o_y}\|_{r,\r}<
%\log\frac{\d}{\d'} 
%\eeqano
\beqa{ineq1}
&&{2r'<r\ ,\quad 2\r'<\r\ ,\quad 2\xi'<\xi}\\
\label{ineq2}
&& {2s'<s\ ,\quad 2\d'<\d\ ,\quad {\cal X}\|\frac{\o_\II}{\o_y}\|_{r,\r}<s-2s'\ ,\quad
{\cal X}\|\frac{\o_\JJ}{\o_y}\|_{r,\r}<
\log\frac{\d}{2\d'} }
\eeqa
and and provided that the  following inequality holds true
\beqa{smallness}
 \widetilde{\tt c}_{n,m}\frac{{\cal X}}{d}   % \|\frac{1}{\o_y}\|_{r,\r}
  \|\frac{\widetilde f}{\o_y}\|_{r,\r,\xi,s,\d} <1\qquad d:=\min\big\{\r'\s', r'\xi', {\d'}^2\big\}
\eeqa
 one can find an operator $$\Phi:\quad {\cal O}_{r,\r,\xi,s,\d}\to \cO_{r_+,\r_+,\xi_+,s_+,\d_+}$$
 with
%  $$r_+:=r-r'\ ,\quad \r_+:=\r-\r'\ ,\quad \xi_+:=\xi-\xi'\ ,\quad s_+:=s-s'-{\cal X}\|\frac{\o_\II}{\o_y}\|_{r,\r}\ ,\quad \d_+:=\d e^{-{\cal X}\|\frac{\o_\JJ}{\o_y}\|_{r,\r}}-\d'$$
\beqano
&&r_+:=r-2r'\ ,\quad \r_+:=\r-2\r'\ ,\quad \xi_+:=\xi-2\xi'\ ,\quad s_+:=s-2s'-{\cal X}\|\frac{\o_\II}{\o_y}\|_{r,\r}\nonumber\\
&& \d_+:=\d e^{-{\cal X}\|\frac{\o_\JJ}{\o_y}\|_{r,\r}}-2\d'
\eeqano
which carries the Hamiltonian  $\HH$ in~\equ{step i} to
$$\HH_+:=\Phi[\HH]=\hh_0+g+\ovl f+f_+$$
where
\beq{bound}\|f_+\|_{r_+,\r_+, \xi_+, s_+,\d_+}\le \widetilde{\tt c}_{n,m}\frac{{\cal X}}{d} \|\frac{\widetilde f}{\o_y}\|_{r,\r,\xi,s,\d}\| f\|_{r,\r,\xi,s,\d}
+ \|\{\phi, g\}\|_{r_1-r',\r_1-\r',\xi_1-\xi',s_1-s',\d_1-\d' }
\eeq
with $$r_1:=r\ ,\quad \r_1:=\r\ ,\quad \xi_1:=\xi\ ,\quad s_1:=s-{\cal X}\|\frac{\o_\II}{\o_y}\|_{r,\r}\ ,\quad \d_1:=\d e^{-{\cal X}\|\frac{\o_\JJ}{\o_y}\|_{r,\r}}$$ for a suitable $\phi\in \cO_{r_1,\r_1,\xi_1,s_1,\d_1}$ verifying \beq{bound on phi}\|\phi\|_{r_1,\r_1,\xi_1,s_1,\d_1 }\le \frac{{\cal X}}{d}    %\|\frac{1}{\o_y}\|_{r,\r} \|\widetilde f\|_{r,\r,\xi,s,\d}
 \|\frac{\widetilde f}{\o_y}\|_{r,\r,\xi,s,\d}
\ .\eeq
Furthermore, 
if
$$(I_+, \f_+, p_+, q_+, y_+, x_+):=\Phi(I, \f, p, q, y, x)$$
the following uniform bounds hold:
\beq{phi close to id}
\max\Big\{s'|I-I_+|,\ \r'|\varphi-\varphi_+|,\ 
\d' |p-p_+|,\ \d |q-q_+|,\ \xi' |y-y_+|,\ r' |x-x_+| \Big\}\leq2\, {\cal X}   % \|\frac{1}{\o_y}\|_{r,\r} \|\widetilde f\|_{r,\r,\xi,s,\d}
 \|\frac{\widetilde f}{\o_y}\|_{r,\r,\xi,s,\d}
\ .
\eeq
\end{lemma}

\proof 
Let $\ovl{\tt c}_{n,m}$ be as in Lemma~\ref{base lemma}. We shall choose $\widetilde{\tt c}_{n,m}$ suitably large with respect to $\ovl{\tt c}_{n,m}$.

\nl
Let $\phi_{khj}$ as in~\equ{solution}. 
Let us fix
\beq{ovl param}0<\ovl r\le r\ ,\quad 0<\ovl\r\le \r\ ,\quad 0<\ovl\xi\le \xi\ ,\quad 0<\ovl s< s\ ,\quad 0<\ovl\d< \d\eeq
and assume that
\beq{ovl param1}{\cal X}\|\frac{\o_\II}{\o_y}\|_{ r,\r} \le s-\ovl s\ ,\qquad {\cal X}\|\frac{\o_\JJ}{\o_y}\|_{ r,\r} \le \log\frac{\d}{\ovl \d}\ .\eeq
Then we have
$$\|\phi_{khj}\|_{\ovl r,\ovl\r,\ovl\xi}\le  %\|\frac{1}{\o_y}\|_{\ovl r,\ovl\r}
 \|\frac{f_{khj}}{\o_y}\|_{\ovl r,\ovl\r,\ovl\xi}\|\int_0^x |e^{-\frac{\l_{khj}}{\o_y}\t}|\|_{\ovl r,\ovl\r,\ovl\xi}d\t\le{\cal X}   %\|\frac{1}{\o_y}\|_{\ovl r,\ovl\r} 
  \|\frac{f_{khj}}{\o_y}\|_{\ovl r,\ovl\r,\ovl\xi} e^{{\cal X}\|\frac{\l_{khj}}{\o_y}\|_{\ovl r,\ovl\r}}\ .$$
%We take, in particular, $\ovl r$, $\ovl\r$, ${\cal X}$ as above and, moreover,
%$\ovl s$, $\ovl\d$ such that
% $$\ovl s+{\cal X}\|\frac{\o_\II}{\o_y}\|_{\ovl r,\ovl\r} \le s\ ,\qquad \ovl \d e^{{\cal X}\|\frac{\o_\JJ}{\o_y}\|_{\ovl r,\ovl\r} }\le \d\ .$$
Since
$$\|\frac{\l_{khj}}{\o_y}\|_{\ovl r,\ovl\r} \le (h+j)\|\frac{\o_\JJ}{\o_y}\|_{\ovl r,\ovl\r} +|k|\|\frac{\o_\II}{\o_y}\|_{\ovl r,\ovl\r} 
$$
we have
$$\|\phi_{khj}\|_{\ovl r,\ovl\r,\ovl\xi}\le {\cal X}   %\|\frac{1}{\o_y}\|_{\ovl r,\ovl\r}  
\|\frac{\widetilde f_{khj}}{\o_y}\|_{\ovl r,\ovl\r,\ovl\xi}
e^{(h+j){\cal X}\|\frac{\o_\JJ}{\o_y}\|_{\ovl r,\ovl\r} +|k|{\cal X}\|\frac{\o_\II}{\o_y}\|_{\ovl r,\ovl\r} }\ .
$$
which yields (after multiplying by $e^{|k|\ovl s}(\ovl\d)^{j+h}$ and summing over $k$, $j$, $h$ with $(k,h-k)\ne (0,0)$) to
$$\|\phi\|_{\ovl r,\ovl\r,\ovl\xi, \ovl s,\ovl\d}\le{\cal X}  %\|\frac{1}{\o_y}\|_{\ovl r,\ovl \r} 
\|\frac{\widetilde f}{\o_y}\|_{\ovl r,\ovl \r,\ovl \xi,\ovl s+{\cal X}\|\frac{\o_\II}{\o_y}\|_{\ovl r,\ovl\r} ,\ovl \d e^{{\cal X}\|\frac{\o_\JJ}{\o_y}\|_{\ovl r,\ovl\r} }}\ . $$
Note that the right hand side is well defined because of~\equ{ovl param1}.
In the case of the choice
\beqano
&&\ovl r=r=:r_1\ ,\quad\ovl\r=\r=: \r_1\ ,\quad \ovl\xi=\xi=:\xi_1\ ,\quad \ovl s=s-{\cal X}\|\frac{\o_\II}{\o_y}\|_{r,\r}=:s_1\nonumber\\
&& \ovl\d=\d e^{-{\cal X}\|\frac{\o_\JJ}{\o_y}\|_{r,\r}}=:\d_1
\eeqano
(which, in view of the two latter inequalities in~\equ{ineq2}, satisfies~\equ{ovl param}--\equ{ovl param1}) the inequality becomes~\equ{bound on phi}.
An application of Lemma~\ref{base lemma},with $r$, $\r$, $\xi$, $s$, $\d$ replaced by $r_1-r'$, $\r_1-\r'$, $\xi_1-\xi'$, $s_1-s'$, $\d_1-\d'$,  concludes  with a suitable choice of $\widetilde{\tt c}_{n,m}>\ovl{\tt c}_{n,m}$ and (by~\equ{f1}) $$f_+:=\Phi_2(\hh_0)+\Phi_1(g)+\Phi_1(f)\ .$$
Observe that the bound~\equ{bound} follows from  Equations~\equ{h power},~\equ{geometric series} and the identities
$$\Phi_2[\hh_0]=\sum_{j=2}^\infty \frac{\cL^j_\phi(\hh_0)}{j!}=\sum_{j=1}^\infty \frac{\cL^{j+1}_\phi(\hh_0)}{(j+1)!}=-\sum_{j=1}^\infty \frac{\cL^{j}_\phi(\widetilde f)}{(j+1)!}$$
$$\Phi_1[g]=\sum_{j=1}^\infty \frac{\cL^j_\phi(g)}{j!}=\sum_{j=0}^\infty \frac{\cL^{j+1}_\phi(g)}{(j+1)!}=-\sum_{j=0}^\infty \frac{\cL^{j}_\phi(g_1)}{(j+1)!}$$
with $g_1:=\cL_\phi(g)=\{\phi, g\}$.
The bounds in~\equ{phi close to id} are a consequence of equalities of the kind
$$I_+-I=\sum_{j=0}^\infty\frac{\cL^{j+1}_\phi(I)}{(j+1)!}=\sum_{j=0}^\infty\frac{\cL^{j}_\phi(-\partial_\f\phi)}{(j+1)!}$$
(and similar).
 $\quad\square$

\nl
The proof of the   Normal Form Lemma goes through iterate applications of Lemma~\ref{iterative lemma}. At this respect, we premise the following

\begin{remark}\label{stronger iterative lemma}\rm
Replacing conditions in~\equ{ineq2} with the stronger ones
%\beqno2s'<s\ ,\quad 2\d'<\d\ ,\quad {\cal X}\|\frac{\o_\II}{\o_y}\|_{r,\r}<s'\ ,\quad
%{\cal X}\|\frac{\o_\JJ}{\o_y}\|_{r,\r}<
%\frac{\d'}{\d} \eeqno

\beq{new cond}
{3s'<s\ ,\quad 3\d'<\d\ ,\quad {\cal X}\|\frac{\o_\II}{\o_y}\|_{r,\r}<s'\ ,\quad
{\cal X}\|\frac{\o_\JJ}{\o_y}\|_{r,\r}<
\frac{\d'}{\d}} \eeq

\nl
(and keeping~\equ{ineq1},~\equ{smallness} unvaried)  one can take, for $s_+$, $\d_+$,  $s_1$, $\d_1$ the simpler expressions
%$$s_{+\rm new}=s-2s'\ ,\quad \d_{+\rm new}=\d-2\d'\ ,\quad s_{1\rm new}:=s-s'\ ,\quad \d_{1\rm new}=\d-\d'\ $$

$${s_{+\rm new}=s-3s'\ ,\quad \d_{+\rm new}=\d-3\d'\ ,\quad s_{1\rm new}:=s-s'\ ,\quad \d_{1\rm new}=\d-\d'\ }$$
(while keeping $r_+$, $\r_+$, $\xi_+$, $r_1$, $\r_1$, $\xi_1$ unvaried).
Indeed, since $1-e^{-x}\le x$ for all $x$,
$$\d_1=\d e^{-{\cal X}\|\frac{\o_\JJ}{\o_y}\|_{r,\r}} = \d-\d(1- e^{-{\cal X}\|\frac{\o_\JJ}{\o_y}\|_{r,\r}})\ge \d-{\cal X}\|\frac{\o_\JJ}{\o_y}\|_{r,\r}\ge \d-\d' =\d_{1\rm new}\ .$$
This also implies $\xi_+=\d_1-\d'\ge \d-2\d'=\xi_{+\rm new}$. That $s_+\ge s_{+\rm new}$, $s_1\ge s_{1\rm new}$ is even more immediate.\end{remark}

\nl
Now we can proceed with the

\paragraph{\it Proof of the the   Normal Form Lemma}
Let $\widetilde{\tt c}_{n,m}$ be as in Lemma~\ref{iterative lemma}. We shall choose $\ovl{\tt c}_{n,m}$ suitably large with respect to $\widetilde{\tt c}_{n,m}$.

\nl
We apply Lemma~\ref{iterative lemma} with
$${2}r'=\frac{r}{3}\ ,\quad {2}\r'=\frac{\r}{3}\ ,\quad {2}\xi'=\frac{\xi}{3}\ ,\quad {3}s'=\frac{s}{3}\ ,\quad {3}\d'=\frac{\d}{3}\ ,\quad g\equiv0\ .$$
We make use of the stronger formulation described in Remark~\ref{stronger iterative lemma}.  Conditions in~\equ{ineq1} and the three former conditions in~\equ{new cond} are trivially true. The two latter inequalities in 
\equ{new cond} reduce to 
$${\cal X}\|\frac{\o_\II}{\o_y}\|_{r,\r}<\frac{s}{9}\ ,\quad
{\cal X}\|\frac{\o_\JJ}{\o_y}\|_{r,\r}<
\frac{1}{9} $$
and they are certainly satisfied by assumption~\equ{normal form assumptions}, for $N>1$. Since  
$$d=\min\{ \r' s', r'\xi', {\d'}^2 \}=\min\{ \r s/{36}, r\xi/{54}, {\d}^2/{81} \}\ge\frac{1}{{81}}
\min\{ \r s, r\xi, {\d}^2 \}=\frac{\tt d}{{81}}
$$
we have that condition~\equ{smallness} is certainly implied by the last inequality in~\equ{normal form assumptions}, once one chooses ${\tt c}_{n,m}>{81} \widetilde{\tt c}_{n,m}$.
By Lemma~\ref{iterative lemma}, it is then possible to conjugate $\HH$ to
$$\HH_1=\hh_0+\ovl f+f_1$$
with $f_1\in \cO_{r^\ppu,\r^\ppu,\xi^\ppu,s^\ppu,\d^\ppu}$, where $(r^\ppu,\r^\ppu,\xi^\ppu,s^\ppu,\d^\ppu):=2/3 (r, \r,\xi, s, \d)$ and
\beq{f1}\|f_1\|_{r^\ppu,\r^\ppu,\xi^\ppu,s^\ppu,\d^\ppu}\le {81}\widetilde{\tt c}_{n,m} \frac{{\cal X}}{\tt d}  % \|\frac{1}{\o_y}\|_{r,\r} 
\|\frac{\widetilde f}{\o_y}\|_{r,\r,\xi,s,\d}\| f\|_{r,\r,\xi,s,\d}\le \frac{\| f\|_{r,\r,\xi,s,\d}}{2}\ .\eeq
since ${\tt c}_{n,m}\ge {162} \widetilde{\tt c}_{n,m}$ and $N\ge1$. Now we aim to apply Lemma~\ref{iterative lemma} $N$ times, each time with parameters
$$r_j'=\frac{r}{6N}\ ,\quad \r_j'=\frac{\r}{6N}\ ,\quad \xi_j'=\frac{\xi}{6N}\ ,\quad s_j'=\frac{s}{{9}N}\ ,\quad \d_j'=\frac{\d}{{9}N}\ .$$
To this end, we let 
\beqano
&&r^{(j+1)}:=r^\ppu-j\frac{r}{3N}\ ,\quad \r^{(j+1)}:=\r^\ppu-j\frac{\r}{3N}\ ,\quad \xi^{(j+1)}:=\xi^\ppu-j\frac{\xi}{3N}\nonumber\\
&&s^{(j+1)}:=s^\ppu-j\frac{s}{3N}\ ,\quad \d^{(j+1)}:=\d^\ppu-j\frac{\d}{3N}\nonumber\\
&& r_1^\ppj:=r^\ppj\ ,\quad \r_1^\ppj:=\r^\ppj\ ,\quad \xi_1^\ppj:=\xi^\ppj\ ,\quad s_1^\ppj:=s^\ppj-\frac{s}{9N}\ ,\nonumber\\
&&\d_1^\ppj:=\d^\ppj-\frac{\d}{9N}\ ,\qquad {\cal X}_j:=\sup\{|x|:\ x\in \Xi_{\xi_j}\}
\eeqano
with $1\le j\le N$.

\nl
We assume that for a certain $1\le i\le N$ and all $1\le j\le i$, we have $\HH_j\in \cO_{r^\ppj, \r^\ppj, \xi^\ppj, s^\ppj, \d^\ppj}$ of the form
\beqa{Hi}
&&\HH_j=\hh_0+g_{j-1}+f_j\ , \quad g_{j-1}\in \cN_{r^\ppj, \r^\ppj, \xi^\ppj, s^\ppj, \d^\ppj}\ ,\quad g_{j-1}-g_{j-2}=\ovl f_{j-1}\\
&&\label{i+1 ineq} \|f_j\|_{r^\ppj, \r^\ppj, \xi^\ppj, s^\ppj, \d^\ppj}\le \frac{\|f_1\|_{r^\ppu, \r^\ppu, \xi^\ppu, s^\ppu, \d^\ppu}}{2^{j-1}}\eeqa
with  $g_{-1}\equiv0$,  $g_0=f_0=\ovl f$.
If $i=N$, we have nothing more to do. If $i<N$, we want to prove that
Lemma~\ref{iterative lemma} can be applied so as to conjugate $\HH_i$ to a suitable $\HH_{i+1}$ such that~\equ{Hi}--\equ{i+1 ineq}
are true with $j=i+1$.
To this end, we have to check
\beqa{smallness0}
&&{\cal X}_i\|\frac{\o_\II}{\o_y}\|_{r_i,\r_i}<s'_i%=\frac{s}{6N}
\ ,\quad
{\cal X}_i\|\frac{\o_\JJ}{\o_y}\|_{r_i,\r_i}<
\frac{\d'_i}{\d_i}\\
\label{smallness i}
&& \widetilde{\tt c}_{n,m}\frac{{\cal X}_i}{d_i} % \|\frac{1}{\o_y}\|_{r_i,\r_i}
\|\frac{f_i}{\o_y}\|_{r_i, \r_i, \xi_i, s_i, \d_i}<1\ .
\eeqa
where
$d_i:=\min\{ \r_i' s_i', r_i'\xi_i', {\d'}_i^2 \}$.
Conditions~\equ{smallness0} are certainly  verified, since in fact they are implied by the definitions above 
(using also $\d_i\le \frac{2}{3}\d$, ${\cal X}_i\le {\cal X}$) and the two former inequalities in~\equ{normal form assumptions}.
 To check the validity of~\equ{smallness i}, we firstly observe that
$$d_i=\min\{r'_j\xi'_j,\ \r'_js'_j,\ (\d'_j)^2\}\ge \frac{\tt d}{{81}N^2}\ .$$
Using then ${\tt c}_{n,m}>{162} \widetilde{\tt c}_{n,m}$,${\cal X}_i<{\cal X}$, Equation~\equ{f1}, the inequality in~\equ{i+1 ineq} with $j=i$ and the last inequality in~\equ{normal form assumptions}, we easily conclude
\beqa{last but one}
&&\|f_i\|_{r_i, \r_i, \xi_i, s_i, \d_i}\le \|f_1\|_{r^\ppu, \r^\ppu, \xi^\ppu, s^\ppu, \d^\ppu}\le {81}\widetilde{\tt c}_{n,m} \frac{{\cal X}}{\tt d}   %\|\frac{1}{\o_y}\|_{r,\r} 
\|\frac{ f}{\o_y}\|_{r,\r,\xi,s,\d}\|{ f}\|_{r,\r,\xi,s,\d}
 \nonumber\\
 &&\le
 \frac{1}{\widetilde{\tt c}_{n,m}}\frac{\tt d}{{81} N^2}\frac{1} {{\cal X}}  (\|\frac{1}{\o_y}\|_{r,\r})^{-1}\le  \frac{1}{\widetilde{\tt c}_{n,m}}\frac{d_i} {{\cal X}_i}  (\|\frac{1}{\o_y}\|_{r_i,\r_i})^{-1}\eeqa
 which is just~\equ{smallness i}.

\nl
 Then the Iterative Lemma is applicable to $\HH_i$, and Equations~\equ{Hi} with $j=i+1$ follow from it. The proof that also~\equ{i+1 ineq} holds (for a possibly larger value of ${\tt c}_{n,m}$)  when $j=i+1$ proceeds along the same lines  as in~\cite[proof of the Normal Form Lemma, p. 194--95]{poschel93} and therefore is omitted. The same for the proof of the first inequality in~\equ{thesis}, for $g_N:=\HH_1$ and~\equ{phi close to id***}.
  $\quad\square$
 
 \subsection{Proof of Proposition~\ref{NFP}}\label{proof of NFP}
 
Pick a positive number $c$ satisfying
$${0\le c<\D+\d}%{\ ,\quad \frac{a}{2}c^2\le \| f\|_{\r, s, \D+\d}}
\ .$$
Then apply Lemma~\ref{NFL} with
$$m=1\ ,\quad \hh_0=\hh(I)+\frac{\o_0}{2}y^2\ ,\quad f_0=\frac{\o_0}{2}x^2+f\ ,\quad (p, q)=\emptyset$$
and $\Upsilon$, $\Xi$, $r$, $\xi$ to be, respectively,
$$\Upsilon=\tilde\Upsilon:=\big\{y:\quad \frac{\d}{2}<  |y|< \D+\frac{\d}{2}\big\}\ ,\quad \Xi=\tilde\Xi:=\big\{x:\quad |x|< \frac{c}{2}\big\}\ ,\quad r=\frac{\d}{2}\ ,\quad \xi=\frac{c}{2}\ .$$
We check~\equ{normal form assumptions}. The second condition does not apply in this case because $\hh_0$ does not depend on $\JJ$. We check the first and the third condition.
We find:
$${\tt d}=\min\{\r s, \frac{\d c}{4}\}\ ,\quad \chi=c\ ,\quad \|\o_y\|=|\o_0||y|\ge \frac{a\d}{2}\ .$$
Then
$$4N\chi\|\frac{\o_\II}{\o_y}\|\le 8Nc\frac{|\o_\II|}{a\d}<s$$
by~\equ{first cond}.
Moreover, using
$$\|f_0\|_{\r, s, r, \xi}\le \frac{M_0}{2} c^2+\EE$$
we have, for any ${\tt c}_*\ge {\tt c}_n:={\tt c}_{n,1}$,
\beqano {\tt c}_{n}N\frac{{\cal X}}{{\tt d}}  % \|\frac{1}{\o_y}\|_{r,\r} 
\| \frac{f_0}{\o_y}\|_{r,\r,\xi,s,\d}&\le&\max\left\{
{\tt c}_*\left(% \|\frac{1}{\o_y}\|_{r,\r} 
 {\frac{c^3M_0}{a\d\r s} }+\frac{2{c}\EE}{a{\d\r s}} \right)N\ ,\quad
 {\tt c}_*\left(% \|\frac{1}{\o_y}\|_{r,\r} 
 {4\frac{c^2M_0}{a\d^2} }+\frac{8\EE}{a{\d^2}} \right)N
 \right\}\nonumber\\
 &<&1 \eeqano
 so Lemma~\ref{NFL} applies.
 By the thesis of Lemma~\ref{NFL}, we then find a real--analytic canonical transformation
\beqano
\tilde\phi:\quad\cI_{\r/3}\times {\mathbb T}^n_{s/3}\times  {\tilde\Upsilon}_{\d/6}\times{\tilde\Xi}_{c/6}&\to& \cI_{\r}\times {\mathbb T}^n_{s}\times {\tilde\Upsilon}_{\d/2}\times{\tilde\Xi}_{c/2}\nonumber\\
(\tilde \II, \tilde\f, \tilde y, \tilde x)&\to&(\II,\f, y, x)
\eeqano
verifying~\equ{transformation estimates}
which carries $\HH$ to
\beqa{tildeH}\tilde\HH:=\HH\circ\tilde\phi=\frac{\o_0}{2}(\tilde x^2+\tilde y^2)+\ovl f(\tilde y, \tilde \II, \tilde x)+\tilde g(\tilde y, \tilde \II, \tilde x)+\tilde f_*(\tilde y, \tilde \II, \tilde x, \tilde \f)\eeqa
where
\beqano
\|\tilde g\|_{\r/3, s/3, \d/6, c/6}&\le&\max\left\{
{\tt c}_n\left(% \|\frac{1}{\o_y}\|_{r,\r} 
 {\frac{c^3M_0}{a\d\r s} }+\frac{2{c}\EE}{a{\d\r s}} \right)\ ,\quad
 {\tt c}_n\left(% \|\frac{1}{\o_y}\|_{r,\r} 
 {4\frac{c^2M_0}{a\d^2} }+\frac{8\EE}{a{\d^2}} \right)
 \right\}\nonumber\\
 &&\text{\tiny$\times$}\left(\frac{M_0}{2} c^2+\EE\right) \nonumber\\
 \|\tilde f_*\|_{\r/3, s/3, \d/6, c/6}&\le& \left(\frac{M_0}{2} c^2+\EE\right)2^{-N}\ .
\eeqano

\noi
Now, if
$$R(\theta):=\left(
\begin{array}{lrr}
\cos\theta&-\sin\theta\\
\sin\theta&\cos\theta
\end{array}
\right)\qquad \theta\in {\mathbb T}$$
the map 
\beqano
r_\theta:\quad && \left(
\begin{array}{lrr}
y\\
x
\end{array}
\right)\to %\in \Upsilon_{1,\d/6} \times \Xi_{1, c/2}
 \left(
\begin{array}{lrr}
y'\\
x'
\end{array}
\right):=R(\theta)\left(
\begin{array}{lrr}
y\\
x
\end{array}
\right)\nonumber\\\nonumber\\%\in \cR_{\d, c}(\theta):=R_\theta (\Upsilon_{1,\d/6} \times \Xi_{1, c/2})
&& (\II', \f')=(\II, \f)
\eeqano
is canonical, therefore so is the map
$$\tilde\phi_\theta:=r_{-\theta}\circ\tilde\phi\circ r_\theta\ .$$
It is possible to choose $\theta_1$, $\cdots$, $\theta_p\in {\mathbb T}$ such that the collection of $\{\tilde\phi_i:=\tilde\phi_{\theta_i}\}_{i=1, \cdots, p}$ is the desired atlantis.  An immediate geometric argument shows that one can bound the number $p$ as in~\equ{p*}. $\quad\square$

\nl
%Assume now, in addition, that~\equ{second smallness} holds.  We first perform the change of coordinates
%$$\hat\phi:\qquad \hat x=\sqrt{2\hat \JJ}\cos \hat\psi\ ,\qquad \hat y=\sqrt{2\hat \JJ}\sin \hat\psi$$
%with $(\hat\JJ, \hat\psi)\in \hat\cA_{\frac{\d}{6}}\times {\mathbb T}_\s$, where
%$$ \frac{\d}{2}\le |\hat\JJ|\le \D+\frac{\d}{2}\ ,$$
%with a suitable $\s>0$.
%This carries the Hamiltonian $\tilde\HH$ in~\equ{tildeH} to
%$$\hat\HH=\o_0\hat\JJ+\hat f(\hat\JJ, \hat\psi, \tilde\II)+\hat f_*(\hat\JJ, \hat\psi, \tilde\II, \tilde\f)$$
%where
%$$\hat f:=(\ovl f+\tilde g)\circ \hat\phi\ ,\quad \hat f_*:=\tilde f_*\circ\hat\phi$$
%verify, respectively,
%$$\|\hat f\|_{\hat\cA_{\frac{\d}{6}}\times {\mathbb T}_{\s}\times\cI_{\r}\times {\mathbb T}^n_{s/3}}\le 2\EE\ ,\quad \|\hat f_*\|_{\hat\cA_{\frac{\d}{6}}\times {\mathbb T}_{\s}\times\cI_{\r}\times {\mathbb T}^n_{s/3}}\le 2^{-N}\EE\ .$$
%Letting the term $\hat f_*$, which is still exponentially small aside, we look for a real--analytic transformation
%\beqano
%\breve\phi:\quad
%\hat\cA_{\frac{\d}{12}}\times {\mathbb T}_{\s/2}\times\cI_{\r/6}\times {\mathbb T}^n_{s/6}&\to&\hat\cA_{\frac{\d}{6}}\times {\mathbb T}_{\s}\times\cI_{\r}\times {\mathbb T}^n_{s/3}\nonumber\\
%(\breve \JJ, \breve\psi, \breve\II, \breve\f)&\to&(\hat \JJ, \hat\psi, \tilde\II, \tilde\f)
%\eeqano
%which carries $\hat\HH_\#:=\o_0\hat\JJ+\hat f(\hat\JJ, \hat\psi, \tilde\II)$ into the form
%$$\breve\HH:=\hat\HH\circ\breve\phi=\o_0\breve\JJ+\ovl{\ovl{\hat f}}(\breve\JJ,  \breve\II)+\breve g(\breve\JJ,  \breve\II)+\breve f_*(\hat\JJ, \hat\psi, \tilde\II, \tilde\f)$$
%where

\newpage\section{A revisited analysis of the two--centre problem. The Euler integral}\label{integration 2CP}

The two--centre problem is the $\dd$--degrees of freedom (with $\dd=2$, $3$)  system of one particle interacting with two fixed masses via Newton Law. If $\pm{\tt v}_0\in {\mathbb R}^\dd$ are the position coordinates of the centres, 
${\tt v}$, with  ${\tt v}\ne \pm{\tt v}_0$, the position coordinate of the moving particle and ${\tt u}=\dot{\tt v}$ its velocity,
the Hamiltonian of the system is
\beqa{2Cold}\ovl\JJ({\tt u}, {\tt v}; {\tt v}_0,\mm_+,\mm_-)=\frac{\|{\tt u}\|^2}{2}-\frac{\mm_+}{\|{\tt v}+{\tt v}_0\|}-\frac{\mm_-}{\|{\tt v}-{\tt v}_0\|}\ ,\eeqa
with $\|\cdot\|$ being the Euclidean distance in ${\mathbb R}^\dd$. The integrability  of $\ovl\JJ$ consists of the existence of $\dd-1$ independent first integrals of motion for $\ovl\JJ$ in involution. When $\dd=3$, there is a ``trivial'' first integral, related to the invariance of $\ovl\JJ$ by rotations around the axis ${\tt v_0}$, given by the projection
$$\Theta= {\tt M}\cdot \frac{{\tt v}_0}{\|{\tt v}_0\|}$$
of the angular momentum ${\tt M}={\tt v}\times {\tt y}$ along the direction ${\tt v}_0$. The existence of the following non--trivial constant of motion, which we shall refer to as {\it Euler integral}:

\beq{G1}\ovl\EE=\|{\tt v}\times {\tt u}\|^2+({\tt v}_0\cdot {\tt u})^2+2 {\tt v}\cdot {\tt v}_0\big(\frac{\mm_+}{\|{\tt v}+{\tt v}_0\|}-\frac{\mm_-}{\|{\tt v}-{\tt v}_0\|}\big)\eeq
was shown by Euler, in the XVIII century.

\nl
In view of our application to the three--body problem, we rewrite the two--centre Hamiltonian in the form~\equ{newH2C}, which differs from~\equ{2Cold} for the position of the centers at $\eufm 0$ and ${\tt x}'$ and the introduction of some mass parameters.

\nl
In this case,  the Euler integral takes the form in~\equ{EEE}--\equ{CL}; see Appendix~\ref{2centres} for a derivation. In the next section, we describe an initial set of coordinates we are going to use for our analysis.

% Note that when the second attracting center collapses (i.e., ${\tt x}'=0$), $\EE$ reduces to $\|{\tt M}\|^2$; 
%when it is switched off ($\m=0$), it
%reduces to $\EE$, which is a combination of first integrals to $\JJ_0$.

\subsection{${\cal K}$ --coordinates}

In this section we
describe a system of canonical coordinates, denoted as $\cK$, which we use for our analysis of the two--centre Hamiltonian~\equ{newH2C}. First of all,
we consider, in the region of  phase space where $\JJ_0$ in~\equ{kepler} takes negative values, the ellipse with initial datum $({\tt y}, {\tt x})$. Denote as: 
 \begin{itemize}
 \item[{\tiny\textbullet}]
 $a$  the {\it semi--major axis}; 
  \item[{\tiny\textbullet}]
 $\ee$  the {\it eccentricity}; 
  \item[{\tiny\textbullet}] ${\tt P}$, with $\|{\tt P}\|=1$, the direction of perihelion;
    \item[{\tiny\textbullet}] $\ell$: the mean anomaly, defined, mod $2\p$, as the area of the elliptic sector spanned by ${\tt x}$ from ${\tt P}$, normalized to $2\p$;
  \item[{\tiny\textbullet}]  the {\it true anomaly} $\n$, defined as $$\n=\arg(\cos\xi-\ee, \sqrt{1-\ee^2}\sin \xi)$$ with 
   \item[{\tiny\textbullet}]  the {\it eccentric anomaly} $\xi$, solving the {\it Kepler equation} $\xi-\ee\sin\xi=\ell$;
  %  \item[{\tiny\textbullet}] $\ee(\L,\GG)=\sqrt{1-\frac{\GG^2}{\L^2}}$ is the {\it eccentricity};
     \item[{\tiny\textbullet}]    the quantity $\varrho=\frac{{1-\ee^2}}{1+\ee\cos\n}=1-\ee\cos\xi$ corresponding to the ratio $\frac{\rr}{a}$.
    \end{itemize}

\nl
Next, we introduce the following notations.
\begin{itemize}
     \item[{\tiny\textbullet}] If ${\tt i}$, ${\tt k}\in {\mathbb R}^3$, with ${\tt i}\perp {\tt k}$, by $\FF\sim({\tt i},\cdot,{\tt k})$, we mean the orthonormal frame  $\FF=(\frac{{\tt i}}{\|{\tt i}\|}, \frac{{\tt k}\times {\tt i}}{\|{\tt k}\times {\tt i}\|},\frac{{\tt k}}{\|{\tt k}\|})$.
     \item[{\tiny\textbullet}] 
Given a couple $(\FF,\FF')$ of orthonormal frames, with $\FF\sim({\tt i},\cdot,{\tt k})$, $\FF'\sim({\tt i}',\cdot,{\tt k}')$, we write $$\FF\to^{(\YY, \XX,  x)}\FF'$$ if ${\tt i}'={\tt k}\times{\tt k}'$ and
$$\ZZ={\tt k}'\cdot \frac{\tt k}{\|{\tt k}\|}\ ,\qquad \XX=\|{\tt k}'\|\ ,\qquad  x=\a_{\tt k}({\tt i},{\tt i'})$$ 
where $\a_{\tt k}({\tt i},{\tt i'})$ is the oriented angle  ${\tt i}$ to ${\tt i}'$, with respect to the counterclockwise orientation established by ${\tt k}$.

\end{itemize}

\nl
We fix an arbitrary  frame $\FF_0\sim({\tt i}_0, \cdot, {\tt k}_0)\subset {\mathbb R}^3$, that we call {\it inertial frame} and, denote as
$${\tt M}={\tt x}\times {\tt y}\ ,\quad {\tt M}'={\tt x}'\times {\tt y}'\ ,\quad {\tt C}={\tt M}'+{\tt M}\ ,$$ 
where ``$\times$'' denotes skew--product in ${\mathbb R}^3$.
Observe the following relations
\beqa{orthogonality}{{\tt x}'}\cdot{{\tt C}}={{\tt x}'}\cdot{\big({\tt M}+{\tt M}'\big)}={{\tt x}'}\cdot{{\tt M}}\ ,\qquad {\tt P}\cdot {\tt M}=0\ ,\quad \|{\tt P}\|=1\ .\eeqa
Then put
$${\tt k}_1={\tt C}\ ,\quad {\tt k}_2={\tt x}'\ ,\quad {\tt k}_3={\tt M}\ ,\quad {\tt k}_4={\tt P}\ ,\quad {\tt i}_{j}:={\tt k}_{j-1}\times {\tt k}_{j}\quad j=1,\ 2,\ 3,\ 4$$
and  assuming
\beqa{nodes}{\tt i}_{j}\ne 0\qquad j=1,\ 2,\ 3,\eeqa
we define

\begin{itemize}
     \item[{\tiny\textbullet}] the frame $\FF_1\sim({\tt i}_1, \cdot, {\tt k}_1)$, that we call {\it invariable  frame};

     \item[{\tiny\textbullet}] the frame $\FF_2\sim({\tt i}_2, \cdot, {\tt k}_2)$, that  we call {\it ${\tt x'}$--frame};

     \item[{\tiny\textbullet}]  the frame $\FF_3\sim({\tt i}_3, \cdot, {\tt k}_3)$, that  we call {\it orbital frame};

     \item[{\tiny\textbullet}]  the frame $\FF_4\sim({\tt i}_4, \cdot, {\tt k}_4)$, that  we call {\it {\tt P}--frame}.
     \end{itemize}

\nl
We then   define the coordinates $$\cK=(\ZZ, \CC, \Theta, \GG, \RR', \L, \zeta, g, \vartheta, {\rm g}, \rr', \ell)$$
via the relations (which take~\equ{orthogonality} into account)
\beqa{p coord**}
&&\FF_0\to^{(\ZZ, \CC, \zeta)}\FF_1\to ^{(\frac{\rr'}{\GG}\Theta, \rr', {g})}\FF_2\to^{ (\Theta, \GG, \vartheta)}\FF_3\to ^{(0, 1, {\rm g})}\FF_4\nonumber\\
&&\RR'=\frac{{\tt y}'\cdot{\tt x}'}{\|{\tt x}'\|}\ ,\quad \L=\mm\sqrt{\cM a}\ ,\quad \ell={\rm mean\ anomaly\ of\ \n}
\eeqa

 % fino qui
    \nl
The canonical character of $\cK$ follows from~\cite{pinzari13}. Indeed, in~\cite{pinzari13}, we considered a set of coordinates for the three--body problem\footnote{An extension to the case of an arbitrary number of planets has been successively worked out in~\cite{pinzari18}.}, thereby denoted as $\cP$, that are related to $\cK$ above via the canonical change
\beqa{planar Delaunay}\cD_{e\ell, \rm pl}:\quad (\L,\GG, \ell, {\rm g})\to (\RR, \Phi, \rr, \f)\eeqa
     usually referred to as {\it planar Delaunay map}, defined as
\beqa{p coord***}
 \arr{
 \RR=\frac{\mm^2\cM}{\L} \frac{\ee\sin\xi}{1-\ee\cos\xi}\\
\Phi=\GG
 }\qquad  \arr{
\rr=a(1-\ee\cos\xi)\\
\f=\n+{\rm g}-\frac{\p}{2}
 }
\eeqa
where $\xi=\xi(\L, \GG, {\rm g})$ is the eccentric anomaly. Since the map $\cD_{e\ell, \rm pl}$ in~\equ{planar Delaunay} and the coordinates $\cP$ of~\cite{pinzari13} are canonical, so is $\cK$. Observe, incidentally, the unusual $\frac{\p}{2}$--shift in 
\equ{p coord***}, due to the fact that, according to the definitions in~\equ{p coord**}, the longitude of $ {\tt P}$ in the orthogonal plane of the frame $\FF_3\sim ({\tt i}_3, \cdot, {\tt k}_3)$ is ${\rm g}-\frac{\p}{2}$, since ${\rm g}$ is the longitude of ${\tt i}_4={\tt M}\times {\tt P}$ in the same plane.
 
\paragraph{The Hamiltonians $\HH$, $\JJ$ and $\EE$ in terms of $\cK$}
We now discuss the main features of the application of the coordinates $\cK$ to the Hamiltonians $\HH$, $\JJ$ and $\EE$, referring to the next section for all details.\\
The utility of using the coordinates $\cK$ for $\HH$, $\JJ$ and $\EE$ relies in the fact that many cyclic coordinates appear. More precisely:

\nl
\begin{itemize}
\item[\tiny\textbullet]
The invariance by rotation exhibited by $\HH$, $\JJ$ and $\EE$ implies that $\ZZ$, $\zeta$ and $g$, conjugated to $\zeta$, $\ZZ$
and $\CC$, are cyclic, because these latter functions identify the angular momentum ${\tt C}$;
\item[\tiny\textbullet] the conservation of ${\tt x}'$ along the motions of  $\JJ$ and $\EE$ implies that $\RR'$ and $\vartheta$ are cyclic for such functions.
\item[\tiny\textbullet]  Therefore, $\HH$ is a function of $(\RR', \L, \GG, \Theta, \rr', \ell, {\rm g}, \vartheta)$ only, while
$\EE$ and $\JJ$ are functions of $(\L, \GG, \Theta, \rr', \ell, {\rm g})$ only.
\end{itemize}
In the case, considered in the paper, of the planar problem, we have a further
simplification. Planar configurations are obtained setting $(\Theta, \vartheta)=(0, \p)$ or $(\Theta, \vartheta)=(0, 0)$. We distinguish three possible planar configurations:
\begin{itemize}
\item[$(\uparrow\uparrow)$:] $(\Theta,\vartheta)=(0,\p)$ and $\CC>\GG$ corresponds to {\it planar motions and prograde motion for} $({\tt x}', {\tt y}')$;
\item[$(\downarrow\uparrow)$:] $(\Theta,\vartheta)=(0,\p)$ and $\CC<\GG$ corresponds to {\it planar   motions  with retrograde motion for}  $({\tt x}', {\tt y}')$;
\item[$(\uparrow\downarrow)$:] $(\Theta,\vartheta)=(0,0)$ corresponds to {\it planar   motions  with retrograde motion for} $({\tt x}', {\tt y}')$.\end{itemize}
 To fix ideas, we consider the case of the planar configuration $(\uparrow\uparrow)$. In this case,  the functions $\JJ$ and $\EE$ in~\equ{newH2C} and~\equ{EEE} have, in terms of $\cK$, the expressions
\beqa{H3B}
\JJ&=&-\frac{\mm^3{\cal M}^2}{2\L^2}-\m\frac{\mm{\cal M}}{\sqrt{{\rr'}^2+2\rr' a\varrho%\sqrt{1-\frac{\Theta^2}{\GG^2}}
 \cos({\rm g}+\n)+{a}^2\varrho^2}}\nonumber\\
\EE&=&\GG^2+\mm^2{\cal M}\rr'%\sqrt{1-\frac{\Theta^2}{\GG^2}}
\sqrt{1-\frac{\GG^2}{\L^2}}\cos{\rm g}\nonumber\\
 & +&\m\mm^2{\cal M}\rr'\frac{{\rr'}+a\varrho%\sqrt{1-\frac{\Theta^2}{\GG^2}} 
 \cos({\rm g}+\n)}{\sqrt{{\rr'}^2+2\rr' a\varrho%\sqrt{1-\frac{\Theta^2}{\GG^2}}
  \cos({\rm g}+\n)+{a}^2\varrho^2}}\nonumber\\
 \HH&=&%\frac{\varepsilon^2{\RR'}^2}{2\mm'}+\frac{\varepsilon^2(\GG-\CC)^2}{2\mm'{\rr'}^2}
-\frac{\mm'{\cal M}'}{\rr'}+\varepsilon\Big(-\frac{\mm^3{\cal M}^2}{2\L^2}-\frac{\m\mm{\cal M}}{\sqrt{{\rr'}^2+2\rr' a\varrho \cos({\rm g}+\n)+{a}^2\varrho^2}}
\Big)\nonumber\\
&+&\varepsilon^2\left(\frac{{\RR'}^2}{2\mm'}+\frac{(\CC-\GG)^2}{2\mm'{\rr'}^2}+\frac{\m}{m_0}\Big(-\RR' \ovl{\tt y}_{2}+\frac{\CC-\GG}{\rr'}\ovl{\tt y}_{  1}\Big)\right)%\nonumber\\
%&=:&-\frac{\mm'{\cal M}'}{\hat\rr'}+
%\varepsilon\JJ(\L, \GG, \ell, {\rm g}, \rr';\m)+\varepsilon^2 f_\cK(\RR', \L, \GG, \rr', \ell, {\rm g};\CC, \m)
\eeqa
where $\ovl{\tt y}_{1}$, $\ovl{\tt y}_{2}$ are the components of the planar impulses, whose analytical expressions, in terms of $\cK$,  will be given in the next Section~\ref{details}; see Equation~\equ{yx}.

 \subsection{Explicit formulae of the ${\cal K}$--map}\label{details}
Let 
\beqa{k incli} i=\cos^{-1}\left(\frac{\ZZ}{\CC}\right)\ ,\quad  i_1=\cos^{-1}\left(\frac{\Theta}{\CC}\right)\ ,\quad i_2=\cos^{-1}\left(\frac{\Theta}{\GG}\right)\eeqa
and 
\beqa{R1R3}
&&R_1(\a):=\left(
\begin{array}{ccc}
1&0&0\\
0&\cos \a&-\sin\a\\
0&\sin\a&\cos\a
\end{array}
\right)\quad  R_3(\a):=\left(
\begin{array}{ccc}
\cos\a&-\sin\a&0\\
\sin\a&\cos\a&0\\
0&0&1
\end{array}
\right)\ .\eeqa
Using the definitions in~\equ{p coord**} and the observation that, if $\FF\to^{(\YY, \XX,  x)}\FF'$, the transformation of coordinates
which relates the coordinates ${\tt x}'$ relatively to $\FF'$ to  the coordinates ${\tt x}$ relatively to $\FF$  is
\beq{general case}{\tt x}=R_3(x)R_1(\iota){\tt x}'\eeq
where $R_1$, $R_3$ are as in~\equ{R1R3}, while $\iota:=\cos^{-1}\frac{\YY}{\XX}$, for the map

 \beqa{k map}
 \phi:\quad {\cal K}=\Big(\ZZ,\CC, \Theta, \GG, \L, \RR', \zeta, \f, \vartheta, {\rm g}, \ell, \rr'\Big)%\in {\mathbb R}^5\times{\mathbb T}^5\times {\mathbb R}\times {\mathbb R}_+
 \to (\underline{\tt y}, \underline{\tt x})=({\tt y}', {\tt y}, {\tt x}', {\tt x})\ .
 \eeqa
we find the following analytical expression
 \beqa{yx}
\phi:\qquad \left\{
\begin{array}{ll}
{\tt x}= R_3(\zeta)R_1(i)R_3(g)R_1(i_1)R_3(\vartheta)R_1(i_2)\ovl{\tt x}(\L,\GG,\ell, {\rm g})\\ \\
{\tt y}= R_3(\zeta)R_1(i)R_3(g)R_1(i_1)R_3(\vartheta)R_1(i_2)\ovl{\tt y}(\L,\GG,\ell, {\rm g})
\\\\
{\tt x}'=\rr' R_3(\zeta)R_1(i)R_3(g)R_1(i_1){\tt k}\\ \\
 {\tt y}'=\frac{{\rm R}'}{{\rm r}'}{\tt x'}+\frac{1}{{\rm r}'^2}{\tt M}'\times{\tt x}'
\end{array}
\right.
\eeqa
with
\beqa{xy}
{\tt k}&=&\left(
\begin{array}{ccc}
0\\
0\\
1
\end{array}
\right)\nonumber\\
\ovl{\tt x}&=&%a(\L)\varrho(\L,\GG, \ell)\cR_{3}(-\frac{\p}{2})\left(
%\begin{array}{ccc}
%\cos({\rm g}+\n(\L,\GG, \ell))\\
%\sin({\rm g}+\n(\L,\GG, \ell))\\
%0
%\end{array}
%\right)\nonumber\\
%&=&
\frac{\L^2}{\mm^2{\cal M}}\cR_3({\rm g}-\p/2) \left(
\begin{array}{ccc}
\cos\xi(\L,\GG,\ell)-\sqrt{1-\frac{\GG^2}{\L^2}}\\
\frac{\GG}{\L}\sin\xi(\L,\GG,\ell)\\
0
\end{array}
\right)\nonumber\\
\ovl{\tt y}&=&\frac{\mm^2{\cal M}}{\L}\cR_3({\rm g}-\p/2) \left(
\begin{array}{ccc}
-\sin\xi(\L,\GG,\ell)\\
\frac{\GG}{\L}\cos\xi(\L,\GG,\ell)\\
0
\end{array}
\right)\nonumber\\
{\tt C}&=&\CC R_3(\zeta)R_1(i){\tt k}\nonumber\\
{\tt M}&=&\GG R_3(\zeta)R_1(i)R_3(g)R_1(i_1)R_3(\vartheta)R_1(i_2){\tt k}\\
{\tt M}'&=&{\tt C}-{\tt M}\eeqa
where $a(\L)$ is as in~\equ{p coord**} and, if  
\beq{ej}{\rm e}(\L,\GG)=\sqrt{1-\frac{\GG^2}{\L^2}}\ ,\eeq
then
\beqa{rhonu}
\ell&=&\xi(\L,\GG,\ell)-\ee(\L,\GG)\sin\xi(\L,\GG,\ell)\nonumber\\
\varrho(\L,\GG,\ell)&:=&1-\ee\cos\xi(\L,\GG, \ell)=\frac{1-\ee(\L,\GG)^2}{1+\ee(\L,\GG)\cos\n(\L,\GG, \ell)}\nonumber\\
 \n(\L,\GG,\ell)&:=&\arg\Big(\cos\xi(\L,\GG,\ell)-\ee(\L,\GG), \frac{\GG}{\L}\sin \xi(\L,\GG,\ell)\Big)
\eeqa

\nl
%\begin{remark}\rm

  \paragraph{Planar case}\label{Planar configurations} 
  
Differently from what happens when one uses the Jacobi reduction of the nodes, in terms of the ${\rm k}$--coordinates,  planar configurations 
are {\it regular}. They can be obtained 
  setting the couple $(\Theta, \vartheta)$ to a particular values. Indeed, planar configurations are obtained taking $(\Theta, \vartheta)=(0,k\p)$, with $k=0$, $1$. Indeed, for $\Theta=0$, one has $i_1=i_2=\frac{\p}{2}$. Since $R_3(g){\tt k}={\tt k}$, it follows from the formulae 
\beqa{Ctot} {\tt C}=\CC \cR_3(\zeta)\cR_1(i){\tt k}\ ,\quad {\tt M}={\tt x}\times {\tt y}=\GG \cR_3(\zeta)\cR_1(i)\cR_3(g)\cR_1(i_1)\cR_3(\vartheta)\cR_1(i_2) {\tt k}\ .\eeqa
 that
 $$(\Theta,\vartheta)=(0,\p)\quad \iff\quad {\tt M}\parallel {\tt C}$$ while, $$(\Theta,\vartheta)=(0,0)\quad \iff\quad 
(-{\tt M})\parallel {\tt C}\ .$$ Therefore, we distinguish the three planar configurations $(\uparrow\uparrow)$, $(\downarrow\uparrow)$ and $(\uparrow\downarrow)$ mentioned in the previous section.

\nl
In such  cases, the formulae~\equ{yx} reduce to
\beqa{yy'}
\left\{
\begin{array}{lll}
\dst {\tt x}=R_3(\zeta)R_1(i)R_3(g)(\ovl{\tt x}_1{\tt i}+\ovl{\tt x}_2{\tt j})\\\\
\dst {\tt y}=R_3(\zeta)R_1(i)R_3(g)(\ovl{\tt y}_1{\tt i}+\ovl{\tt y}_2{\tt j})\\\\
\dst {\tt x}'=-\rr'R_3(\zeta)R_1(i)R_3(g){\tt j}\\\\
\dst {\tt y}'=-\RR'R_3(\zeta)R_1(i)R_3(g){\tt j}+\frac{\CC-\s\GG}{\rr'}R_3(\zeta)R_1(i)R_3(g){\tt i}
\end{array}
\right.
\eeqa
with
$${\tt i}=\left(
\begin{array}{lll}
1\\
0\\
0
\end{array}
\right)\ ,\qquad {\tt j}=\left(
\begin{array}{lll}
0\\
1\\
0
\end{array}
\right)\ ,\quad \s=\arr{+1\quad{\rm for}\quad (\uparrow\uparrow) \\
-1\quad {\rm otherwise}\ .
}$$
Taking $\ell=0$ in the definition of ${\tt x}$ and normalizing, we find the perihelion direction
$${\tt P}=R_3(\zeta)R_1(i)R_3(g)(-(\sin{\rm g}){\tt i}+(\cos{\rm g}){\tt j})$$
which shows, as anticipated in the introduction, that ${\tt x}'$ and ${\tt P}$ form a convex angle equal to $|\p-{\rm g}|$.

\paragraph{Derivation of~\equ{H3B}}
Using the general formulae in~\equ{yx}, we find
%${\tt x}'\cdot {\tt x}$ can be written as
\beqa{inner0}{\tt x}'\cdot {\tt x}%&=&%\big(\rr' R_3(\zeta)R_1(i)R_3(g)R_1(i_1){\tt k}\big)\cdot
%\big(R_3(\zeta)R_1(i)R_3(g)R_1(i_1)R_3(\vartheta)R_1(i_2)\ovl{\tt x}(\L,\GG,\ell, {\rm g})\big)\nonumber\\
&=&{\tt k}\cdot R_1(i_2)\ovl{\tt x}(\L,\GG,\ell, {\rm g})=-\rr'a\varrho\sqrt{1-\frac{\Theta^2}{\GG^2}}\cos({\rm g}+\n)\eeqa
where we have used $R_3^{\rm t}(\vartheta){\tt k}={\tt k}$, the relation
$$\sin i_2=\sqrt{1-\frac{\Theta^2}{\GG^2}}$$
(which is implied by the definition of $i_2$ in~\equ{k incli}) and the expression for $\ovl{\tt x}$ in~\equ{xy}
%$\ovl{\tt x}$ in~\equ{yx} in the form
Equations~\equ{inner0},~\equ{xy} and the definition of $\rr'=\|{\tt x}'\|$
 then imply that
the Euclidean distance between ${\tt x}'$ and ${\tt x}$ has the expression
\beqa{distance}\|{\tt x}'-{\tt x}\|^2={{\rr'}^2+2\rr' a\varrho\sqrt{1-\frac{\Theta^2}{\GG^2}}
\cos({\rm g}+\n)+a^2\varrho^2}\ .\eeqa
The expression of ${\tt P}$ is obtained from the one for ${\tt x}$
in~\equ{yx} taking $\n=\ell=0$ and normalizing. Namely, 
$${\tt P}=\frac{\ovl{\tt x}}{a\varrho}=R_3(\zeta)R_1(i)R_3(g)R_1(i_1)R_3(\vartheta)R_1(i_2)\ovl{\tt P}$$
%upon replacing $\ovl{\tt x}$ in~\equ{xy} 
with $$\ovl{\tt P}=\left(
\begin{array}{ccc}
\sin{\rm g}\\
-\cos{\rm g}\\
0
\end{array}
\right)$$
Then, analogously to~\equ{inner0}, 
 we find, for the inner product ${\tt x}'\cdot {\tt P}$ the expression
\beqa{inner}{\tt x}'\cdot {\tt P}=-\rr'\sqrt{1-\frac{\Theta^2}{\GG^2}}\cos{\rm g}\eeqa
%The formulae~\equ{distance} and~\equ{inner} will be used in the next section.
%\end{remark}
Using the formulae in~\equ{distance},~\equ{inner}, the definition of $\GG=\|\CC\|$,
 we find that
the functions ${\JJ}$, $\EE$ in~\equ{cal G new}, written terms of the coordinates ${\cal K}$, are given by
\beqa{h1k}
\JJ(\L, \GG, \Theta, \rr', \ell, {\rm g})&=&-\frac{\mm^3{\cal M}^2}{2\L^2}-\m\frac{\mm{\cal M}}{\sqrt{{\rr'}^2+2\rr' a\varrho\sqrt{1-\frac{\Theta^2}{\GG^2}} \cos({\rm g}+\n)+{a}^2\varrho^2}}\nonumber\\
%&=:&\JJ_{0,\cK}+\m \JJ_{1, \cK}\nonumber\\
\EE(\L, \GG, \Theta, \rr', \ell, {\rm g})&=&\GG^2+\mm^2{\cal M}\rr'\sqrt{1-\frac{\Theta^2}{\GG^2}}\sqrt{1-\frac{\GG^2}{\L^2}}\cos{\rm g}\nonumber\\
 & +&\m\mm^2{\cal M}\rr'\frac{{\rr'}+a\varrho\sqrt{1-\frac{\Theta^2}{\GG^2}} \cos({\rm g}+\n)}{\sqrt{{\rr'}^2+2\rr' a\varrho\sqrt{1-\frac{\Theta^2}{\GG^2}} \cos({\rm g}+\n)+{a}^2\varrho^2}}%\nonumber\\
%&=:&\EE_{0,\cK}+\m\EE_{1,\cK}
\eeqa
In the planar cases, letting $\Theta=0$, one has the two former equations in~\equ{H3B}. The third equation is easily obtained from~\equ{yy'}. %${\tt M}'$ reduces to
%$${\tt M}'=(\CC-\GG)R_3(\zeta)R_1(i){\tt k}$$
\newpage\section{Action--angle coordinates in the planar case}\label{2centres3}

\nl The purpose of the present section is to give a qualitative picture of the zones in phase space where action--angle coordinates do exist,  for values of $\m$ sufficiently small, leaving aside the question of the explicit expression of the action--angle coordinates. We do this only in the case of  the planar case ($\Theta\equiv0$).

\nl
We introduce the following notations, that will be used below.

\begin{itemize}
\item[{\tiny\textbullet}]  $\II:\quad (\L, \GG, \ell, {\rm g})\to \II(\L, \GG, \ell, {\rm g})=(\JJ(\L, \GG, \ell, {\rm g}), \EE(\L, \GG, \ell, {\rm g}))$ will be called {\it integral map};
\item[{\tiny\textbullet}]  The manifolds $$\cM_\m({\tt J}, {\tt E},\rr'):=\Big\{(\L, \GG, \ell, {\rm g}):\ \JJ(\L, \GG, \rr', \ell, {\rm g})={\tt J}, \  \EE(\L, \GG, \rr', \ell, {\rm g})={\tt E}\Big\}$$  will be named {\it reduced level sets}.  The {\it parameters} ${\tt I}:=({\tt J}, {\tt E})$ are allowed to vary in a certain set $\P_0(\rr')$ that we call {\it parameter space} that we shall specify below.

%\item[{\tiny\textbullet}] the union $\cM_\m(\rr', \Theta)$  of all $\cM_\m({\tt J}, {\tt E},\rr', \Theta)$'s;

\item[{\tiny\textbullet}] The sets \beqa{projected}  {\cal M}_0({\tt J}, \rr')=\bigcup_{{\tt E}:\ ({\tt J}, {\tt E})\in \P_0(\rr')}\cM({\tt J}, {\tt E}, \rr')\eeqa
 which will be called {\it ${\tt J}$--leaves of reduced phase space}.  

\item[{\tiny\textbullet}] The sets \beqa{M0mu}  {\cal M}_0(\rr')=\bigcup_{-\frac{\mm\cM}{\rr'}<{\tt J}<0}\cM({\tt J}, \rr')=\bigcup_{({\tt J}, {\tt E})\in \P_0(\rr')}\cM({\tt J}, {\tt E}, \rr')\eeqa
  given by the union of all of the  $\cM_\m({\tt J}, {\tt E},\rr')$'s,  will be called, as said, {\it reduced phase space}.   %which are {\it smooth, connected and compact}.
\item[{\tiny\textbullet}]   We choose
the parameter space $\P_0(\rr')$ in~\equ{M0mu}, as follows
\beqa{Pi0}
\P_0(\rr')&:=&\bigcup_{-\frac{\mm\cM}{\rr'}<{\tt J}<0} \P_0({\tt J}, \rr')
\eeqa
with  
\beqa{leafofparameterset}
\P_0({\tt J}, \rr')&:=&\Big\{{\tt E}:\  -\mm^2{\cal M}\rr'\le {\tt E}\le -\frac{\mm^3{\cal M}^2}{2{\tt J}}\left(1+\frac{{\rr'}^2{\tt J}^2}{\mm^2\cM^2}\right)\Big\} \eeqa
{\it the ${\tt J}$--leaf of $\P_0(\rr')$}.

\item[{\tiny\textbullet}] 
We also define, for any fixed $-\frac{\mm\cM}{\rr'}<{\tt J}<0$, the  {\it  separatrices in the parameter space}:
$${\Sigma}_{\rm 0, 1}(\rr'):=\bigcup_{-\frac{\mm\cM}{\rr'}<{\tt J}<0}{\Sigma}_{\rm 0, 1}({\tt J}, \rr')
$$
where
\beqa{J1}
%{\cal M}_{-}(\rr')&:=&\Big\{({\tt J}, {\tt E})\in{\mathbb R}_-\times {\mathbb R}:\quad \frac{{\tt E}}{{\tt L}_0({\tt J})^2}=-\d({\tt J},\rr')\Big\}\nonumber\\
\Sigma_{\rm 0}({\tt J}, \rr'):=\Big\{({\tt J}, \mm^2{\cal M}\rr')%:\quad \exists\ {\tt J}:\ -\frac{\mm\cM}{\rr'}<{\tt J}<0\ ,\quad {\tt E}=\mm^2{\cal M}
%\rr'
\Big\}\ ,\quad \Sigma_{\rm 1}({\tt J}, \rr'):=\left\{\left({\tt J}, -\frac{\mm^3{\cal M}^2}{2{\tt J}}\right)%:\quad -\frac{\mm\cM}{\rr'}<{\tt J}<0\ ,\quad {\tt E}=-\frac{\mm^3{\cal M}^2}{2{\tt J}}
\right\}
%{\cal M}_{+}(\rr')&:=&\Big\{({\tt J}, {\tt E})\in{\mathbb R}_-\times {\mathbb R}:\quad \frac{{\tt E}}{{\tt L}_0({\tt J})^2}=1+\frac{\d({\tt J},\rr')^2}{4}\Big\}\nonumber\\
%{\cal M}_0^*(\rr')&:=&{\cal M}_0\setminus\Big({\cal M}_{-}\cup {\cal M}_{+}\cup \Sigma_{\rm 0}\cup {\cal M}_{1}\Big)
\eeqa
are {\it  their ${\tt J}$--leaves}.
%the  {\it separatrices in the reduced phase space}:
%\beqa{separatrices}
%{\cal S}_{\rm 0}(\rr'):=\bigcup_{({\tt J}, {\tt E})\in \Sigma_{\rm 0}}\cM({\tt J}, {\tt E}, \rr')\ ,\quad {\cal S}_{\rm 1}(\rr'):=\bigcup_{({\tt J}, {\tt E})\in \Sigma_{\rm 1}}\cM({\tt J}, {\tt E}, \rr')\ .\eeqa
%These latter sets can be also written as the union, for all ${\tt J}$ as in~\equ{J1}, of the 
\item[{\tiny\textbullet}]  Define similarly the {\it ${\tt J}$--leaves of the separatrices in the reduced phase space}
\beqano
&&{\cal S}_{\rm 0, 1}({\tt J}, \rr'):=\cM_0({\tt J}, {\tt E}, \rr'):\quad ({\tt J}, {\tt E})\in \Sigma_{\rm 0, 1}({\tt J}, \rr') %\nonumber\\
%&& {\cal S}_{\rm 1}({\tt J}, \rr'):=\cM_0({\tt J}, {\tt E}, \rr'):\quad ({\tt J}, {\tt E})\in \Sigma_{\rm 1}({\tt J}, \rr')
\eeqano
\item[{\tiny\textbullet}]  The union of all  {\it ${\tt J}$--leaves of the separatrices in the reduced phase space}
$${\cal S}_{\rm 0, 1}(\rr'):=\bigcup_{-\frac{\mm\cM}{\rr'}<{\tt J}<0}{\cal S}_{\rm 0, 1}({\tt J}, \rr')
$$
will be called {\it separatrices in the reduced phase space}.
 \end{itemize}
 
\nl
Let us comment the choice of  $\P_0(\rr')$.
\begin{itemize}\rm

\item[\tiny\textbullet] Let
\beqa{delta***}
a:=-\frac{\mm\cM}{2{\tt J}}=\frac{\L^2}{\mm^2\cM}\ ,\quad \d:=-\frac{2\rr'{\tt J}}{\mm\cM}=\frac{\rr'}{a}\ ,\quad \widehat{\tt E}:=\frac{{\tt E}}{\mm^2\cM a}=\frac{{\tt E}}{\L^2}
\eeqa
with $a$ being the semi--major axis of the 
 instantaneous ellipse through $({\tt x}_\cK, {\tt y}_\cK)$.  In terms of the definitions in~\equ{delta***}, the interval for ${\tt J}$ in the definition of $\P_0(\rr')$ in~\equ{J1}
is just  \beqa{coll case}0<\d<2\ .\eeqa
In this situation,
a  proper choice of $(\L, \GG, \ell, {\rm g})$ leads to a collision between ${\tt x}'_\cK$ and ${\tt x}_\cK$, which is instead not possible under the the complemental situation condition 
\beqa{non coll case}
\d\ge 2
\eeqa
Also the nature of the equilibrium $(0,0)$ for the function $\EE_0$ is determined by the which among~\equ{coll case}  or~\equ{non coll case} is verified: it  is hyperbolic under~\equ{coll case}; elliptic under~\equ{non coll case}.  Therefore, {\it the choice of the interval of values for ${\tt J}$ corresponds to the set of values of ${\tt J}$ such that  collisions are possible or, equivalently, $(0,0)$ is an hyperbolic equilibrium to $\EE_0$,} with the projected separatrix ${\tt S}_{\rm 0}({\tt J}, \rr')$ corresponding to be the projected level set through such equilibrium.

\item[\tiny\textbullet]  The extremal values of ${\tt E}$ in the definition of $\P_0(\rr')$ in~\equ{J1} are chosen so as to coincide with maximum and the minimum of $\EE_0$ as a function of $(\GG, {\rm g})$. This can be easily seen changing the names as in~\equ{delta***},
dividing equation~\equ{P0} by $\mm^2\cM a=\L^2$, and checking that the maximum and minimum of $\widehat{\tt E}$ as a function of $\widehat\GG:=\GG/\L$ are $-\d$ and $1+\d^2/4$ (the details are given in Section~\ref{proof of phase portrait}).

 \end{itemize}

 \begin{definition}\label{def: reduced aa}\rm 
 We say that the couple of sets $({\cal W}^\ppj_\m(\rr')$, ${\cal M}^\ppj_\m(\rr'))$ is a couple of  {\it Liouville--Arnold domains} if 
    ${\cal M}^\ppj_\m(\rr')$  is a open and  connected subset of $ {\cal M}_\m(\rr')$   foliated by smooth, compact and connected reduced level sets $\cM_\m({\tt J}, {\tt E},\rr')$, which  is maximal with this properties and there exists a diffeomorphism
 
 \beqa{phij}
\begin{array}{llllll}\hat{\!\!\!\cA}:\quad &{\cal W}_\m^\ppj(\rr')&\to&\cM^\ppj_{\m}(\rr')\times{\mathbb T}^2\\\\
%\hat\cK=
&( \L, \GG,  \ell, {\rm g}) &\to%\hat{\!\!\!\cA}(\hat\cK)=
&\big(\cL, \cG,\l, \g\big)
 \end{array}
\eeqa
called {\it reduced action--angle coordinates}, which preserves the reduced symplectic form for any fixed $\rr'$, i.e., 
$$d\cL\wedge d\l+d\cG\wedge d\g=d\L\wedge d\ell+d\GG\wedge d{\rm g}\ \qquad \forall\rr'$$
and
such that $(\cL, \cG)$ are first integrals to $\EE$, $\JJ$.

     \end{definition}

\begin{definition}\label{full aa}\rm
We say that $\cA=(\RR', \cL, \cG, \rr', \l, \g)$ are {\it full action--angle coordinates} to $\JJ$ if it is possible to determine a diffeomorphism

\beqa{aa}
\begin{array}{llllll}{\!\!\!\cA}:\quad &{\cal W}_\m^\ppj&\to&\cM^\ppj_{\m}\times{\mathbb T}^2\\\\
%\hat\cK=
&(\RR', \L, \GG, \hat\rr',  \ell, {\rm g}) &\to%\hat{\!\!\!\cA}(\hat\cK)=
&\big(\hat\RR', \cL, \cG,\hat\rr', \l, \g\big)
 \end{array}
\eeqa
where
\beqano
&&{\cal W}^\ppj_\m:=\Big\{(\hat\rr', \cL, \cG):\quad \hat\rr'\in {\mathbb R}_+\ ,\quad (\cL, \cG)\in {\cal W}^\ppj(\hat\rr')\Big\}\nonumber\\
&&{\cal M}^\ppj_\m:=\Big\{(\rr', \L, \GG, \ell, {\rm g}):\quad \rr'\in {\mathbb R}_+\ ,\quad (\L, \GG, \ell, {\rm g})\in \cM^\ppj_{\m}(\rr')\Big\} 
\eeqano
which preserves the  symplectic form
$$d\hat\RR'\wedge d\hat\rr'+d\cL\wedge d\l+d\cG\wedge d\g=d\RR'\wedge d\rr'+d\L\wedge d\ell+d\GG\wedge d{\rm g}\ $$
 and such that  the projection of $\cA_\m^\ppj$ on the coordinates $\cL, \cG, \l, \g$ coincides with ${\cA^\ppj_\m}(\rr')$ in~\equ{phij}.

 \end{definition}

\nl
In the next sections we study Liouville--Arnold domains and full action--angle coordinates. The case $\m=0$ is explicit.

%\subsection{The case $\m=0$}
%\subsubsection{Phase portrait}
\subsection{Case $\m=0$}

Using the definitions in~\equ{delta***}, we rewrite the parameter set $\P_0(\rr')$  in the more compact form 

\beqano
&&\P_0(\rr')=\left\{({\tt J}, {\tt E}):\quad 0<\d<2\ ,\quad -\d\le \widehat{\tt E}\le 1+\frac{\d^2}{4}\right\}
\eeqano
Moreover, from the definitions of $\Sigma_{\rm 0}({\tt J}, \rr')$, $\Sigma_{\rm 1}({\tt J}, \rr')$ it follows that
\beqano
&&({\tt J}, {\tt E})\in \Sigma_{\rm 0}(\rr')\quad \Leftrightarrow\quad 0<\d<2\ ,\quad  \widehat{\tt E}=\d\nonumber\\
&&({\tt J}, {\tt E})\in\Sigma_{\rm 1}(\rr')\quad \Leftrightarrow\quad0<\d<2\ ,\quad  \widehat{\tt E}=1%{\cal M}_{+}(\rr')&:=&\Big\{({\tt J}, {\tt E})\in{\mathbb R}_-\times {\mathbb R}:\quad \frac{{\tt E}}{{\tt L}_0({\tt J})^2}=1+\frac{\d({\tt J},\rr')^2}{4}\Big\}\nonumber\\
%{\cal M}_0^*(\rr')&:=&{\cal M}_0\setminus\Big({\cal M}_{-}\cup {\cal M}_{+}\cup \Sigma_{\rm 0}\cup {\cal M}_{1}\Big)
\eeqano
Therefore, $\Sigma_{\rm 0}(\rr')$, $\Sigma_{\rm 1}( \rr')$ are one--dimensional subsets  $\P_0(\rr')$. As such, when $\d\ne 1$, they divide the two--dimensional parameter space $\P_0(\rr')$ in three open and connected components, given by
\beqano
&&\P^\ppu_0(\rr')=\Big\{({\tt J}, {\tt E}):\quad 0<\d<2\ ,\quad -\d< \widehat{\tt E}<\min\{\d,\ 1\}\Big\}\nonumber\\
&&\P^\ppd_0(\rr')=\Big\{({\tt J}, {\tt E}):\quad 0<\d<2\ ,\quad \min\{\d,\ 1\}< \widehat{\tt E}<\max\{\d,\ 1\}\Big\}\nonumber\\
&&\P^\ppt_0(\rr')=\Big\{({\tt J}, {\tt E}):\quad 0<\d<2\ ,\quad \max\{\d,\ 1\}< \widehat{\tt E}< 1+\frac{\d^2}{4}\Big\}
\eeqano
%When $\d=1$, the connected components are just two, $\P^\ppu_0(\rr')$ and $\P^\ppt_0(\rr')$. 
We let%, in both cases,
\beqano\cM^\ppj_0(\rr')=\bigcup_{({\tt J}, {\tt E})\in \P_0^\ppj(\rr')} \cM_0({\tt J}, {\tt E}, \rr')\ .\eeqano

\begin{proposition}\label{prop: aa}
It is possible to find  two functions $\cG_-(\cL_0, \rr')$, $\cG_+(\cL_0, \rr')$, smooth for $\cL_0\ne\sqrt{\mm^2\cM{\rr'}}$,
such that $({\cal W}^\ppj_0(\rr'), {\cal M}^\ppj_0(\rr'))$
are Liouville--Arnold domains, with
\beqa{Wi}
&&{\cal W}^\ppu_0(\rr')=\Big\{(\cL_0, \cG_0):\quad \cL_0>\sqrt{\frac{\mm^2\cM\rr'}{2}}\ ,\ 0< \cG_0< \cG_-(\cL_0, \rr')\Big\}\nonumber\\
&&{\cal W}^\ppd_0(\rr')=\Big\{(\cL_0, \cG_0):\quad \cL_0>\sqrt{\frac{\mm^2\cM\rr'}{2}}\ ,\ \cG_-(\cL_0, \rr')< \cG_0< \cG_+(\cL_0, \rr')\Big\}\nonumber\\
&&{\cal W}^\ppt_0(\rr')=\Big\{(\cL_0, \cG_0):\quad \cL_0>\sqrt{\frac{\mm^2\cM\rr'}{2}}\ ,\ \cG_+(\cL_0, \rr')< \cG_0< \cL_0\Big\}
\eeqa
%For $\d=1$, $\cG_-(\cL_0, \rr')=\cG_+(\cL_0, \rr')=:{\tt G}_{\rm mid}(\cL_0, \rr')$ and the Liouville--Arnold domains reduce to
%${\cal W}^\ppu_0(\rr')$ and ${\cal W}^\ppt_0(\rr')$.
\end{proposition}

%\nl
%We prove this proposition only in the case $0<\d<2$, $\d\ne 1$. The case $\d=1$ is similar.

\nl
When $\m=0$, the equations for the level sets $\cM_0({\tt J}, {\tt E},\rr')$ reduce to
\beqa{level curves}
\arr{\dst{\JJ}_0(\L)=-\frac{\mm^3{\cal M}^2}{2\L^2}={\tt J}\\
\dst\EE_0(\L, \GG, {\rm g};\rr')=\GG^2+\mm^2{\cal M}\rr'\sqrt{1-\frac{\GG^2}{\L^2}}\cos{\rm g}={\tt E}}
\eeqa

\nl
Such equations show that each level set $\cM_0({\tt J}, {\tt E},\rr')$ is the product  
\beqa{M0}\cM_0({\tt J}, {\tt E},\rr')=\ovl\cC_{0, 1}({\tt J})\times\ovl\cC_{0, 2}({\tt J}, {\tt E},\rr')\eeqa
with
$$\ovl\cC_{0, 1}({\tt J})=\{{\tt L}_0({\tt J})\}\times {\mathbb T}$$
where 
\beqa{L_0(E0)}{\tt L}_0({\tt J})=\sqrt{-\frac{\mm^3{\cal M}^2}{2{\tt J}}}\ ,\qquad {\tt J}<0\eeqa
while   $\ovl\cC_{0, 2}({\tt J}, {\tt E},\rr')$ is the set of $(\GG, {\rm g})$
such that
\beqa{P0}%\ovl\EE_0(\GG, {\rm g}, {\tt J},\rr'):=
\EE_0=\GG^2+\mm^2{\cal M}\rr'\sqrt{1-\frac{\GG^2}{\L^2}}\cos{\rm g}={\tt E}\eeqa
with $\L$ replaced by ${\tt L}_0({\tt J})$.  $\ovl\cC_{0, 1}({\tt J})$, $\ovl\cC_{0, 2}({\tt J}, {\tt E},\rr')$ will be called {\it projected level sets} or also {\it base circles}  with $\m=0$.

\nl
It follows that any ${\tt J}$--leaf of the phase space~\equ{projected}
is
$$\cM({\tt J}, \rr')=\ovl\cC_{0, 1}({\tt J})\times \ovl\cC_{0, 2}({\tt J}, \rr')$$
with
\beqa{projected phase space}\ovl\cC_{0, 2}({\tt J}, \rr'):=\bigcup_{{\tt E}\in \P_0({\tt J}, \rr')}\ovl\cC_{0, 2}({\tt J},{\tt E},  \rr')\eeqa
the {\it   the projected phase space}.
 %}
%&&\cN_\eta:= \arr{\{\eta\}\quad\quad\  {\rm if}\quad \eta\in (0,1)\\
%\{1, \eta\}\quad {\rm if}\quad \eta\in [1,2)\label{N}
%Let
%\beqa{L_0(E0)} a_0({\tt J}):=\frac{{\tt L}_0({\tt J})^2}{{\cal M}\mm^2}\ ,\qquad \d({\tt J},\rr'):=-2\frac{\rr'{\tt J}}{\mm\cM}= \frac{\rr'}{a_0({\tt J})}
%\eeqa
%and take
%$\d({\tt J},\rr')\in (0,2)$.

\nl
%Let $\d(h, \rr')$  and ${\tt L}_0(h)$ be as in~\equ{L_0(E0)}.
%Then  let% and $\eta\in (0,2)$,

\nl
By~\equ{M0} also the ${\tt J}$--leaves of the separatrices in the reduced phase space are 

\beqa{leafs}
&&{\cal S}_{\rm 0, 1}({\tt J}, \rr'):=\{{\tt L}_0({\tt J})\}\times{\mathbb T}\times {\tt S}_{\rm 0, 1}({\tt J},\rr')%\ ,\nonumber\\
%&& {\cal S}_{\rm 1}({\tt J}, \rr'):=\{{\tt L}_0({\tt J})\}\times{\mathbb T}\times {\tt S}_1({\tt J},\rr')
\eeqa
with
\beqano
&&{\tt S}_{\rm 0, 1}({\tt J},\rr'):=\ovl\cC_{0, 2}({\tt J}, {\tt E}, \rr'):\quad ({\tt J}, {\tt E})\in \Sigma_{\rm 0, 1}({\tt J}, \rr')%\nonumber\\
%&& {\tt S}_{\rm 1}({\tt J},\rr'):=\ovl\cC_{0, 2}({\tt J}, {\tt E}, \rr'):\quad ({\tt J}, {\tt E})\in \Sigma_{\rm 1}({\tt J}, \rr')
\eeqano
being the {\it projected separatrices}. 

\nl
Observe that the  {\it projected level set ${\tt S}_0$} is the level curve for $\EE_0$ in the space $(\GG, {\rm g})$ through the origin, where
that $\EE_0$ has an extremal point at $(\GG, {\rm g})=(0,0)$:
$$\partial_\GG\EE_0|_{(\GG, {\rm g})=(0,0)}=\partial_{\rm g}\EE_0|_{(\GG, {\rm g})=(0,0)}=0\ ,\qquad \EE_0|_{(\GG, {\rm g})=(0,0)}=\mm^2\cM\rr'\ .$$

 \begin{figure}
 \includegraphics[height=4.0cm, width=7.0cm]{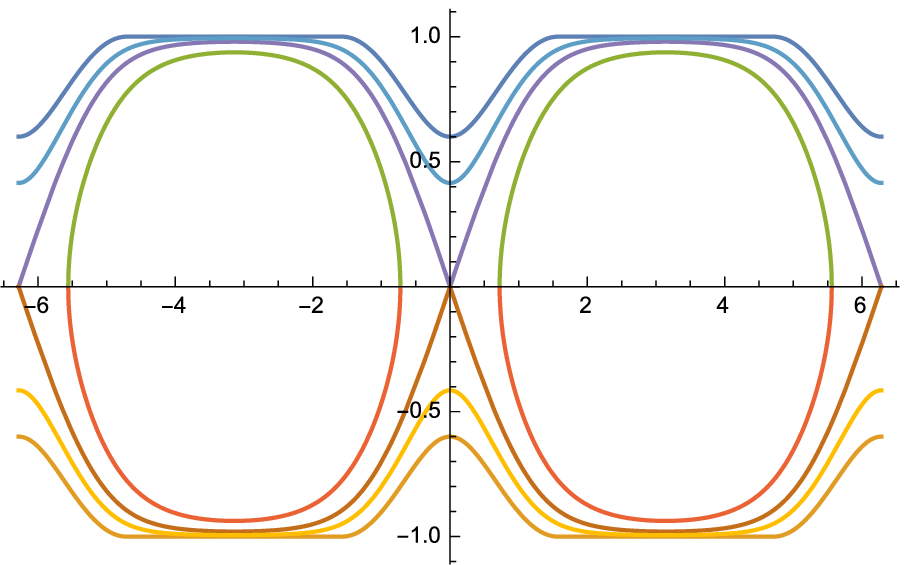} \includegraphics[height=4.0cm, width=7.0cm]{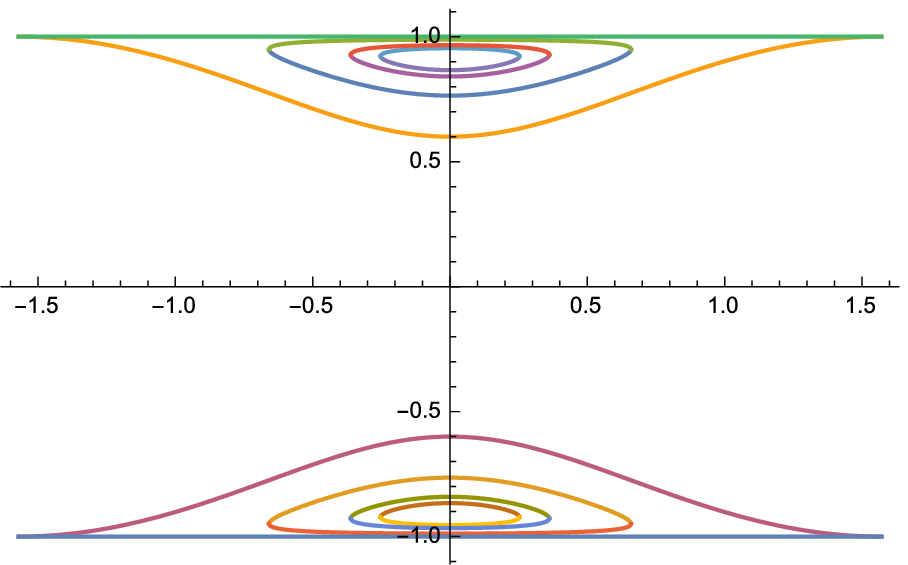} %
\caption{Case $0<\d<1$. $
{\tt S}_{\rm 0}({\tt J}, \rr')$ (left) is inner to $
{\tt S}_{\rm 1}({\tt J}, \rr')$ (right).}\label{figure12}
% \end{figure}
 %  \begin{figure}
 \center{\includegraphics[height=4.0cm, width=7.0cm]{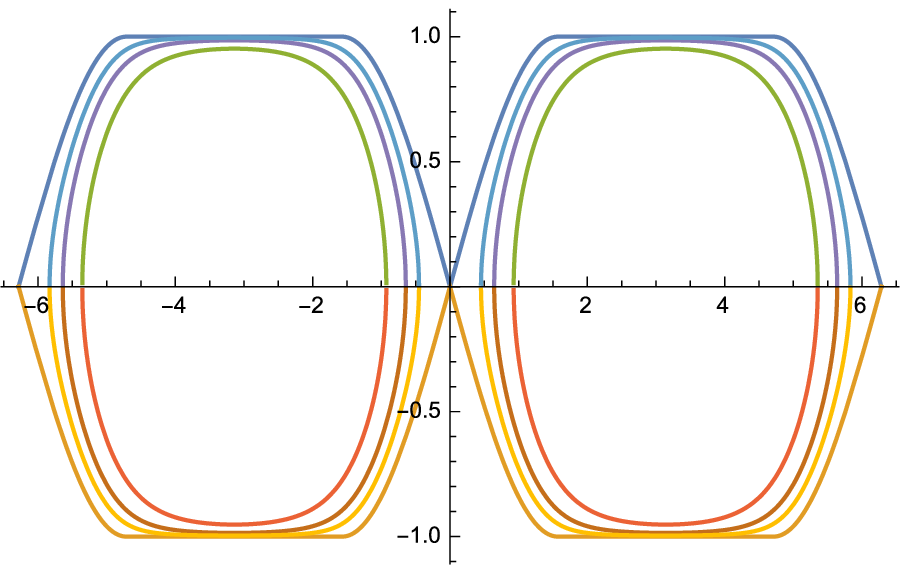}\includegraphics[height=4.0cm, width=7.0cm]{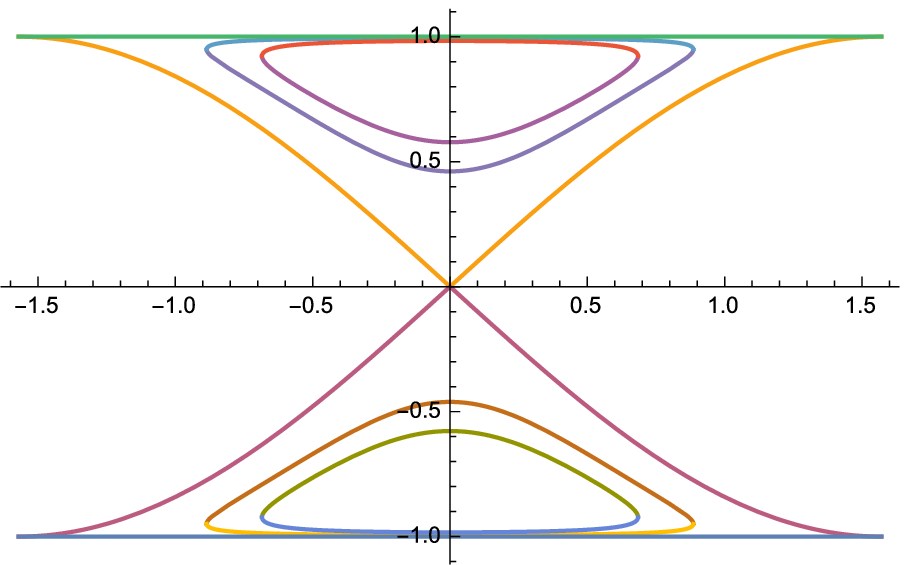}}  \caption{
Case $\d=1$. $
{\tt S}_{\rm 0}({\tt J}, \rr')$ (left) and $
{\tt S}_{\rm 1}({\tt J}, \rr')$ (right) coincide.}\label{figure5}
% \end{figure}
% \begin{figure}
 \includegraphics[height=4.0cm, width=7.0cm]{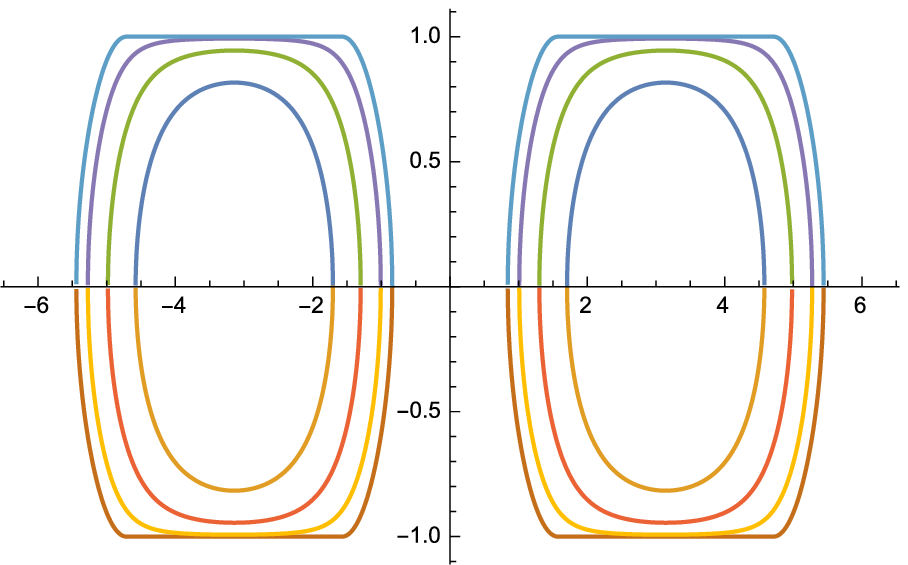}\includegraphics[height=4.0cm, width=7.0cm]{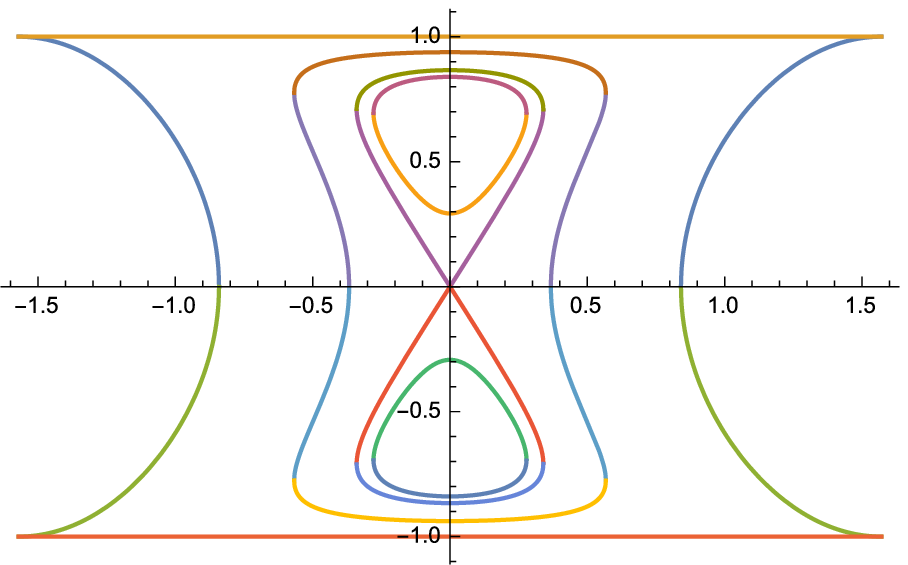}%
 \caption{Case $1<\d<2$. $
{\tt S}_{\rm 1}({\tt J}, \rr')$(left) is inner to $
{\tt S}_{\rm 0}({\tt J}, \rr')$ (right).}\label{figure34}
 \end{figure}

%so that the set ${\cal M}_0^*$ in~\equ{J} corresponds to be ${\cal M}_0^*={\cal M}_0\setminus\{{\cal M}_{-}\ ,\ \Sigma_{\rm 0}\ ,\ \Sigma_{\rm 1}\ ,\ {\cal M}_{+}\}$.

%The  {\it separatrix} corresponds to be the level curve with $\frac{{\tt E}}{{\tt L}_0({\tt J})^2}=\d({\tt J},\rr')$.

\nl
The proof of the  proposition below is deferred to Section~\ref{proof of phase portrait}.  %It is a more detailed statement of the first thesis of Proposition~\ref{prop: aa}.

\begin{proposition}\label{prop: phase portrait}
The reduced phase space
${\cal M}_0(\rr')$ is includes three  open and connected subsets   foliated by compact, connected and smooth reduced level sets 
given by
$${\cal M}^\ppj_0(\rr')=\bigcup_{({\tt J}, {\tt E})\in \P_0^\ppj(\rr')}\ovl\cC_{0, 1}({\tt J})\times \ovl\cC_{0, 2}({\tt J}, {\tt E}, \rr')\}\ .$$ 
%such that
% $$\cP_0(\rr')=\bigcup_{-\frac{\mm\cM}{\rr'}<{\tt J}<0}\ovl\cC_{0, 2}({\tt J},  \rr')=[0, \L]\times {\mathbb T}$$  
The ${\cal M}^\ppj_0(\rr')$ are  
delimited by the two separatrices in the reduced phase space
$
{\cal S}_{\rm 0}(\rr')$, ${\cal S}_{\rm 1}(\rr')$, and are in turn foliated by~\equ{leafs}. The projected separatrix $
{\tt S}_{\rm 0}({\tt J}, \rr')$ in the projected phase space $
\cP_{\rm 0}({\tt J}, \rr')$   is inner, outer to the projected separatrix $
{\tt S}_{\rm 1}({\tt J}, \rr')$, accordingly to wether $0<\d< 1$ or $1< \d<2$, respectively. For $\d=1$,
${\tt S}_{\rm 0}({\tt J}, \rr')$ and ${\tt S}_{\rm 1}({\tt J}, \rr')$ coincide.  The projected separatrix  $
{\tt S}_{\rm 0}({\tt J}, \rr')$ coincides with the $\EE_0$--level set to the saddle $(\GG, {\rm g})=(0,0)$
(see  Figures~\ref{figure12},~\ref{figure5} and~\ref{figure34}).
 \end{proposition}
 
 \begin{definition}\label{action map}\rm
 We say that
 $${\tt A}(\rr'):\qquad {\tt I}=({\tt J}, {\tt E})\in \P^\ppj(\rr')\to {\tt A}(\rr')({\tt I})=\big({\tt L}({\tt J}, {\tt E}), {\tt G}({\tt J}, {\tt E})\big)\in {\cal W}^\ppj(\rr')$$
 is an action map to a  reduced action--angle coordinates
\equ{phij}
  if ${\tt A}$ is a diffeomorphism and 
 $$
 {\tt A}\circ\II=(\cL, \cG)
 $$
 \end{definition}

\nl
To prove the second thesis of Proposition~\ref{prop: aa}, we first need to show the following result.

\begin{proposition}\label{prop: continuity} It is possible to find  reduced action--angle coordinates 
 \beqa{phij0}
\begin{array}{llllll}\ \ {{\hat{\!\!\!\cA}}_0}(\rr'):\quad &{\cal W}_0^\ppj(\rr')\times{\mathbb T}^2&\to \cM^\ppj_{0}(\rr')\\\\
&{\hat{\!\!\!\cA}}_0(\rr')=(\cL_0, \cG_0,\l_0,\g_0) &\to \hat\cK=( \L, \GG,  \ell, {\rm g})
 \end{array}
\eeqa
equipped with an action map of the form
$${\tt A}_0(\rr')({\tt I})= ({\tt L}_0({\tt J}), {\tt G}_0({\tt J}, {\tt E}, \rr'))$$
where  ${\tt E}\in\P_0({\tt J}, \rr')\to{\tt G}_0({\tt J}, {\tt E},\rr')$ is continuous and increasing on all of $\P_0({\tt J}, \rr')$.
\end{proposition}

\nl
The main point of this proposition is the continuity of the function ${\tt E}\to{\tt G}_0({\tt J}, {\tt E},\rr')$ for ${\tt E}\in\P_0({\tt J}, \rr')$, namely, also for  ${\tt E}\in \Sigma_0({\tt J}, \rr')$, or $\Sigma_1({\tt J}, \rr')$. 

\nl
Referring to Section~\ref{proof of 4.3} for the technical details, let us remark, here, the main idea that allows to obtain this continuity. The proof consists of two steps. At the first step, we check that  Arnold's scheme is well defined to this case. As well known, by~\cite{arnold63a},  
one first constructs a map
\beqa{JELG}\hat{\tt A}_0:\quad {\tt I}=({\tt J}, {\tt E})\in \P^\ppj_0(\rr')\to (\hat{\tt L}_0({\tt J}), \hat{\tt G}_0({\tt J}, {\tt E}, \rr'))\in {\cal W}^\ppj_0(\rr')\eeqa
where
 $\hat{\tt L}_0({\tt J})$ $\hat{\tt G}_0({\tt J}, {\tt E}, \rr')$, which we call {\it Arnold actions}, are defined, in terms of the integrals ${\tt J}$ and ${\tt E}$ as
$$\hat{\tt L}_0({\tt J})={\tt In}[\ovl\cC_{0, 1}({\tt J})]\ ,\qquad \hat{\tt G}_0({\tt J}, {\tt E}, \rr')={\tt In}[\ovl\cC_{0, 2}({\tt J}, {\tt E}, \rr')]$$ 
where  ${\tt In}[\cC]$ denotes  the area of the inner region delimited by  $\cC$, divided by $2\p$. The first equation immediately gives
$\hat{\tt L}_0({\tt J})={\tt L}_0({\tt J})=\L$, so the problem reduces to the determination of angle coordinate, which we call {\it Arnold angle}, associated to the one--degree of freedom Hamiltonian $(\GG, {\rm g})\to \EE_0$ in~\equ{P0}, which, accordingly to~\cite{arnold63a}, is given by

\beqa{g0}\hat\g_0(\L, \GG,  {\rm g})=2\p\left.\frac{t({\tt J}, {\tt E}, \rr';\GG)}{T({\tt J}, {\tt E}, \rr')}\right|_{({\tt J}, {\tt E})=(\JJ_0(\L), \EE_0(\L, \GG, {\rm g}))}\eeqa
where, for a given $\GG$  such that $(\GG, {\rm g})\in$ $\ovl\cC_{0, 2}({\tt J}, {\tt E}, \rr')$ for some ${\rm g}$,
$t({\tt J}, {\tt E}, \rr';\GG)$ is the time needed to reach $(\GG, {\rm g})$ on $\ovl\cC_{0, 2}({\tt J}, {\tt E}, \rr')$ and $T({\tt J}, {\tt E}, \rr')$ is the period associated to  $\ovl\cC_{0, 2}({\tt J}, {\tt E}, \rr')$. Even though
the explicit computation is forbidden, since it involves elliptic integrals, however,  it is standard (see Section~\ref{proof of 4.3**} for the details) to show that
  \begin{proposition}\label{propo: invertibility}
For all $j=1$, $2$, $3$, the Arnold map~\equ{JELG} is a action map to a reduced action--angle coordinate
\beqa{Arnold}\hat{\hat{\!\!\!\cA}}_0:\quad(\hat\cL_0, \hat\cG_0, \hat\l_0, \hat\g_0)\in {\cal W}^\ppj_0(\rr')\times {\mathbb T}^2\to (\L, \GG, \ell, {\rm g})\in {\cal M}^\ppj_0(\rr')\eeqa
to $\JJ_0$, where $\hat\g_0$ is as in~\equ{g0}.
\end{proposition}
To avoid confusion with the symbols  defined below, we  shall refer to $\hat{\tt A}_0$ as {\it Arnold action map} and to $\  \hat{\hat{\!\!\!\cA}}_0$ as {\it Arnold action--angle coordinates}.
Indeed, as a the second step for the proof of Proposition~\ref{prop: continuity}, we introduce a modification of Arnold's action--angle coordinates, redefining
%Proposition~\ref{inversion} guarantees that the function ${\tt E}\in\P_0({\tt J}, \rr')\to {\tt G}_0({\tt J}, {\tt E}, \rr')$ is increasing in any connected component of the set $\P_0({\tt J}, \rr')$ in~\equ{leafofparameterset},
%but  ${\tt E}\in\P_0({\tt J}, \rr')\to {\tt G}_0({\tt J}, {\tt E}, \rr')$ needs not to be continuous. We now show that the coordinates~\equ{phij0} can be chosen so that the function ${\tt E}\in\P_0({\tt J}, \rr')\to {\tt G}_0({\tt J}, {\tt E}, \rr')$ is, besides increasing, also continuous.
%
%\nl
$${\tt L}_0({\tt J})=\hat{\tt L}_0({\tt J})$$
and ${\tt G}_0({\tt J}, {\tt E}, \rr')$ as follows.  If $0<\d\le 1$:
\beqa{G01}{\tt G}_0({\tt J}, {\tt E}, \rr'):=\left\{
\begin{array}{lrr}
{\tt In}\Big[\ovl\cC_{0, 2}({\tt J}, {\tt E}, \rr')\Big] \quad\textrm{ for } -\d< \widehat{\tt E}<\d\ \textrm{ or }\d< \widehat{\tt E}<1\\\\
{\tt Ext}\Big[\ovl\cC_{0, 2}({\tt J}, {\tt E}, \rr')\Big] \quad\textrm{ for }  1< \widehat{\tt E}<1+\frac{\d^2}{4}
\end{array}
\right.
\eeqa
 If $1<\d<2$:

\beqa{G02}
{\tt G}_0({\tt J}, {\tt E}, \rr'):=\left\{
\begin{array}{lrr}
{\tt In}\Big[\ovl\cC_{0, 2}({\tt J}, {\tt E}, \rr')\Big] \quad\textrm{ for }  -\d< \widehat{\tt E}<1\\\\
{\tt Ext}\Big[\ovl\cC_{0, 2}({\tt J}, {\tt E}, \rr')\Big] \quad\textrm{ for }1< \widehat{\tt E}<\d\quad {\rm or}\quad 
\d<\widehat{\tt E}<1+\frac{\d^2}{4}
\end{array}
\right.\eeqa

\nl
 where   ${\tt Ext}[\ovl\cC_{0, 2}({\tt J}, {\tt E}, \rr')]$
denote the area, normalized to $2\p$, of  the {\it complement} of  the  inner region, with respect to the strip $[0, \L]$. It is evident from the definition that 

\begin{proposition}\label{prop: continuity***}
The function ${\tt E}\to {\tt G}_0({\tt J}, {\tt E}, \rr')$ is continuous on $\P_0({\tt J}, \rr')$.
\end{proposition}
The main point that allows this new definition is the relation
\beqa{Ext}
%{\tt In}\Big[\ovl\cC_{0, 2}({\tt J}, {\tt E}, \rr')\Big]&=&\frac{1}{2\p}\int_0^{2\p}\GG_0({\tt J}, {\tt E}, \psi_2, \rr') \partial_{\psi_2}{\rm g}_0({\tt J}, {\tt E}, \psi_2, \rr') d\psi_2\nonumber\\
{\tt In}\Big[\ovl\cC_{0, 2}({\tt J}, {\tt E}, \rr')\Big]+{\tt Ext}\Big[\ovl\cC_{0, 2}({\tt J}, {\tt E}, \rr')\Big]=\L= {\tt L}_0({\tt J})
\eeqa
thanks to which we obtain a new set of coordinates
$$\hat{\!\!\!\cA}_0=(\cL_0, \cG_0, \l_0, \g_0)$$
such  that
 the coordinates $(\l_0, \g_0)$, naturally and canonically associated to the new actions, are angles too, since they are related to $(\hat\l_0, \hat\g_0)$ via linear relations with integer coefficients.

\nl
We finally turn to full action--angle coordinates to $\JJ_0$. Let ${\cal W}^\ppj_0(\rr')$, ${\cal M}^\ppj_0(\rr')$ be as in Proposition~\ref{prop: aa} and let ${\cal W}^\ppj_0$, ${\cal M}^\ppj_0$ be as in
Definition~\ref{full aa}. A standard procedure (see Section~\ref{full with m=0} for details) allows to prove that

\begin{proposition}\label{phiA0}
It is possible to find full action--angle coordinates $\cA_0$ to $\JJ_0$ with ${\cal W}^\ppj_0(\rr')$, ${\cal M}^\ppj_0(\rr')$ as in Proposition~\ref{prop: aa} and $\cA_0$ having the form
\beqa{lift}&&\GG={\tt G}_{0}(\cL_0, \cG_0, {\rm g}_0, \hat\rr_0')\ ,\quad {\rm g}={\tt g}_{0}(\cL_0, \cG_0, {\rm g}_0, \hat\rr_0')\nonumber\\
&&\L=\cL_0\ ,\quad \ell=\l_0+\f_0(\cL_0, \cG_0, {\rm g}_0, \hat\rr_0')\nonumber\\
&&\RR'={\hat\RR}_0'+\r_{0}(\cL_0, \cG_0, {\rm g}_0, \hat\rr_0')\ ,\quad \rr'={\hat\rr}_0'
\eeqa
\end{proposition}

\subsection{Case $\m>0$, small}\label{mu>0}

The following proposition is proved in Section~\ref{proof of 4.13}

% DA QUI
\begin{proposition}\label{Liouville Arnold3bis}
For all $j=1$, $2$, $3$ and any compact  set %$\cR\subset {\mathbb R}_+$ compact, 
${\tt K}\subset \P^\ppj_0(\rr')$ 
%
% any map $\rr'\in \cR\to {\cal M}^*(\rr')$  with ${\cal M}^*(\rr')\subset {\cal M}_0^*(\rr')$ compact,  
one can find $\m_0({\tt K})$ such that, for all $({\tt J}, {\tt E})\in {\tt K}$, all $0<\m<\m_0({\tt K})$, 
$\cM_\m({\tt J}, {\tt E}, \rr')$ is smooth, connected and compact. 
\end{proposition}

\nl
The previous proposition has the following consequence. Recall the definition of the sets ${\cal W}^\ppj_0(\rr')$ in~\equ{Wi}. For a given set $D\subset{\mathbb R}^n$, denote
$$D_{-\d_0}:=\big\{{\tt P}\in {\mathbb R}^n:\ B^n_{\d_0}({\tt P})\subset D\big\}$$

\begin{proposition}\label{act aa mu pos}
Fix $j=1$, $2$, $3$; $\d_0>0$ so small that $\P^\ppj(\rr')_{-2\d_0}\ne \emptyset$, and a compact set ${\tt K}\subset \P^\ppj(\rr')_{-2\d_0}$. Then it is possible to find $\m_1(\d_0, {\tt K})>0$ and ${\cal W}^\ppj_{\m, {\tt K}}(\rr')\subset {\cal W}^\ppj_0(\rr')$ such that, if $0<\m<\m_1(\d_0, {\tt K})$ and $$\cM^\ppj_{\m, {\tt K}}(\rr'):=\bigcup_{({\tt J}, {\tt E})\in {\tt K}}\cM_{\m, {\tt K}}({\tt J}, {\tt E}, \rr')\ ,$$ $({\cal W}^\ppj_{\m, {\tt K}}(\rr'), \cM^\ppj_{\m, {\tt K}}(\rr'))$ are Liouville--Arnold domains for $\JJ$.
\end{proposition}

\proof By Proposition~\ref{Liouville Arnold3bis}, if $0<\m<\m_0({\tt K})$, one can find  ${\cal W}^\ppj_{\m, {\tt K}}(\rr')$ such that $({\cal W}^\ppj_{\m, {\tt K}}(\rr'), \cM^\ppj_{\m, {\tt K}}(\rr'))$ are Liouville--Arnold domains for $\JJ$. But also, by Proposition~\ref{prop: aa}, one can find  $\tilde{\cal W}^\ppj_{0, {\tt K}}(\rr')\subset {\cal W}^\ppj_0(\rr')$ such that $(\tilde{\cal W}^\ppj_{0, {\tt K}}(\rr'), \cM^\ppj_{0, {\tt K}}(\rr'))$ are Liouville--Arnold domains for $\JJ_0$. 
Since ${\tt K}\subset \P^\ppj_{-2\d_0}(\rr')$, there  exists $\d_1$ depending on $\d_0$  such that
$\tilde{\cal W}^\ppj_{0, {\tt K}}(\rr')\subset {\cal W}^\ppj_{0, -2\d_1}(\rr')\subset {\cal W}^\ppj_0(\rr')$. But $\tilde{\cal W}^\ppj_{0, {\tt K}}(\rr')$ is $\m$--close to ${\cal W}^\ppj_{\m, {\tt K}}(\rr')$, so, for a suitable $0<\m_1(\d_0, {\tt K})<\m_0({\tt K})$, 
${\cal W}^\ppj_{\m, {\tt K}}(\rr')\subset {\cal W}^\ppj_{0, -\d_1}(\rr')$,
for all $0<\m<\m_1(\d_0, {\tt K})$. Since ${\cal W}^\ppj_{0, -\d_1}(\rr')\subset {\cal W}^\ppj_{0}(\rr')$, the theorem is proved. \quad $\square$

\nl
The following Proposition is proved in Section~\ref{proof of 4.13}.
 
\begin{proposition}\label{Liouville Arnold3quater}

Under the assumptions of Proposition~\ref{Liouville Arnold3bis}, a diffeomorphism
$$\phi_\m({\tt J}, {\tt E}, \rr'):\quad \psi=(\psi_1, \psi_2)\in{\mathbb T}^2\to \cM_\m({\tt J}, {\tt E}, \rr') $$
can be chosen of the form
%and can be represented as union of graphs 
\beqano\phi_\m({\tt J}, {\tt E}, \rr'):\qquad \arr{\dst
\L=\L_\m({\tt J}, {\tt E}, \rr', \psi_1, \psi_2)\\
\GG=\GG_\m({\tt J}, {\tt E}, \rr', \psi_1, \psi_2)\\
\ell=\psi_1\\
{\rm g}={\rm g}_\m({\tt J}, {\tt E}, \rr', \psi_1, \psi_2)
}%\qquad \arr{\dst
%\L=\L_\m({\tt J}, {\tt E}, \rr', \ell, {\rm g})\\
%\GG=-\GG_\m({\tt J}, {\tt E}, \rr', -\ell, -{\rm g})\\
%\ell\in {\mathbb T}\\
%{\rm g}\in \cD_\m({\tt J}, {\tt E}, \rr')
%}
\eeqano
where ${\rm g}_\m$ verifies
$${\rm g}_\m({\tt J}, {\tt E}, \rr', \psi_1, 0)=\frac{1-\s({\tt J}, {\tt E})}{2}\p\ ,\quad \forall\ \psi_1\in {\mathbb T}$$ 
with $\s({\tt J}, {\tt E}):=\arr{
-1\quad -\d<\widehat{\tt E}<1\\
+1\quad 1<\widehat{\tt E}<1+\frac{\d^2}{4}\ .\\
}$
%with $\cD_\m({\tt J}, {\tt E}, \rr')\subset [0,2\p]$ mod $2\p$, compact.
\end{proposition}

%By construction, the union of the graphs~\equ{FG1} is a graph of the form of the first formula  in~\equ{graphs}, with $\cD_\m$ replaced by $\cD^\circ:=\cup_i\cD_i$.
%\\
%Since $\cD_i$'s, ${\cD'}_i$'s  can be chosen so that the set $\cD^\circ$, ${\cD'}^\circ:=\cup_i\cD_i$ is an arbitrary punctured neighborhood of a finite number of points where, using
%the graphs~\equ{FG}, they it be extended with regularity, one obtains a
%graph of the form of the first formula  in~\equ{graphs}. The second formula follows by symmetry.
\nl
We define the {\it base circles} $$\cC_{1, \m}:=\phi_\m({\tt J}, {\tt E}, \rr')({\mathbb T}\times \{0\}_{\rm mod \ 2\p})\ ,\quad \cC_{2, \m}:=\phi_\m({\tt J}, {\tt E}, \rr')(\{0\}_{\rm mod\ 2\p}\times {\mathbb T})$$
which have parametric equation 
$$\cC_{1,\m}({\tt J}, {\tt E}, \rr'):\quad \arr{\dst
\L=\L_{\m, 1}({\tt J}, {\tt E}, \rr', \psi_1)\\
\GG=\GG_{\m, 1}({\tt J}, {\tt E}, \rr',  \psi_1)\\
\ell=\psi_1\\
{\rm g}= \frac{1-\s}{2}\p
}\qquad \psi_1\in {\mathbb T}$$
$$\cC_{2,\m}({\tt J}, {\tt E}, \rr'):\quad \arr{\dst
\L=\L_{\m, 2}({\tt J}, {\tt E}, \rr', \psi_2)\\
\GG=\GG_{\m, 2}({\tt J}, {\tt E}, \rr',  \psi_2)\\
\ell=0\\
{\rm g}={\rm g}_{\m, 2}({\tt J}, {\tt E}, \rr',  \psi_2)
}\qquad \psi_2\in {\mathbb T}$$
Then we define, analogously to the case $\m=0$, the {\it Arnold actions}
\beqano
&&\hat{\tt L}_\m:={\tt In}[\cC_{1, \m}]=
\frac{1}{2\p}\int_{\cC_1}(\L d\ell+{\rm G}d{\rm g})=\frac{1}{2\p}\int_{\ovl\cC_{1,\m}}\L d\ell=
{\tt In}[\ovl\cC_{1, \m}]\nonumber\\
&&\hat{\tt G}_\m:={\tt In}[\cC_{2, \m}]=\frac{1}{2\p}\int_{\cC_2}(\L d\ell+{\rm G}d{\rm g})=\frac{1}{2\p}\int_{\ovl\cC_{2,\m}}{\rm G}d{\rm g}=
{\tt In}[\ovl\cC_{2,\m}]
\eeqano
where we have introduced the {\it projected curves} $\ovl\cC_{1, \m}$, $\ovl\cC_{2, \m}$
$$\ovl\cC_{1, \m}:\quad \arr{\L=\L_{\m, 1}({\tt J}, {\tt E}, \rr', \psi_1)\\
\ell=\psi_1}\qquad \ovl\cC_{2,\m}:\quad \arr{\GG=\GG_{\m, 2}({\tt J}, {\tt E}, \rr', \psi_2)\\
{\rm g}={\rm g}_{\m, 2}({\tt J}, {\tt E}, \rr',  \psi_2)
}$$ and  we  have used the fact that ${\rm g}$ (respectively, $\ell$) is constant along $\cC_{1,\m}$ (respectively, $\cC_{2,\m}$).

\nl
Similarly to the case $\m=0$, we define the
{\it Arnold  map}
\beqa{JELG*}\hat{\tt A}:\quad ({\tt J}, {\tt E})\in\P^\ppj_{\m, {\tt K}}(\rr')\to (\hat{\tt L}_\m,\hat{\tt G}_\m)\in {\cal W}^\ppj_{\m, {\tt K}}(\rr'):=\hat{\tt A}(\P^\ppj_{\m, {\tt K}}(\rr'))\eeqa

\nl
The following  statement   is of the same kind of an analogue result (Proposition~\ref{propo: invertibility})  given for $\m=0$, but it is actually weker.: here continuity in ${\tt E}$ is not discussed, since the involved  domains ${\cal W}^\ppj_{\m, {\tt K}}$, ${\cal M}^\ppj_{\m, {\tt K}}$ are deprived of a neighborhood of their boundaries. However, there is continuity in $\m$. The proof is omitted, since it is an immediate consequence of the Implicit Function Theorem.

\begin{proposition}
For all ${\tt K}$, all $j=1$, $2$, $3$, the Arnold map~\equ{JELG*} is a action map to a reduced action--angle coordinate
 \beqano
  {{{\!\!\!\cA}_\m}}:\quad (\cL, \cG,  \l,\g)\in {\cal W}^\ppj_{\m, {\tt K}}(\rr')\times{\mathbb T}^2&\to (\L, \GG, \ell, {\rm g})\in {\cal M}^\ppj_{\m, {\tt K}}(\rr')
  \eeqano
to $\JJ$,  which reduce to $\ {\!\!\!\cA}_0$ as $\m\to 0$.

\end{proposition}
 
\nl
We finally turn to full action--angle coordinates. For given $j$, $\d_0$, ${\tt K}$, we denote ${\cal W}^\ppj_{\m, {\tt K}}=\cup_{\rr'>0}{\cal W}^\ppj_{\m, {\tt K}}(\rr')$, ${\cal M}^\ppj_{\m, {\tt K}}=\cup_{\rr'>0}{\cal M}^\ppj_{\m, {\tt K}}(\rr')$. The following proposition is proved in Section~\ref{proof of 4.13}.

\begin{proposition}\label{cor: action-angle} It is possible to find full action--angle coordinates $\cA_\m:\ {\cal W}^\ppj_{\m, {\tt K}}\to {\cal M}^\ppj_{\m, {\tt K}}$,  having the form
\beqa{action angle full}
&&\L={\tt \L}_\m( \cL, \cG, \l, \g, {\hat\rr}')\ ,\quad \GG={\tt G}_\m( \cL, \cG, \l, \g, {\hat\rr}')\nonumber\\
&&\ell={\tt l}_\m( \cL, \cG, \l, \g, {\hat\rr}')\ ,\quad {\rm g}={\tt g}_\m( \cL, \cG, \l, \g, {\hat\rr}')\nonumber\\
&&\RR'={\hat\RR}'+\hat\r( \cL, \cG, \l, \g, {\hat\rr}')\ ,\quad \rr'={\hat\rr}'\eeqa
 \end{proposition}

\begin{remark}\rm We remark that, since $\cA_\m$ is $\m$--close to the transformation $\cA_0$, then the function $\JJ_{\cA_\m}:=\JJ\circ\cA_\m$ has the form
 \beqa{HE}\JJ_{\cA_\m}(\cL, \cG, \hat\rr';\m)=-\frac{\mm^3{\cal M}^2}{2\cL^2}+\m\UU(\cL, \cG, \hat\rr';\m)\ .\eeqa
 We shall use this in the next Section~\ref{setup}.
\end{remark}

 \nl
We conclude this section with  providing, for completeness, the analytical expression of $\ovl\cC_{1,\m}$ and $\ovl\cC_{2, \m}$, even though it will be not used in the paper. The  proof of the following proposition is given Section~\ref{proof of 4.13}.
\begin{proposition}\label{Liouville Arnoldter}
$\ovl\cC_{1, \m}({\tt J}, {\tt E}, \rr')$ is the projection in the plane $(\L, \ell)$
of the curve in the space $(\L, \GG, \ell)$ defined by equations
\beqa{graph curves1}
\arr{\dst-\frac{\mm^3{\cal M}^2}{2\L^2}-\m\frac{\mm{\cal M}}{\sqrt{{\rr'}^2-2\rr' \s a \varrho\cos\n+{a}^2\varrho^2}}=\JJ\\\\
\dst\GG^2-\s\mm^2{\cal M}\rr'\sqrt{1-\frac{\GG^2}{\L^2}}+\m\mm^2{\cal M}\rr'\frac{{\rr'}-\s a \varrho\cos\n}{\sqrt{{\rr'}^2-2\rr' \s a \varrho\cos\n+{a}^2\varrho^2}}=\EE
}
\eeqa
$\ovl\cC_{2,\m}({\tt J}, {\tt E}, \rr')$ 
 is the projection in the plane $({\rm g}, \GG)$
of the curve in the space $(\L, \GG, {\rm g})$ defined by equations
  \beqa{graph curves}
   \arr{
 \dst
 -\frac{\mm^3{\cal M}^2}{2\L^2}-\m\frac{\mm{\cal M}}{\sqrt{{\rr'}^2+2\rr' a(1-\ee) \cos{\rm g}+{a}^2(1-\ee)^2}} ={\tt J}\\\\
\dst\GG^2+\mm^2{\cal M}\rr'\sqrt{1-\frac{\GG^2}{\L^2}}\cos{\rm g}+\m\mm^2{\cal M}\rr'\frac{{\rr'}+a(1-\ee) \cos{\rm g}}{\sqrt{{\rr'}^2+2\rr' a(1-\ee) \cos{\rm g}+{a}^2(1-\ee)^2}}\\
\qquad\qquad\qquad\qquad\qquad={\tt E}}
\eeqa
 \end{proposition}
 
\subsection{Proofs}

% DA QUI
 \subsubsection{Proof  of Proposition~\ref{prop: phase portrait}}\label{proof of phase portrait}

The first part of the proof of Proposition~\ref{prop: phase portrait}  consists in proving that
\begin{proposition}\label{prop: phase portrait1} 
For any $j=1$, $2$, $3$, any ${\tt E}\in \P_{0}^\ppj({\tt J}, \rr')$, the level sets $\ovl\cC_{0, 2}({\tt J}, {\tt E}, \rr')$ defined by equation~\equ{P0}  are smooth, connected and compact curves, consisting of the union of graphs
\beqa{ovlgampm}
\ovl\cC_{2, 0, \pm}({\tt J}, {\tt E}, \rr'):\quad\arr{\dst
\L={\tt L}_0({\tt J})\\
{\rm g}=\pm\ovl{\rm g}_0({\tt J}, {\tt E}, \rr',{\rm G}')\quad \mod\ 2\p\\
\ell\in {\mathbb T}\\
{\rm G}'\in \ovl\cF_{0}({\tt J}, {\tt E}, \rr')
}\eeqa
%The projected phase space $\cC_{0, 2}({\tt J}, \rr')$ in~\equ{projected phase space}  is the strip ${\mathbb T}\times [0,{\tt L}_0({\tt J})]$  in the plane $({\rm g}, \GG)$. 
\end{proposition}

 \proof 
%Since equations~\equ{level curves} do not depend on $\ell$ and the former  can be  trivially inverted with respect to $\L$ as in~\equ{L_0(E0)}, 
%the study of the level sets~\equ{level curves} reduces to the level sets
%$$\GG^2+\mm^2{\cal M}\rr'\sqrt{1-\frac{\GG^2}{{\tt L}_0({\tt J})^2}}\cos{\rm g}={\tt E}$$
%in the plane $({\rm g}, \GG)$.
 To simplify notations, we
divide  this equation  by ${\tt L}_0({\tt J})^2$, and we rewrite it as
\beq{G0pl}\widehat\EE_0(\widehat\GG, {\rm g})=\widehat\GG^2+\d \sqrt{1-\widehat\GG^2}\cos{\rm g}=\widehat{\tt E}\ , \eeq
where
\beqa{delta}
&&\widehat\EE_0(\widehat\GG, {\rm g}):=\frac{\EE_0({\tt L}_0({\tt J})^2\widehat\GG, {\rm g})}{{\tt L}_0({\tt J})^2}\ ,\quad \widehat{\tt E}:=\frac{{\tt E}}{{\tt L}_0({\tt J})^2}\ ,\quad \widehat\GG:=\frac{\GG}{{\tt L}_0({\tt J})}\nonumber\\
&& \d=\mm^2\cM\frac{\rr'}{{\tt L}_0({\tt J})^2}\eeqa
and we study the rescaled level sets~\equ{G0pl} in the plane $({\rm g}, \widehat\GG)$.
 For  $\d\in (0,2)$, $\widehat\EE_0$ has a minimum, a saddle and a maximum, respectively at
$$\hat{\tt P}_{-}=(\p,0)\ ,\qquad \hat{\tt P}_0=(0,0)\ ,\qquad  \hat{\tt P}_+=(0,\sqrt{1-\frac{\d^2}{4}})$$
where it takes the values, respectively,
$$\widehat\EE_{0-}=-\d\ ,\qquad \widehat\EE_{0\rm sad}=\d\ ,\quad \widehat\EE_{0+}=1+\frac{\d^2}{4}\ .$$
Thus, we study  the level sets~\equ{delta} for \beqa{cal J}\widehat{\tt E}\in %[\widehat\EE_{0-},\widehat\EE_{0+}]=
\left[-\d, 1+\frac{\d^2}{4}\right]\ .\eeqa  We solve for ${\rm g}$:
\beqa{cosg}{\rm g}={\rm g}_\pm=\pm\cos^{-1}\left(\frac{\widehat{\tt E}-\widehat\GG^2}{\d \sqrt{1-\widehat\GG^2}}\right)\quad \mod\ 2\p\ .\eeqa
Using
\beqa{Gpm}1-\left(\frac{\widehat{\tt E}-\widehat\GG^2}{\d \sqrt{1-\widehat\GG^2}}\right)^2=\frac{\d^2-\widehat{\tt E}^2-2(\frac{\d^2}{2}-\widehat{\tt E})\widehat\GG^2-\widehat\GG^4}{\d^2(1-\widehat\GG^2)}=\frac{(\widehat\GG^2-\widehat\GG_{-}^2)(\widehat\GG_{+}^2-\widehat\GG^2)}{\d^2(1-\widehat\GG^2)}\eeqa
with
\beqa{eq: Gpm}\widehat\GG^2_\pm&=&\widehat{\tt E}-\frac{\d^2}{2}\pm\sqrt{\big(\widehat{\tt E}-\frac{\d^2}{2}\big)^2
+\d^2-\widehat{\tt E}^2
}\nonumber\\
&=&\widehat{\tt E}-\frac{\d^2}{2}\pm\d\sqrt{1+\frac{\d^2}{4}-\widehat{\tt E}}
\eeqa
one sees the the equality~\equ{cosg} is well defined for
 \beq{limits}\widehat\GG_{\rm min}\le \widehat\GG\le \widehat\GG_{\rm max}\eeq
where
$$\widehat\GG^2_{\rm min}:=\max\{\widehat\GG^2_{-}, 0\}\ ,\qquad \widehat\GG^2_{\rm max}:=\min\{\widehat\GG^2_{+}, 1\}\ .$$
Note  that, when $\widehat{\tt E}$ takes its maximum value $1+\frac{\d^2}{4}$, one has $\widehat\GG_+^2=\widehat\GG_-^2=1-\frac{\d^2}{4}$. Therefore,   by~\equ{cosg} and~\equ{limits}, the level set
with  $\widehat{\tt E}=1+\frac{\d^2}{4}$
 reduces to the maximum point $(0, \pm\sqrt{1-\frac{\d^2}{4}})$.
Writing $$\widehat\GG_-^2=\frac{\widehat{\tt E}^2-\d^2}{\widehat{\tt E}-\frac{\d^2}{2}+\d\sqrt{1+\frac{\d^2}{4}-\widehat{\tt E}}}$$
and noticing that
$$1-\widehat\GG_+^2=1-\left(\widehat{\tt E}-\frac{\d^2}{2}+\d\sqrt{1+\frac{\d^2}{4}-\widehat{\tt E}}\right)=\left(\frac{\d}{2}-\sqrt{1+\frac{\d^2}{4}-\widehat{\tt E}}\right)^2\ge 0\ ,$$
one finds that
\beqa{obser}
\widehat\GG_{\rm min}=\arr{
0\qquad {\rm if}\quad -\d\le \widehat{\tt E}\le \d\\
\widehat\GG_-\quad {\rm if}\quad \widehat{\tt E}>\d
}\ ,\qquad \widehat\GG_{\rm max}=\widehat\GG_+\ .
\eeqa
Observe that
\beqa{continuity}\lim_{\widehat{\tt E}\to \d}\GG^2_{\rm min}=\lim_{\widehat{\tt E}\to \d}\GG^2_{-}=0\ ,\quad \lim_{\widehat{\tt E}\to \d}\GG^2_{\rm max}=\lim_{\widehat{\tt E}\to \d}\GG^2_{\rm +}=\d(2-\d)\eeqa
and
\beqa{Gmax}\lim_{\widehat{\tt E}\to 1}\GG^2_{\rm max}=\lim_{\widehat{\tt E}\to 1}\GG^2_{+}=1\ ,\quad \lim_{\widehat{\tt E}\to 1}\widehat\GG_-^2=1-\d^2\ ,\quad \lim_{\widehat{\tt E}\to 1}\widehat\GG_{\rm min}^2=\max\{1-\d^2, 0\}\ ,\eeqa
which are obtained using
$$\lim_{\widehat{\tt E}\to \d}\widehat\GG_\pm^2=\d-\frac{\d^2}{2}\pm\d\left(1-\frac{\d}{2}\right)\ ,\quad \lim_{\widehat{\tt E}\to 1}\widehat\GG^2_\pm=1-\frac{\d^2}{2}\pm \frac{\d^2}{2}$$
in turn implied  by~\equ{eq: Gpm}.
\\
In particular, $\GG_{\rm min}$, is  continuous for $\widehat{\tt E}=\d$.
%
%$$\widehat\GG^2_{\rm min}=\arr{
%0\qquad {\rm if}\quad \d\ge 1\\
%1-\d^2\quad {\rm if}\quad 0<\d<1
%}\ ,\qquad \widehat\GG_{\rm max}=1\ \qquad{\rm for}\quad \widehat{\tt E}=1\ .$$
%
%
%
The inequality~\equ{limits} defines a ``symmetric'' domain of $\widehat\GG$ with respect to the origin, consisting of the union
\beqa{union}\widehat\cD=\widehat\cD_-\cup\widehat\cD_+\eeqa
 of two ``symmetric'' intervals
$$\widehat\cD_-=\left[-\widehat\GG_{\rm max}\, -\widehat\GG_{\rm min}\right]\ ,\qquad \widehat\cD_+=\left[\widehat\GG_{\rm min}\, \widehat\GG_{\rm max}\right]\ .$$
Observe that,  for  $\widehat{\tt E}>\d$ and $\widehat{\tt E}\ne 1$, the union~\equ{union} is disjoint, since, in this case, 
$\widehat\GG_{\rm min}>0$ (se~\equ{obser}).
The functions
  ${\rm g}_\pm$ in~\equ{cosg} are even functions of $\widehat\GG$ on such ``symmetric'' domain. 
  By construction, $\cM_0({\tt J}, {\tt E}, \rr')$ can be recovered as the union of  graphs of ${\rm g}_\s$ for $\widehat\GG\in\widehat\cD_{\s'}$, where $\s$, $\s'=\pm$. We denote such graphs as $\cF_{\s\s'}$.
  The equality~\equ{Gpm} implies that
$$\partial_{\rm g}(\widehat\EE_0-{\tt E})\Big|_{{\cF}_{\s\s'}}=-\d\sqrt{1-\widehat\GG^2}\sin{\rm g}_\s=-\s\sqrt{(\widehat\GG^2-\widehat\GG_{-}^2)(\widehat\GG_{+}^2-\widehat\GG^2)}\ .$$
It vanishes only at the extremal points of $\widehat\cD_{\s'}$. Therefore,
denoting as $\widehat\cF ^\circ_{\s\s'}$ the restriction of $\widehat\cF_{\s\s'}$ to a pre--fixed compact sub--domain $\widehat\cD_{\s'}^\circ\subset \widehat\cD_{\s'}$ which does not include such extremal points, condition~\equ{G graphs} is immediately met by
$\widehat\cF ^\circ_{\s\s'}$. 
%%%
%%%

\vskip.2in
\noi
{\bf Completion of the proof of Proposition~\ref{prop: phase portrait}}
We now turn  to study the curves in~\equ{G0pl} 
in the plane $({\rm g}, \widehat\GG)$, for $\widehat{\tt E}$ as in~\equ{cal J}. By symmetry, we limit to study the behavior of  ${\rm g}_+$  for $\widehat\GG\in \widehat\cD_+$. We denote as
\beqa{ovlundlg1}\underline{\rm g}:=\cos^{-1}\left(\frac{\widehat{\tt E}-\widehat\GG_{\rm min}^2}{\d \sqrt{1-\widehat\GG_{\rm min}^2}}\right)\ ,\quad \overline{\rm g}:=\cos^{-1}\left(\frac{\widehat{\tt E}-\widehat\GG_{\rm max}^2}{\d \sqrt{1-\widehat\GG_{\rm max}^2}}\right)\eeqa
 the values that ${\rm g}_+$ takes at the extrema of $\widehat\cD_+$.
 The explicit value of $\underline{\rm g}$, $\ovl{\rm g}$ is
\beqa{ovlundlg2}
\underline{\rm g}=\arr{0\qquad \qquad\ \ {\rm if}\qquad\qquad \widehat{\tt E}>\d\\
\cos^{-1}\frac{\widehat{\tt E}}{\d}\qquad {\rm if}\quad -\d\le \widehat{\tt E}\le \d
}\ ,\qquad \overline{\rm g}=\arr{
\p\quad {\rm if}\quad\widehat{\tt E}<1\\
\frac{\p}{2}\quad{\rm if}\quad \widehat{\tt E}=1\\
0\quad {\rm if}\quad\widehat{\tt E}>1
}
\eeqa
This follows from  the definitions in~\equ{eq: Gpm} and~\equ{obser}.
In particular, from~\equ{eq: Gpm} one finds, for $(\s,\widehat{\tt E})\ne(+,1)$
  \beqano
\frac{\widehat{\tt E}-\widehat\GG_{\s}^2}{\d \sqrt{1-\widehat\GG_{\s}^2}}&=&\frac{\widehat{\tt E}-\left(\widehat{\tt E}-\frac{\d^2}{2}+\s\d\sqrt{1+\frac{\d^2}{4}-\widehat{\tt E}}\right)}{\d\sqrt{1-\left(\widehat{\tt E}-\frac{\d^2}{2}+\s\d\sqrt{1+\frac{\d^2}{4}-\widehat{\tt E}}\right)}}\nonumber\\
&=&\sign\left(\frac{\d}{2}-\s\sqrt{1+\frac{\d^2}{4}-\widehat{\tt E}}\right)\nonumber\\
&=&\arr{+1\quad {\rm for}\quad \s=-\\
-1\quad {\rm for}\quad \s=+\ \&\ \widehat{\tt E}<1\\
%\phantom{-}0\quad {\rm for}\quad \s=+\ \&\ \widehat{\tt E}=1\\
+1\quad {\rm for}\quad \s=+\ \&\ \widehat{\tt E}>1
}
\eeqano
while, for $(\s,\widehat{\tt E})=(+,1)$,
 \beqano
\frac{\widehat{\tt E}-\widehat\GG_{+}^2}{\d \sqrt{1-\widehat\GG_{+}^2}}&=&
\frac{\sqrt{1-\widehat\GG_{+}^2}}{\d}=\frac{\sqrt{1-\left(1-\frac{\d^2}{2}+\d\sqrt{1+\frac{\d^2}{4}-1}\right)}}{\d}=0\eeqano
%In summary,
%$$\frac{\widehat{\tt E}-\widehat\GG_{\s}^2}{\d \sqrt{1-\widehat\GG_{\s}^2}}=\arr{+1\quad {\rm for}\quad \s=-\\
%-1\quad {\rm for}\quad \s=+\ \&\ \widehat{\tt E}<1\\
%\phantom{-}0\quad {\rm for}\quad \s=+\ \&\ \widehat{\tt E}=1\\
%+1\quad {\rm for}\quad \s=+\ \&\ \widehat{\tt E}>1
%}$$
%
%
Let us study the graph of ${\rm g}_+$ as a function of $\widehat\GG$, for $\widehat\GG\in \widehat\cD_+$.
From the formula
\beqa{g derivative}\partial_{\widehat\GG}{\rm g}_+=\frac{\widehat\GG}{\sqrt{(\widehat\GG^2-\widehat\GG^2_{-})(\widehat\GG^2_{\rm max}-\widehat\GG^2)}}\frac{2-\widehat{\tt E}-
\widehat\GG^2}{1-\widehat\GG^2}\ .
\eeqa
%This expression shows that the tangency is vertical for $\widehat\GG=\widehat\GG_{\rm max}$
%
% QUIQUIQUI
one sees that 
 $\widehat\GG=\widehat\GG_0:=\sqrt{2-\widehat{\tt E}}\notin \widehat\cD_+$ is an extremal point, as soon as $\widehat\GG_0\in \widehat\cD_+$.
Using
\beqa{GMax}
\widehat\GG_0^2-\widehat\GG_{\rm max}^2&=&2-\widehat{\tt E}-\left(\widehat{\tt E}-\frac{\d^2}{2}+\d\sqrt{1+\frac{\d^2}{4}-\widehat{\tt E}}\right)\nonumber\\
&=&\sqrt{1+\frac{\d^2}{4}-\widehat{\tt E}}\left(2\sqrt{1+\frac{\d^2}{4}-\widehat{\tt E}}-\d\right)\nonumber\\
&=&2\frac{\sqrt{1+\frac{\d^2}{4}-\widehat{\tt E}}}{\sqrt{1+\frac{\d^2}{4}-\widehat{\tt E}}+\d}(1-\widehat{\tt E})
\eeqa
and
\beqano
\widehat\GG_0^2-\widehat\GG_{\rm min}^2&\ge&2-\widehat{\tt E}-\left(\widehat{\tt E}-\frac{\d^2}{2}-\d\sqrt{1+\frac{\d^2}{4}-\widehat{\tt E}}\right)\nonumber\\
&=&\sqrt{1+\frac{\d^2}{4}-\widehat{\tt E}}\left(2\sqrt{1+\frac{\d^2}{4}-\widehat{\tt E}}+\d\right)\ge0\ .\eeqano
we see that
$${\rm g}_0\arr{\ge \widehat\GG_{\rm max}\quad {\rm for}\quad \widehat{\tt E}< 1\\
\in \widehat\cD_+\quad {\rm for}\quad \widehat{\tt E}\ge 1
}\ .$$
As a consequence,
\begin{itemize}
\item[{\tiny\textbullet}] For $\widehat{\tt E}<1$, $\widehat\GG_0>\widehat\GG_{\rm max}$ and hence ${\rm g}_+$ increases,
 in $\widehat\cD_+$,
 from $\underline{\rm g}$ to $\overline{\rm g}$.
%In particular, the case $\d<1$ and $\d<\widehat{\tt E}<1$ corresponds to rotational motions, while, for $-\d<\widehat{\tt E}<\d$ the motions are librational.
\item[{\tiny\textbullet}]   For $\widehat{\tt E}>1$,  ${\rm g}_+$ increases from $\underline{\rm g}$  to ${\rm g}_0$ for $\widehat\GG_{\rm min}\le \widehat\GG\le \widehat\GG_0$ and decreases from ${\rm g}_0$ to  $\overline{\rm g}$, for $\widehat\GG_0\le \widehat\GG\le \widehat\GG_{\rm max}$.
\end{itemize}
Collecting these informations,  the phase portrait of $\widehat{\tt E}$ in the plane $({\rm g}, \widehat\GG)$ can be summarized  as follows
 (see also Figures~\ref{figure12},~\ref{figure5} and~\ref{figure34}).
%$$\left(\pm \sqrt{2-\widehat{\tt E}}, \ \pm\left( \p-\cos^{-1}\left(\frac{2}{\d}\sqrt{\widehat{\tt E}-1}\right) \right)\right)$$
% \item[3.2] 
\begin{itemize}
\item[\rm 1.] For $0<\d<1$:
\begin{itemize}
\item[\rm 1.1] For $-\d\le\widehat{\tt E}<\d$ ${\rm g}$ ``librates'' around $\p$, with  maximum elongation in $[\cos^{-1}\frac{\widehat{\tt E}}{\d}, 2\p-\cos^{-1}\frac{\widehat{\tt E}}{\d}]$;
\item[\rm 1.2]  $\widehat{\tt E}=\d$ is the level set through the saddle ({\it separatrix});
\item[\rm 1.3] For $\d<\widehat{\tt E}<1$, ${\rm g}$ ``rotates'', namely takes all the values in  ${\mathbb T}$.
\item[\rm 1.4] The curve $\widehat{\tt E}=1$ splits into $\widehat\GG=1$ and $\widehat\GG=\sqrt{1-\d^2\cos^2{\rm g}}$, with ${\rm g}\in {\mathbb T}$. Such two branches glue smoothly at 
$(\pm \frac{\p}{2}, 1)$, with ${\rm g}$ mod $2\p$.
\item[\rm 1.5]  For $1<\widehat{\tt E}\le1+\frac{\d^2}{4}$, ${\rm g}$ librates around $0$, with maximum elongation $$[-\frac{2}{\d}\sqrt{\widehat{\tt E}-1}, \frac{2}{\d}\sqrt{\widehat{\tt E}-1}]\ .$$
    \end{itemize} 
    \item[\rm 2.] For $\d=1$:
\begin{itemize}
\item[\rm 2.1] For $-1\le\widehat{\tt E}<1$ ${\rm g}$ ``librates'' around $\p$, with  maximum elongation in $[\cos^{-1}\widehat{\tt E}, 2\p-\cos^{-1}\widehat{\tt E}]$;
\item[\rm 2.2]  $\widehat{\tt E}=1$ is the level set through the saddle ({\it separatrix}). It splits into $\widehat\GG=1$ and $\widehat\GG=|\sin{\rm g}|$, with ${\rm g}\in {\mathbb T}$. Such two branches glue smoothly at 
$(\pm \frac{\p}{2}, 1)$, with ${\rm g}$ mod $2\p$.
\item[\rm 2.3]  For $1<\widehat{\tt E}<\frac{5}{4}$, ${\rm g}$ librates around $0$, with maximum elongation $$[-2\sqrt{\widehat{\tt E}-1}, 2\sqrt{\widehat{\tt E}-1}]\ .$$
    \end{itemize} 
    \item[\rm 3.] For $1<\d<2$:
\begin{itemize}
\item[\rm 3.1] For $-\d\le\widehat{\tt E}<1$ ${\rm g}$ ``librates'' around $\p$, with  maximum elongation in $[\cos^{-1}\frac{\widehat{\tt E}}{\d}, 2\p-\cos^{-1}\frac{\widehat{\tt E}}{\d}]$;
\item[\rm 3.2] the curve  $\widehat{\tt E}=1$ splits into $\widehat\GG=1$ and $\widehat\GG=\sqrt{1-\d^2\cos^2{\rm g}}$, for $-\p\le {\rm g}\le \cos^{-1}\frac{1}{\d}$. Such two branches glue smoothly at 
$(\pm \frac{\p}{2}, 1)$, with ${\rm g}$ mod $2\p$.
\item[\rm 3.3] For $1<\widehat{\tt E}<\d$, ${\rm g}$ ``librates'' around $0$ with maximum elongation $$[-\frac{2}{\d}\sqrt{\widehat{\tt E}-1}, \frac{2}{\d}\sqrt{\widehat{\tt E}-1}]\ .$$
\item[\rm 3.3] $\widehat{\tt E}=\d$ is the level set through the saddle $(0,0)$ ({\it separatrix})
\item[\rm 3.4]  For $\d<\widehat{\tt E}\le1+\frac{\d^2}{4}$, ${\rm g}$ librates around $0$, with maximum elongation $[-\frac{2}{\d}\sqrt{\widehat{\tt E}-1}, \frac{2}{\d}\sqrt{\widehat{\tt E}-1}]$.
    \end{itemize}

\nl
Finally, the analysis of $\partial_{\widehat\GG}{\rm g}_+$ in~\equ{g derivative} allows to infer that the curves in~\equ{G0pl} are smooth for $\widehat{\tt E}\notin\{\pm \d,\ 1,\ 1+\frac{\d^2}{4}\}$, as claimed.
        \end{itemize}
        
\subsubsection{Proof of Proposition~\ref{propo: invertibility}} \label{proof of 4.3**}

By Proposition~\ref{prop: phase portrait} and the Liouville--Arnold theorem, for any $j=1$, $2$, $3$, any $ \cM_0({\tt J}, {\tt E}, \rr')\subset \cM^\ppj_0(\rr')$ one finds a diffeomorphism
$$\phi_0({\tt J}, {\tt E}, \rr'):\quad (\psi_1, \psi_2)\in {\mathbb T}^2\to \cM_0({\tt J}, {\tt E}, \rr')$$
between the 2--torus and $ \cM_0({\tt J}, {\tt E}, \rr')$.
Equations~\equ{M0} show that $\phi_0$
splits as the direct product 
$$\phi_0({\tt J}, {\tt E}, \rr')=\phi_{01}({\tt J})\otimes\phi_{02}({\tt J}, {\tt E}, \rr') $$
of two diffeomorphisms on the circle given by
\beqano
&&\phi_{01}({\tt J}):\quad \ell\in{\mathbb T}\to({\tt L}_0({\tt J}),\ \ell)\in \ovl\cC_{0, 1}({\tt J}) \nonumber\\
&&\phi_{02}({\tt J}, {\tt E}, \rr'):\quad \psi_2\in{\mathbb T}\to\big(\GG_0({\tt J}, {\tt E}, \psi_2, \rr')\ ,\ {\rm g}_0({\tt J}, {\tt E}, \psi_2, \rr')\big)\in \ovl\cC_{0, 2}({\tt J}, {\tt E}, \rr')
\eeqano
Then the action coordinates are defined, by~\cite{arnold63a}, as

\beqa{cL0}
\hat{\tt L}_0({\tt J})&=&\frac{1}{2\p}\int_0^{2\p}{\tt L}_0({\tt J}) d\psi_1={\tt L}_0({\tt J})=\sqrt{-\frac{\mm^3{\cal M}^2}{2{\tt J}}}=\L\eeqa
and 
\beqa{cG0}
\hat{\tt G}_0({\tt J}, {\tt E}, \rr')&=&\frac{1}{2\p}\int_0^{2\p}\GG_0({\tt J}, {\tt E}, \psi_2, \rr') \partial_{\psi_2}{\rm g}_0({\tt J}, {\tt E}, \psi_2, \rr') d\psi_2\nonumber\\
&=&-\frac{1}{2\p}\int_{\ovl\cC_{0, 2}({\tt J}, {\tt E}, \rr')}{\rm g}d\GG'
\nonumber\\
&=&
{\tt In}\Big[\ovl\cC_{0, 2}({\tt J}, {\tt E}, \rr')\Big]
\eeqa
where 
  $\ovl\cC_{0, 2}({\tt J}, {\tt E}, \rr'):=\ovl\cC_{2, 0, +}({\tt J}, {\tt E}, \rr')\cup\ovl\cC_{2, 0, -}({\tt J}, {\tt E}, \rr')$.

\nl
We are now ready to prove that the map~\equ{JELG} is a diffeomorphism. Since the map
\beqa{L0}{\tt J}\to \hat{\tt L}_0({\tt J})={\tt L}_0({\tt J})\eeqa
is trivially a diffeomorphism, we only need to show that so is the map
$${\tt E}\in\P_0({\tt J}, \rr')\to \hat{\tt G}_0({\tt J}, {\tt E},\rr')\ .$$
This follows from the following proposition.
\begin{proposition}\label{inversion1}
For all $j=1$, $2$, $3$, the function ${\tt E}\in\P^\ppj_0({\tt J}, \rr')\to \partial_{\tt E}\hat{\tt G}_0({\tt J}, {\tt E},\rr')$ is  finite and sign definite.
\end{proposition}

\nl
For the proof of this proposition, as well as for other proofs below, we need  the explicit expression of  the formulae in~\equ{G01}--\equ{G02}. They are as follows.
We let $\widehat\cG_0(\widehat{\tt E};\d):=\frac{\cG_0}{{\tt L}_0({\tt J})^2}$ and, as in the previous sections, $\hat\GG:=\frac{\GG}{{\tt L}_0({\tt J})^2}$. The formulae for $\widehat\cG_0$ corresponding to the definitions in~\equ{G01}--\equ{G02} are:

\nl
-- if
 $0<\d<1$:\beqa{A1}&&\widehat\cG_0(\widehat{\tt E};\d)=2\cdot\nonumber\\
&&\left\{
\begin{array}{lll}\widehat\GG_{\rm max}%-\widehat\GG_{\rm min}
-\frac{1}{\p}\int_{\widehat\GG_{\rm min}}^{\widehat\GG_{\rm max}}\cos^{-1}\frac{\widehat{\tt E}-\widehat\GG^2}{\d \sqrt{1-\widehat\GG^2}}d\widehat\GG\quad&{\rm if}\quad& -\d< \widehat{\tt E}< \d\ \&\ \d< \widehat{\tt E}< 1\\%\\
%\widehat\GG_{\rm max}-\frac{1}{\p}\int_{\widehat\GG_{\rm min}}^{\widehat\GG_{\rm max}}\cos^{-1}\frac{\widehat{\tt E}-\widehat\GG^2}{\d \sqrt{1-\widehat\GG^2}}d\widehat\GG&{\rm if}& \d< \widehat{\tt E}< 1\\
1-%(\widehat\GG_{\rm max}-\widehat\GG_{\rm min})+
\frac{1}{\p}\int_{\widehat\GG_{\rm min}}^{\widehat\GG_{\rm max}}%\left(\p-
\cos^{-1}\frac{\widehat{\tt E}-\widehat\GG^2}{\d \sqrt{1-\widehat\GG^2}}%\right)
d\widehat\GG&{\rm if}& 1< \widehat{\tt E}< 1+\frac{\d^2}{4}
\end{array}
\right.
\eeqa
-- if $1<\d<2$:
\beqa{A2}
&&\widehat\cG_0(\widehat{\tt E};\d)=2\cdot\nonumber\\
&&\left\{
\begin{array}{lll}\widehat\GG_{\rm max}%-\widehat\GG_{\rm min}
-\frac{1}{\p}\int_{\widehat\GG_{\rm min}}^{\widehat\GG_{\rm max}}\cos^{-1}\frac{\widehat{\tt E}-\widehat\GG^2}{\d \sqrt{1-\widehat\GG^2}}d\widehat\GG\quad&{\rm if}\quad& -\d< \widehat{\tt E}<  1\\
1-%(\widehat\GG_{\rm max}-\widehat\GG_{\rm min})+
\frac{1}{\p}\int_{\widehat\GG_{\rm min}}^{\widehat\GG_{\rm max}}%\left(\p-
\cos^{-1}\frac{\widehat{\tt E}-\widehat\GG^2}{\d \sqrt{1-\widehat\GG^2}}%\right)
d\widehat\GG
%1-\frac{1}{\p}\int_{\widehat\GG_{\rm min}}^{\widehat\GG_{\rm max}}\left(\cos^{-1}\frac{\widehat{\tt E}-\widehat\GG^2}{\d \sqrt{1-\widehat\GG^2}}\right)d\widehat\GG
&{\rm if}& \begin{array}{llll}1< \widehat{\tt E}< \d\  \& \ \d< \widehat{\tt E}< 1+\frac{\d^2}{4}\\
%\widehat{\tt E}\ne \d
\end{array}
\end{array}
\right.
\eeqa

\proof {\bf of Proposition~\ref{inversion1}}
 Let us  compute the derivative
$\partial_{\widehat{\tt E}}\widehat\cG_0$. We aim to check that
\beqa{A'}\partial_{\widehat{\tt E}}\widehat\cG_0&=&
 -\frac{2}{\p}\int_{\widehat\GG_{\rm min}}^{\widehat\GG_{\rm max}}\partial_{\widehat{\tt E}}
\cos^{-1}\frac{\widehat{\tt E}-\widehat\GG^2}{\d \sqrt{1-\widehat\GG^2}} d\widehat\GG\nonumber\\
&=&
 %-\frac{2}{\p}
 %\int_{\widehat\GG_{\rm min}}^{\widehat\GG_{\rm max}}\partial_{\widehat{\tt E}}{\rm g}_+d\widehat\GG
% \nonumber\\
% &=&
\frac{2}{\p}\int_{\widehat\GG_{\rm min}}^{\widehat\GG_{\rm max}}\frac{d\widehat\GG}{\sqrt{(\widehat\GG^2-\widehat\GG^2_{-})(\widehat\GG^2_{\rm max}-\widehat\GG^2)}}\ .\eeqa%for all $1<\d<2$, and all $-\d\le\widehat{\tt E}\le 1+\frac{\d^2}{4}$.
Observe, once we shall have checked this formula, the thesis follows observing that that the integral looses its meaning only when $\widehat{\tt E}=\pm\d$, or $\widehat{\tt E}=1+\frac{\d^2}{4}$, %or $\widehat{\tt E}=1$, 
because $\widehat\GG_-=0$ in the former case, $\widehat\GG_-=\widehat\GG_+$ in the latter.

\nl
Using the formula (recall the definitions of $\underline{\rm g}$, $\ovl{\rm g}$ in~\equ{ovlundlg1})
\beqano
\partial_{\widehat{\tt E}}\int_{\widehat\GG_{\rm min}}^{\widehat\GG_{\rm max}}\cos^{-1}\frac{\widehat{\tt E}-\widehat\GG^2}{\d \sqrt{1-\widehat\GG^2}} d\widehat\GG-
\int_{\widehat\GG_{\rm min}}^{\widehat\GG_{\rm max}}%\frac{d\widehat\GG}{\sqrt{(\widehat\GG^2-\widehat\GG^2_{-})(\widehat\GG^2_{\rm max}-\widehat\GG^2)}}
\partial_{\widehat{\tt E}}
\cos^{-1}\frac{\widehat{\tt E}-\widehat\GG^2}{\d \sqrt{1-\widehat\GG^2}} d\widehat\GG
=\ovl{\rm g}\partial_{\widehat{\tt E}}\widehat\GG_{\rm max}-\underline{\rm g}\partial_{\widehat{\tt E}}\widehat\GG_{\rm min}
\eeqano
we obtain that the quantity
$$B:=\partial_{\widehat{\tt E}}\widehat\cG_0(\widehat{\tt E};\d)+\frac{2}{\p}\int_{\widehat\GG_{\rm min}}^{\widehat\GG_{\rm max}}\partial_{\widehat{\tt E}}
\cos^{-1}\frac{\widehat{\tt E}-\widehat\GG^2}{\d \sqrt{1-\widehat\GG^2}} d\widehat\GG$$
takes the following values.
For $0<\d<1$, 
\beqano
%&&
B=
2\left\{
\begin{array}{lll} +
\left(1-\frac{\ovl{\rm g}}{\p}\right)\partial_{\widehat{\tt E}}\widehat\GG_{\rm max}+%\left(
\frac{\underline{\rm g}}{\p}%-1\right)
\partial_{\widehat{\tt E}}\widehat\GG_{\rm min}
\ &{\rm for}\ & -\d< \widehat{\tt E}< \d\ \&\ \d< \widehat{\tt E}< 1\\\\
 %+
%\left(1-\frac{\ovl{\rm g}}{\p}\right)\partial_{\widehat{\tt E}}\widehat\GG_{\rm max}+\frac{\underline{\rm g}}{\p}\partial_{\widehat{\tt E}}\widehat\GG_{\rm min}
%&{\rm for}& \d< \widehat{\tt E}< 1\\
%
 -\frac{\ovl{\rm g}}{\p}\partial_{\widehat{\tt E}}\widehat\GG_{\rm max}+\frac{\underline{\rm g}}{\p}\partial_{\widehat{\tt E}}\widehat\GG_{\rm min}&{\rm for}& 1< \widehat{\tt E}< 1+\frac{\d^2}{4}
\end{array}
\right.
\eeqano
while, for $1\le\d<2$:
\beqano
B
=2
\left\{
\begin{array}{lll} +
\left(1-\frac{\ovl{\rm g}}{\p}\right)\partial_{\widehat{\tt E}}\widehat\GG_{\rm max}+%\left(
\frac{\underline{\rm g}}{\p}%-1\right)
\partial_{\widehat{\tt E}}\widehat\GG_{\rm min}\ &{\rm for}\ & -\d< \widehat{\tt E}<  1\\\\
 -\frac{\ovl{\rm g}}{\p}\partial_{\widehat{\tt E}}\widehat\GG_{\rm max}+\frac{\underline{\rm g}}{\p}\partial_{\widehat{\tt E}}\widehat\GG_{\rm min}
%1-\frac{2}{\p}\int_{\widehat\GG_{\rm min}}^{\widehat\GG_{\rm max}}\left(\cos^{-1}\frac{\widehat{\tt E}-\widehat\GG^2}{\d \sqrt{1-\widehat\GG^2}}\right)d\widehat\GG
&{\rm for}& %\left.
\begin{array}{llll}1< \widehat{\tt E}\le \d\ \& \  \d< \widehat{\tt E}\le 1+\frac{\d^2}{4}\\
%\widehat{\tt E}\ne \d
\end{array}
%\right\}
\end{array}
\right.
\eeqano
Since $\widehat\GG_{\rm min}=0$ for $\widehat{\tt E}<\d$; 
$\underline{\rm g}=0$ for $\widehat{\tt E}>\d$; $\ovl{\rm g}=0$ for $\widehat{\tt E}>1$; $\ovl{\rm g}=\p$ for $\widehat{\tt E}<1$ (see~\equ{obser} and~\equ{ovlundlg2}),  one finds  $B\equiv0$, hence~\equ{A'}. \qquad $\square$

\nl
We are now ready for the
\paragraph{Proof of Proposition~\ref{inversion1}} It follows from Proposition~\ref{inversion} and the formulae~\equ{cL0},~\equ{G01},~\equ{G02},~\equ{Ext}.  \qquad $\square$

\nl
To complete the proof of Proposition~\ref{propo: invertibility}, we need to define the angles $\hat\l_0$, $\hat\g_0$ and prove the differentiability of the map~\equ{Arnold}.
By~\cite{arnold63a}, the construction of the angles $\hat\l_0$, $\hat\g_0$ goes as follows.
%\beqano
%\hat\G_\pm(\hat\cL_0, \hat\cG_0):\quad\arr{\dst
%\L=\hat\cL_0\\
%{\rm g}=\pm{\rm g}_0(\hat\cL_0, \hat\cG_0, \rr',{\rm G}')\\
%\ell\in {\mathbb T}\\
%{\rm G}'\in \cF_{0}(\hat\cL_0, \hat\cG_0, \rr')
%}\eeqano
%
Denote as  $\cC_{0, 2}(\hat\cL_0, \hat\cG_0, \rr')$ the composition of $\ovl\cC_{0, 2}({\tt J}, {\tt E}, \rr')$ with the inverse of the map~\equ{JELG}. Fix $\hat{\tt P}=(\hat{\rm g}, \hat\GG)\in \cC_{0, 2}(\hat\cL_0, \hat\cG_0, \rr')$. For a fixed $\GG$ such that there exists  ${\tt P}=({\rm g}, \GG)\in \cC_{0, 2}(\hat\cL_0, \hat\cG_0, \rr')$, choose ${\tt P}(\GG)=(\ovl{\rm g}, \GG)$ so that  ${\tt P}(\GG)\in \cC_{0, 2}(\hat\cL_0, \hat\cG_0, \rr')$ and $\GG\to {\tt P}(\GG)$ is continuous.
 Consider then the generating function
\beqa{hatS}\hat S(\hat\cL_0, \hat\cG_0, \ell, \GG)=\hat\cL_0\ell-\int_{\cC_{0, 2}(\hat\cL_0, \hat\cG_0, \rr')_{\hat{\tt P}}^{{\tt P}(\GG)}}{\rm g}d\GG'\ ,\eeqa
where, given a smooth plane curve $\cC$, and two points $\hat{\tt P}$, ${\tt P}\in \cC$, we denote as $\int_{\cC_{\hat{\tt P}}^{\tt P}}$  the  integral, in the counterclockwise direction, along  $\cC$ with stating point $\hat{\tt P}$ and endpoint ${\tt P}$. 
Then $\hat S$ gives
\beqa{angles}
\arr{\hat\l_0(\cL_0, \cG_0, \ell, \GG)=\ell-\partial_{\hat\cL_0}\int_{\cC_{0, 2}(\hat\cL_0, \hat\cG_0, \rr')_{\hat{\tt P}}^{{\tt P}(\GG)}}{\rm g}d\GG'\\
 \hat\g_0(\cL_0, \cG_0, \GG)=-\partial_{\hat\cG_0}\int_{\cC_{0, 2}(\hat\cL_0, \hat\cG_0, \rr')_{\hat{\tt P}}^{{\tt P}(\GG)}}{\rm g}d\GG'
 }
\eeqa

\nl
 The following proposition easily implies  the invertibility and differentiability of the map~\equ{Arnold},  and hence concludes the proof of Proposition~\ref{propo: invertibility} (which is the first step of the proof of Proposition~\ref{prop: continuity}).

\begin{proposition}\label{angles}
The map
$$(\ell, \GG)\to \big(\l_0(\cL_0, \cG_0, \ell, \GG), \g_0(\cL_0, \cG_0,  \GG)\big)$$
is invertible.
\end{proposition}
\proof
 The latter equation in~\equ{angles}, independent of $\ell$, is nothing else than the definition of the angular coordinate for the one--dimensional Hamiltonian $(\GG, {\rm g})\to\EE$ in~\equ{P0}, which, by the chain rule and~\equ{cG0}, can be written as
\beqano
\hat\g_0(\L, \GG, {\rm g})&=&\left.\frac{-\partial_{{\tt E}}\int_{\ovl\cC_{0, 2}({\tt J}, {\tt E}, \rr')_{\hat{\tt P}}^{{\tt P}(\GG)}}{\rm g}d\GG'}{\partial_{\tt E}\hat{\tt G}_0({\tt J}, {\tt E}, \rr')}\right|_{(\L, \GG, {\rm g})}=2\p\left.\frac{-\partial_{{\tt E}}\int_{\ovl\cC_{0, 2}({\tt J}, {\tt E}, \rr')_{\hat{\tt P}}^{{\tt P}(\GG)}}{\rm g}d\GG'}{-\partial_{{\tt E}}\int_{\ovl\cC_{0, 2}({\tt J}, {\tt E}, \rr')}{\rm g}d\GG'}\right|_{(\L, \GG, {\rm g})}\nonumber\\
&=&2\p\left.\frac{t({\tt J}, {\tt E}, \rr';\GG)}{T({\tt J}, {\tt E}, \rr')}\right|_{(\L, \GG, {\rm g})}
\eeqano
where $t({\tt J}, {\tt E}, \rr';\GG)=-\partial_{{\tt E}}\int_{\ovl\cC_{0, 2}({\tt J}, {\tt E}, \rr')_{\hat{\tt P}}^{{\tt P}(\GG)}}{\rm g}d\GG'$ is the time needed to reach $\GG$ on $\ovl\cC_{0, 2}({\tt J}, {\tt E}, \rr')$ and $T({\tt J}, {\tt E}, \rr')=t({\tt J}, {\tt E}, \rr';\hat\GG)$ is the period associated to  $\ovl\cC_{0, 2}({\tt J}, {\tt E}, \rr')$ and $f({\tt J}, {\tt E})\big|_{(\L, \GG, {\rm g})}$
is a short for $f({\tt J}, {\tt E})\big|_{{\tt J}=\JJ_0(\L), {\tt E}=\EE_0(\L, \GG, {\rm g})}$
. So the function
$$\GG\to  \hat\g_0(\cL_0, \cG_0,  \GG)$$
is invertible 
by the Liouvile--Arnold theorem applied to such one--dimensional system. The inversion of the full system~\equ{angles} reduces to invert the former after expressing $\GG$ as a function of $\cL_0, \cG_0,  \hat\g_0$ via the latter. But this is trivial, because such equation is linear. \quad $\square$

\subsubsection{Proof of Proposition~\ref{prop: continuity}}\label{proof of 4.3} The following proposition is proved in Section~\ref{proof of 4.3**}
\begin{proposition}\label{inversion}
For all $j=1$, $2$, $3$, the function  ${\tt E}\in\P^\ppj_0({\tt J}, \rr')\to \partial_{\tt E}{\tt G}_0({\tt J}, {\tt E},\rr')$ is  finite and positive. It is infinite on $\partial\P_0({\tt J}, \rr')$.
\end{proposition}

\nl
The following result, combined with Proposition~\ref{prop: continuity***}, completes the second step and hence the proof of Proposition~\ref{prop: continuity} .

\begin{proposition}\label{2cases}
It is possible to find a  linear change with integer coefficients 
such that $\cG_0$, expressed in terms of $({\tt J}$, ${\tt E}$, $\rr')$, coincides with the function in~\equ{G01}--\equ{G02}.
\end{proposition}

\proof We distinguish two cases.

\nl
{\it Case 1: $[0<\d\le 1$ \& $(-\d<\hat{\tt E}<1$ or $\d<\hat{\tt E}<1)]$ or $[1<\d<2$ \& $-\d<\hat{\tt E}<1]$}

\nl
In this case
${\tt G}_0({\tt J}, {\tt E}, \rr')={\tt In}[\ovl\cC_{0, 2}({\tt J}, {\tt E}, \rr')]$. Then one can take $(\hat\cL_0, \hat\cG_0, \hat\l_0, \hat\g_0)=(\cL_0, \cG_0, \l_0, \g_0)$
and there is nothing else to prove.

\vskip.2in
\noi
{\it Case 2:$[0<\d\le 1$ \& $1<\hat{\tt E}<1+\frac{\d^2}{4}]$ or $[1<\d<2$ \& $(1<\hat{\tt E}<\d$ or $\d<\hat{\tt E}<1+\frac{\d^2}{4})]$}

\nl
In this case ${\tt G}_0({\tt J}, {\tt E}, \rr')={\tt Ext}[\ovl\cC_{0, 2}({\tt J}, {\tt E}, \rr')]$. Since
\beqano{\tt G}_0({\tt J}, {\tt E}, \rr')+\hat{\tt G}_0({\tt J}, {\tt E}, \rr')={\tt In}[\ovl\cC_{0, 2}({\tt J}, {\tt E}, \rr')]+{\tt Ext}[\ovl\cC_{0, 2}({\tt J}, {\tt E}, \rr')]={\tt L}_0({\tt J})=\hat\cL_0\ ,\eeqano
one can consider the canonical transformation generated by
$$S(\cL_0, \cG_0, \ell, \GG):=\hat S(\cL_0, \cL_0-\cG_0, \ell, \GG)=\cL_0\ell-\int_{\cC_{0, 2}(\cL_0, \cL_0-\cG_0, \rr')_{\hat{\tt P}}^{{\tt P}(\GG)}}{\rm g} d\GG'$$
with $\hat S(\hat\cL_0, \hat\cG_0, \ell, \GG)$ as in~\equ{hatS}. This is equivalent to take
$$(\cL_0, \cG_0, \l_0, \g_0):= (\hat\cL_0, \hat\cL_0-\hat\cG_0, \hat\l_0+\hat\g_0, -\hat\g_0)$$
which is  a  linear change with integer coefficients, as claimed. \qquad $\square$

\nl

\subsubsection{Completion of the proof of Proposition~\ref{prop: aa}} We shall use the following result, which (specularly to the case of Proposition~\ref{propo: invertibility}), follows from the invertibility of the map~\equ{L0} combined with Proposition~\ref{inversion}.

\begin{proposition}
The map \beqa{JELG***}({\tt J}, {\tt E})\in \P^\ppj_0(\rr')\to ({\tt L}_0({\tt J}), {\tt G}_0({\tt J}, {\tt E}, \rr'))\in{\cal W}^\ppj_0(\rr')\eeqa
is invertible for all $j=1$, $2$, $3$.
\end{proposition}

\nl The sets ${\cal W}_0(\rr')$, ${\cal W}^\ppj_0(\rr')$ are the image under the transformation
~\equ{JELG***} of the sets $\P_0(\rr')$ and   $\P^\ppj_0(\rr')$. This image can be explicitly computed using that
 ${\tt J}\to {\tt L}_0({\tt J})$ is explicitly given in~\equ{cL0} and  the function ${\tt E}\in \P_0({\tt J}, \rr')\to{\tt G}_0({\tt J}, {\tt E}, \rr')$ is increasing and continuous, with the minimum $0$
and the maximum ${\tt L}_0({\tt J})$. Then equations~\equ{Wi} follow, with
 $$\cG_-(\cL_0, \rr'):={\rm min}\{{\tt G}_{0}(\cL_0, \rr'), {\tt G}_{1}(\cL_0, \rr')\}\ ,\ \cG_+(\cL_0, \rr'):=\max\{{\tt G}_{0}(\cL_0, \rr'), {\tt G}_{1}(\cL_0, \rr')\}$$
where
$${\tt G}_{0, 1}(\cL_0, \rr'):={\tt In}\left[{\tt S}_{0, 1}\left(-\frac{\mm^3\cM^2}{2\cL_0^2}, \rr'\right)\right]$$
is the area of the inner region delimited by the separatrices in the parameter space, written in terms of $\cL_0$. Observe that the continuity of the map  ${\tt E}\in \P_0({\tt J}, \rr')\to{\tt G}_0({\tt J}, {\tt E}, \rr')$ was crucial in the proof. $\quad \square$

\subsubsection{Proof of Proposition~\ref{phiA0}
}\label{full with m=0}

As in the proof of Proposition~\ref{2cases}, we distinguish two cases.

\nl
{\it Case 1: $[0<\d\le 1$ \& $(-\d<\hat{\tt E}<1$ or $\d<\hat{\tt E}<1)]$ or $[1<\d<2$ \& $-\d<\hat{\tt E}<1]$}

\nl
We look at the canonical transformation generated by
$$S_{\rm full, 1}(\hat\RR', \cL_0, \cG_0, \rr', \ell, \GG):=\hat\RR_0'\rr'+\cL_0\ell-\int_{\cC_{0, 2}(\cL_0, \cG_0, \rr')_{\hat{\tt P}}^{{\tt P}(\GG)}}{\rm g}d\GG'$$

\vskip.2in
\noi
{\it Case 2: $[0<\d\le 1$ \& $1<\hat{\tt E}<1+\frac{\d^2}{4}]$ or $[1<\d<2$ \& $(1<\hat{\tt E}<\d$ or $\d<\hat{\tt E}<1+\frac{\d^2}{4})]$}

\nl
In this case, we consider
$$S_{\rm full, 2}(\hat\RR_0', \cL_0, \cG_0, \rr', \ell, \GG):=\hat\RR_0'\rr'+\cL_0\ell-\int_{\cC_{0, 2}(\cL_0, \cL_0-\cG_0, \rr')_{\hat{\tt P}}^{{\tt P}(\GG)}}{\rm g}d\GG'$$
In both cases, we obtain a transformation of the form~\equ{lift}. \quad $\square$

 \subsubsection{Proof of Propositions~\ref{Liouville Arnold3bis},~\ref{Liouville Arnold3quater},~\ref{cor: action-angle} and~\ref{Liouville Arnoldter}}\label{proof of 4.13}
 
 \nl
We rewrite the manifolds $\cM_0({\tt J},{\tt E},\rr')$ that we have studied in the previous section as the set 
of solutions of
 \beqa{F0}\FF_0(\L,\GG, \ell, {\rm g}, {\tt J}, {\tt E}):=({\JJ}_0(\L)-{\tt J}, \EE_0(\L, \GG,  {\rm g})-{\tt E})=0\eeqa
and we observe that
  \begin{lemma}\label{graphs lemma}
For all $j=1$, $2$, $3$, all $({\tt J}, {\tt E})\in {\cal M}_0^\ppj(\rr')$, there exists a chain of graphs
 $${\cF }_{01}\quad {\cF}'_{01}\quad \cdots \quad {\cF }_{0N}\quad {\cF}'_{0N}$$
 given by
 $$%(\a, z_{0i}(\a))=
{\cF}_{0i}({\tt J},{\tt E}):=\Big\{\Big(\L_{0}({\tt J}), {\rm G}_{0i}({\rm g}, {\tt J},{\tt E}), \ell,  {\rm g}\Big)\ :\quad (\ell, {\rm g})\in {\mathbb T}\times\cD_i\Big\}
$$
 $$%(\a, z_{0i}(\a))=
{\cF}'_{0i}({\tt J},{\tt E}):=%\big\{(\a, z_{0i}(\a))\big\}=
\Big\{\Big(\L_{0}({\tt J}), \GG, \ell,  {\rm g}_{0i}(\GG, {\tt J},{\tt E})\Big)\ :\quad (\ell, \GG)\in{\mathbb T}\times \cD'_i\Big\}$$
%$$(\b , w_{0j}(\b))=\Big(({\tt J},{\tt E},{\rm g}, \ell), \L_j({\tt J},{\tt E},{\rm g}, \ell), {\rm G}_j({\tt J},{\tt E},{\rm g}, \ell)\Big)$$
for suitable compact sets $\cD'_i({\tt J}, {\tt E})\subset {\mathbb R}_+$, $\cD_i({\tt J}, {\tt E})\subset {\mathbb T}$ 
and functions $\L_{0i}({\tt J})$, ${\rm g}_{0i}(\GG, {\tt J}, {\tt E})$, ${\rm G}_{0i}({\rm g}, {\tt J}, {\tt E})$ ($i=1$, $\cdots$, $N$),
 such that any two consecutive graphs in the chain are partially overlapping and
$$\cM_0({\tt J}, {\tt E}, \rr')=\cup_{i=1}^N{\cF }_{0i}\cup_{i=1}^N{\cF}'_{0i}\ .$$
In addition,
the following holds
 \beqa{G graphs}
&&\det\partial_{(\L,{\rm g})}\FF_0(\L,\GG, \ell, {\rm g}, {\tt J}, {\tt E})\Big|_{(\L,\GG, \ell, {\rm g})\in\cF_{0i}({\tt J}, {\tt E})}\ne 0
\nonumber\\
&& \det\partial_{(\L,{\rm G})}\FF_0(\L,\GG, \ell, {\rm g},{\tt J}, {\tt E})\Big|_{(\L,\GG, \ell, {\rm g})\in\cF'_{0i}({\tt J}, {\tt E})}\ne 0%({\tt J},{\tt E})\in {\cal M}_0^*\ ,
%\quad \forall\ {\rm G}\in\cD'_j
\eeqa 
 for all $ i=1,\cdots, N$.
 Finally, the $\cD_i$'s and $\cD'_i$'s can be chosen so that the sets $\cD^\circ:=\cup_i\cD_i$, ${\cD'}^\circ:=\cup_i\cD'_i$ are arbitrary punctured neighborhoods of a finite number of points.
\end{lemma}

\nl
Proposition~\ref{Liouville Arnold3bis} is proved in the following form
 \begin{proposition}\label{Liouville Arnold3} 
 Under the assumptions of Proposition~\ref{Liouville Arnold3bis}, the manifolds $\cM_\m$ $({\tt J}$, ${\tt E}$, $\rr')$
are two--dimensional smooth, connected and compact manifolds $($hence, diffeomorhic to ${\mathbb T}^2)$, given by the union of graphs
\beqa{FG1}
%(\a, z_{i}(\a))=
{\cF}'_{i}({\tt J},{\tt E}, \rr')=\Big\{\Big(\L'_{i}(\ell, {\rm g}, {\tt J},{\tt E}, \rr'), {\rm G}_{i}(\ell, {\rm g}, {\tt J},{\tt E}, \rr'), \ell,  {\rm g}\Big)\ :\ (\ell, {\rm g})\in {\mathbb T}\times\cD_i\Big\}
\eeqa 
and 
\beqa{FG}%(\a, z_{0i}(\a))=
{\cF}_{j}({\tt J},{\tt E}, \rr')=%\big\{(\a, z_{i}(\a))\big\}=
\Big\{\Big(\L_{i}({\rm G}, \ell, {\tt J},{\tt E}, \rr'),\GG, \ell,  {\rm g}_{i}({\rm G},\ell, {\tt J},{\tt E}, \rr')\Big)\ :\ (\ell, \GG)\in {\mathbb T}\times \cD'_i\Big\}
\eeqa
which reduce to ${\cF}'_{i}$, ${\cF}_{j}$ as $\m\to 0$.
%with $\cD_\m({\tt J}, {\tt E}, \rr')\subset [0,2\p]$ mod $2\p$, compact.
\end{proposition}
We start with proving that $\cM_\m({\tt J}, {\tt E}, \rr')$
are two--dimensional smooth, connected and compact manifolds given by the union of graphs~\equ{FG1}--\equ{FG}.
We shall use the Implicit Function Theorem.
 The key point is that, for any ${\tt K}$ as in the assumption, the functions $\JJ$ and $\EE$ are regular.

\proof {\bf of Lemma~\ref{graphs lemma}} The $\cF'_{0i}$'s are completely described in the proof of Proposition~\ref{prop: phase portrait} 
 (compare~\equ{cosg}). For the $\cF'_{0i}$'s, 
 one rewrites equation~\equ{P0} as
 $$\widehat\GG^4-(2\widehat{\tt E}-\widehat\d^2)\widehat\GG^2+\widehat{\tt E}^2-\widehat\d^2=0\ ,$$
 with $\widehat\GG:=\frac{\GG}{{\tt L}_0({\tt J})}$, $\widehat{\tt E}:=\frac{{\tt E}}{{\tt L}_0({\tt J})^2}$, $\widehat\d:=\d({\tt J}, \rr')\cos{\rm g}$.
According to Cartesio rule, this equation has two acceptable solutions in $\widehat\GG^2$ for $\widehat{\tt E}\le-|\widehat\d|$ or
 $\max\big\{|\widehat\d|,\ \frac{\widehat\d^2}{2}\big\}\le \widehat{\tt E}\le1+\frac{\widehat\d^2}{4}$; only one solution is acceptable when  $-|\widehat\d|<\widehat{\tt E}<\min\big\{1+\frac{\widehat\d^2}{4},\ |\widehat\d|\big\}$; none in the other cases. 
The corresponding solutions are
$\GG=\GG_{\s, \s'}({\rm g}, {\tt J}, {\tt E}, \rr')$, where
\beqano
\GG_{\s, \s'}({\rm g}, {\tt J}, {\tt E}, \rr')&=&\s\Big({\tt L}_0({\tt J})\Big[\frac{{\tt E}}{{\tt L}_0({\tt J})^2}-\frac{\d({\tt J},\rr')^2}{2}\cos^2{\rm g}\nonumber\\
&+& \s'\d({\tt J},\rr')|\cos{\rm g}|\Big(1-\frac{{\tt E}}{{\tt L}_0({\tt J})^2}+\frac{\d({\tt J},\rr')^2}{4}\cos^2{\rm g}\Big)^{1/2}\Big]^{1/2}\Big)\ .
\eeqano
choosing $\s\in\{\pm1\}$ and 
$\s'=+1$ or $\s'\in\{\pm1\}$ according to the cases above.
 The fact that the $\cF_{0i}$'s, $\cF'_{0i}$'s can be chosen so as to satisfy~\equ{G graphs} follows from that, since $\JJ_0$ is independent of $\L$, then~\equ{G graphs} is equivalent to condition $(\partial_\GG \FF_0,\partial_{\rm g}\FF_0)\ne (0,0)$ for all $({\rm G}, {\rm g})\in \ovl\cC_{0, 2}({\tt J}, {\tt E},\rr')$, which is certainly satisfied for all $({\tt J}, {\tt E})\in {\cal M}_0^*$, by the definition of ${\cal M}_0^*$.  $\quad\square$

 \nl
\proof {\bf of Proposition~\ref{Liouville Arnold3}} Let ${\tt K}\subset \cM_0(\rr')$ compact and let $({\tt J}, {\tt E})\in {\tt K}$.  We want to show that there exists $\m_0({\tt J}, {\tt E}, \rr')>0$, depending continuously on  $({\tt J}, {\tt E}, \rr')$, such that the set $\cM_\m({\tt J}, {\tt E},\rr')$
 is smooth, connected and compact, with $\m<\m_0({\tt J}, {\tt E}, \rr')$, so that the theorem will be proved with $\m_0({\tt K}):=\min_{({\tt J}, {\tt E}, \rr')\in {\tt K}}\m_0({\tt J}, {\tt E}, \rr')$.
The manifolds $\cM_\m({\tt J}, {\tt E},\rr')$   have equation (see~\equ{H3B})
 \beqa{perturbed level sets}
 \arr{
 \dst\JJ={\JJ}_0+\m{\JJ}_1=
 -\frac{\mm^3{\cal M}^2}{2\L^2}-\m\frac{\mm{\cal M}}{\sqrt{{\rr'}^2+2\rr' a\varrho \cos({\rm g}+\n)+{a}^2\varrho^2}} ={\tt J}\\
\dst\EE=\EE_0+\m\EE_1=
\GG^2+\mm^2{\cal M}\rr'\sqrt{1-\frac{\GG^2}{\L^2}}\cos{\rm g}\\
\qquad \dst+\m\mm^2{\cal M}\rr'\frac{{\rr'}+a\varrho \cos({\rm g}+\n)}{\sqrt{{\rr'}^2+2\rr' a\varrho \cos({\rm g}+\n)+{a}^2\varrho^2}}={\tt E}}
\eeqa
We write the equation for $\cM_\m({\tt J}, {\tt E}, \rr')$ as
\beqa{F}\FF(\L,\GG, \ell, {\rm g}, {\tt J}, {\tt E})=({\JJ}(\L,\GG, \ell, {\rm g})-{\tt J}, \EE(\L,\GG, \ell, {\rm g})-{\tt E})=0\ .\eeqa
We aim to apply the Implicit Function Theorem (Lemma~\ref{IFT})
to this
$\FF$,  taking
\beqano
z=(\L, {\rm g})\ ,\quad \a=(\GG, \ell)\ ,\quad A=\cD'_i\times {\mathbb T}\ ,\quad   z_0(\GG, \ell)=({\tt L}_0({\tt J}),{\rm g}_{0i}(\GG, {\tt J}, {\tt E}))
\eeqano
or
\beqano
z=(\L, {\rm G})\ ,\quad \a=(\ell, {\rm g})\ ,\quad A= {\mathbb T}\times \cD_j\ ,\quad   z_0(\ell,{\rm g})=({\tt L}_0({\tt J}),{\rm G}_{0j}({\rm g}, {\tt J}, {\tt E}))
\eeqano
We split $$\FF=\FF_0+\m\FF_1\ . $$
The key point is that, having chosen $({\tt J}, {\tt E})\in{\cal M}^*\subset {\cal M}_0^*$ and ${\cal M}_0^*$ does not intersect $\Sigma_{\rm 1}\cup \Sigma_{\rm 0}$, one can find a neighborhood $\UU$ of  $\cup_i{\cF}_{0i}\cup_i{\cF}'_{0i}$
such that $\FF_1$ is smooth in $\UU$. Namely, there exists $\r_{0}=\r_{0}({\tt J}, {\tt E}, \rr')>0$ independent of $\m$ such that $\FF_0$ and $\FF_1$ are  of class $C^1$ on the domains 
$$\cF_{{0i}\r_{0}}=\big\{
(\L,\GG, \ell,  {\rm g}):\ |\L-{\tt L}_0({\tt J})|\le \r_{0}\ ,\quad |{\rm g}-{\rm g}_{0i}(\GG, {\tt J}, {\tt E})|\le \r_{0}\ ,\quad \GG\in \cD'_i\ ,\quad \ell\in {\mathbb T}
\big\}\ .$$
$$\cF'_{{0i}\r_{0}}=\big\{
(\L,\GG, \ell,  {\rm g}):\ |\L-{\tt L}_0({\tt J})|\le \r_{0}\ ,\quad |{\rm G}-{\rm G}_{0i}(\GG, {\tt J}, {\tt E})|\le \r_{0}\ ,\quad {\rm g}\in \cD_i\ ,\quad \ell\in {\mathbb T}
\big\}\ .$$
Writing then
$$\partial_{\rm g}\FF=\partial_{\rm g}\FF_0+\m\partial_{\rm g}\FF_1\qquad \partial_{\rm G}\FF=\partial_{\rm G}\FF_0+\m\partial_{\rm G}\FF_1$$
one sees that the former condition in~\equ{IFT cond} is satisfied for
$$\mu\le \min_i\Big\{\frac{\min_{\cD'_i}|\partial_{\rm g}\FF_0|}{2\max_{\cD'_i} |\partial_{\rm g}\FF_1|},\ \frac{\min_{\cD_i}|\partial_{\rm G}\FF_0|}{2\max_{\cD_i} |\partial_{\rm G}\FF_1|}\Big\}=:\mu_{01}({\tt J}, {\tt E}, \rr')$$
Denote also $$m^{-1}:=\frac{1}{2}\min_i\Big\{\min_{\cD'_i}|\partial_{\rm g}\FF_0|\ ,\ \min_{\cD_i}|\partial_{\rm G}\FF_0|\Big\}=:\Big(\ovl m({\tt J}, {\tt E}, \rr')\Big)^{-1}\ .$$
 The second condition in~\equ{IFT cond} holds with
$$\r=2\m  \ovl m\max_i\Big\{\sup_{\cD'_i}|\FF_1|, \sup_{\cD_i}|\FF_1|\Big\}=:\m \ovl\r$$
provided that $\m$ satisfies
$$\m\le \frac{\r_0({\tt J}, {\tt E}, \rr')}{ \ovl\r({\tt J}, {\tt E}, \rr')}=:\mu_{02}({\tt J}, {\tt E}, \rr')$$
Finally, the third condition in~\equ{IFT cond} holds provided that
$$\m\le \big(2 \ovl m\ovl\r\big)^{-1}\big(\sup_{\cF_{0i\r_0}}|\partial_\GG^2\FF|,\sup_{\cF'_{0i\r_0}}|\partial_{\rm g}^2\FF|\big)=:\mu_{03}({\tt J}, {\tt E}, \rr')\ .$$
So, assuming $\m\le\m_0({\tt J}, {\tt E}, \rr'):=\min_{i=1,2,3}\m_{0i}({\tt J}, {\tt E}, \rr')$ (which, as desired is a continuous function of $({\tt J}, {\tt E}, \rr')$), Lemma~\ref{IFT} applies. We then obtain that the solutions of Equations in~\equ{F} can be described as union of graphs of the form~\equ{FG1}--\equ{FG}
which are $\m$--close to ${\cF}_{0i}({\tt J},{\tt E},\rr')$, ${\cF}'_{0i}({\tt J},{\tt E},\rr')$, respectively. The fact that the union of such graphs is smooth, connected and compact is a consequence of the uniqueness claimed by the Implicit Function Theorem and the 
fact that the union of the ${\cF}_{0i}({\tt J},{\tt E},\rr')$, ${\cF}'_{0i}({\tt J},{\tt E},\rr')$ is so. \\
\proof {\bf of Proposition~\ref{Liouville Arnold3quater}}
Consider the case $-\d<\widehat{\tt E}<1$.  By Proposition~\ref{Liouville Arnold3}, for any $\ell\in {\mathbb T}$, any $-\d<\widehat{\tt E}<1$, possibly $\widehat{\tt E}\ne \d$, any sufficiently small $\m$ the projection of  $\cM_\m({\tt J}, {\tt E}, \rr')$ on the $({\rm g}, \GG)$--plane is a closed curve encircling $(\p, 0)$. Then we can parametrize such curve as $$\arr{\GG_\m({\tt J}, {\tt E}, \rr', \ell, \psi_2)=\r_\m({\tt J}, {\tt E}, \rr', \ell, \psi_2)\cos\psi_2\\
{\rm g}_\m({\tt J}, {\tt E}, \rr', \ell, \psi_2)-\p=\r_\m({\tt J}, {\tt E}, \rr', \ell, \psi_2)\sin\psi_2
}$$
so ${\rm g}_\m({\tt J}, {\tt E}, \rr', \ell, 0)=\p$. The case  $1<\widehat{\tt E}<1+\frac{\d^2}{4}$ is similar.\quad$\square$
%{\beqa{tau1}\t=-\partial_{\tt E}\oint_{\ovl\cC_{0, 2}({\tt J}, {\tt E},\rr')}^\GG {\rm g} d\GG'=-\oint_{\ovl\cC_{0, 2}({\tt J}, {\tt E},\rr')}^\GG {\rm g}_{\tt E} d\GG' \ ,\eeqa}

\proof {\bf of Proposition~\ref{Liouville Arnoldter}} The equations~\equ{graph curves1} (equations~\equ{graph curves}) correspond to curves along the graphs~\equ{perturbed level sets} obtained taking  $\ell=0$, ($\psi_2=0$) and using that, for $\ell=0$, $\zeta=0$, so $\varrho=1-\ee\cos0=1-\ee$, (for $\psi_2=0$, ${\rm g}=\frac{1-\s}{2}\p$, so $\cos\frac{1-\s}{2}\p=-\s$).  \quad$\square$

\proof {\bf of Proposition~\ref{cor: action-angle}} The proof extends the one given for $\m=0$ (Proposition~\ref{phiA0}). Let ${\tt P}(\ell, {\rm g})$ a  point belonging to $\cM_\m({\tt J}, {\tt E}, \rr')$; $\ovl{\tt P}:={\tt P}(0, \frac{1-\s({\tt J}, {\tt E})}{2}\p)$. 
We choose a curve along $\cM_\m({\tt J}, {\tt E}, \rr')$   connecting  $\ovl{\tt P}$ to ${\tt P}(\ell, {\rm g})$ as follows. Let $$\ovl\cC_{\m}({\tt J}, {\tt E}, \rr')_{\ovl{\tt P}}^{{\tt P}(\ell, {\rm g})}:=\ovl\cC_{1, \m}({\tt J}, {\tt E}, \rr')_{\ovl{\tt P}}^{{\tt P}(\ell, \GG_1)}\cup \ovl\cC_{2, \m}({\tt J}, {\tt E}, \rr')_{{\tt P}(\ell, \GG_1)}^{{\tt P}(\ell, {\rm g})}$$ where $\GG_1$ is the value of $\GG_{\m, 1}$ for  $\psi_1=\ell$.
Let $\hat\cC_{\m}(\cL, \cG, \rr'):=\ovl\cC_{\m}({\tt A}^{-1}(\rr')(\cL, \cG), \rr')$; $\hat{\tt P}(\cL, \cG, \rr'):=\ovl{\tt P}({\tt A}^{-1}(\rr')(\cL, \cG))$.

\nl
{\it Case 1: $\{[0<\d\le 1$ \& $(-\d<\hat{\tt E}<1$ or $\d<\hat{\tt E}<1)]$ or $[1<\d<2$ \& $-\d<\hat{\tt E}<1]\}\cap \P^\ppj$}

\nl
We look at the canonical transformation generated by
$$S_{\rm full, 1}(\hat\RR', \cL, \cG, \rr', \ell, \GG):=\hat\RR'\rr'+\int_{\hat\cC_{\m}(\cL, \cG, \rr')_{\hat{\tt P}}^{{\tt P}(\ell, {\rm g})}}(\L d\ell'-{\rm g}d\GG')$$

\vskip.2in
\noi
{\it Case 2: $\{[0<\d\le 1$ \& $1<\hat{\tt E}<1+\frac{\d^2}{4}]$ or $[1<\d<2$ \& $(1<\hat{\tt E}<\d$ or $\d<\hat{\tt E}<1+\frac{\d^2}{4})]\}\cap\P_\m^\ppj$}

\nl
In this case, we consider
$$S_{\rm full, 2}(\hat\RR', \cL, \cG, \rr', \ell, \GG):=\hat\RR'\rr'+\int_{\hat\cC_{\m}(\cL, \cL-\cG, \rr')_{\hat{\tt P}}^{{\tt P}(\ell, {\rm g})}}(\L d\ell'-{\rm g}d\GG')$$
In both cases, we obtain a transformation of the form~\equ{action angle full}, which is $\m$--close to $\cA_0$. \quad $\square$

\newpage\section{Proof of Theorem~A}\label{setup}

In this section we provide the proof of a more precise statement of Theorem~A. To state it we need some preparation. We consider the three--body problem Hamiltonian~\equ{3BP}, and aim to transform in into the form~\equ{HH}.

\nl
In terms of the coordinates $\cK$, the Hamiltonian~\equ{3BP} is as in~\equ{H3B}.
We rename
\beqa{CC}\hat\CC:=\varepsilon\CC\ .\eeqa

\nl
This change of notation is more appropriate if one wants to consider large values of $\CC$. Our result will actually allow for $\CC\sim\varepsilon^{-1}$. 

\nl
In terms of the coordinates \beqa{hatA}{\!\!\!\cA}=(\hat\RR', \cL, \cG, \hat\rr', \l, \g)\eeqa
defined in Proposition~\ref{cor: action-angle}, %combined with the non--canonical change
%\beqano\hat\phi:\qquad (\hat\RR, \hat\rr)\to(\hat\RR, \tilde\rr)=(\varepsilon^{-1}\hat \RR, \hat\rr)\eeqano
this Hamiltonian becomes
\beqa{model}
\hat\HH&=&%\frac{\varepsilon^2{\RR'}^2}{2\mm'}+\frac{\varepsilon^2(\GG-\CC)^2}{2\mm'{\rr'}^2}
\frac{\varepsilon^2(\hat\RR'+\hat\r)^2}{2\mm'}+\frac{(\hat\CC-\varepsilon\cG-\varepsilon\GG_1)^2}{2\mm'{\hat\rr'}^2}-\frac{\mm'{\cal M}'}{\hat\rr'}+\varepsilon\left(
-\frac{\mm^3{\cal M}^2}{2\cL^2}+\m\UU(\cL, \cG, \hat\rr';\m)
\right)\nonumber\\
&+&\frac{\m\varepsilon}{m_0}\Big(\varepsilon(\hat\RR' +\hat\r){\rm y}_{\cA, 2}-\frac{\hat\CC-\varepsilon\cG-\varepsilon\GG_1}{\hat\rr'}{\rm y}_{\cA, 1}\Big)%\nonumber\\
%&=:&-\frac{\mm'{\cal M}'}{\hat\rr'}+
%\varepsilon\JJ(\L, \GG, \ell, {\rm g}, \rr';\m)+\varepsilon^2 f_\cK(\RR', \L, \GG, \rr', \ell, {\rm g};\CC, \m)
\eeqa
having used Equations~\equ{action angle full},~\equ{HE} and having let ${\tt G}_\m=\cG+\GG_1$ and
 ${\rm y}_{\cA, i}:=\ovl{\tt y}_{  i}\circ \cA$. 
 We manipulate $\hat\HH$ a bit.
 At first, we split
\beqano
&&(\hat\RR'+\hat\r)^2=\hat{\RR'}^2+2\hat\RR'\hat\r+\hat\r^2\nonumber\\\nonumber\\
&&(\hat\CC-\varepsilon\cG-\varepsilon\GG_1)^2=(\hat\CC-\varepsilon\cG)^2-2\varepsilon\GG_1(\hat\CC-\varepsilon\cG)+\varepsilon^2\GG_1^2
\eeqano
Next, we Taylor--expand the function \beqa{V'}V'(\hat\rr')&:=&
-\frac{\mm'{\cal M}'}{{\hat\rr}'}+\frac{(\hat\CC-\varepsilon\cG)^2}{2\mm'{\hat\rr'}^2}
\eeqa
%
%Observe that, when $\varepsilon=0$, $\JJ'$ reduces to the Kepler Hamiltonian
%$$\JJ'_{\rm kep}=\frac{\varepsilon^2{\hat\RR'}^2}{2\mm'}+V'(\hat\rr')\ .$$
%Denoting as
%$$\hat\rr'_0:=\frac{\hat\CC^2}{{\mm'}^2{{\cal M}'}}$$
%the value of $\hat\rr'$ where $V'$ reaches its minimum, we let the coordinate $\hat\rr'$ to vary in the closed disk
%\beqa{radius}|\hat\rr'-\hat\rr'_0|\le \D<\hat\rr'_0\ .\eeqa
%The coordinate $\hat\RR'$ is fixed  to vary in the domain
%\beqa{RR}|\hat\RR'|<\sqrt{2\mm'\EE'+\frac{{\mm'}^4{{\cal M}'}^2}{\hat\CC^2}}\qquad \EE'\ge -\frac{{\mm'}^3{{\cal M}'}^2}{2\hat\CC^2}=:\EE'_0\eeqa
%The domains~\equ{radius} and~\equ{RR} have been chosen to coincide with the values one would have for the dynamics of $\JJ'_{\rm k}$  with energy $\le \EE'$, with a proper relation between $\D$ and $\EE'$. More restrictions on $\D$ and $\hat\EE'$ will be posed below.
%In correspondence of these choices, the function $f$ in~\equ{pert funct} is real--analytic, and satisfies
%$$\sup|f|\le \cc_1 \varepsilon$$
%with $\cc_1$ depending on $\D$, $\EE'$.\\
around its minimum
$$\hat\rr'_0(\cG, \hat\CC, \varepsilon):=\frac{(\hat\CC-\varepsilon\cG)^2}{{\mm'}^2{{\cal M}'}}\ .$$
We obtain
\beqano
V'(\hat\rr')=-\frac{{\mm'}^3{{\cal M}'}^2}{2(\hat\CC-\varepsilon\cG)^2}+
\frac{{\mm'}^7{{\cal M}'}^4}{2(\hat\CC-\varepsilon\cG)^6}(\hat\rr'-\hat\rr'_0)^2+\hat v(\hat\rr', \GG_1; \hat\CC, \varepsilon)
\eeqano
with
$$\hat v=\OO_3\big(\hat\rr'-\hat\rr'_0( \cG, \hat\CC, \varepsilon); \cG, \hat\CC, \varepsilon\big)\ .$$
Finally, we rewrite $\hat\HH$ as

%with $\CC$ as in~\equ{CC},
 %sees that, in  terms of the coordinates $\cA$ described in Proposition~\ref{cor: action-angle}, the Hamiltonian takes the form 
\beqa{H}
\hat\HH&=&\hat\hh(\cL, \cG; \hat\CC, \varepsilon)+\frac{\varepsilon^2{\hat\RR'}^2}{2\mm'}+
\frac{{\mm'}^7{{\cal M}'}^4}{2(\hat\CC-\varepsilon\cG)^6}(\hat\rr'-\hat\rr'_0)^2+\hat f(\hat\RR', \cL, \cG, \hat\rr',\hat\ell, \hat\g; \hat\CC, \varepsilon, \m)\nonumber\\
\eeqa
with

\beqano
\hat\hh(\cL, \cG; \hat\CC, \varepsilon)&:=&-\frac{{\mm'}^3{{\cal M}'}^2}{2(\hat\CC-\varepsilon\cG)^2}-\varepsilon\frac{\mm^3{\cal M}^2}{2\cL^2}\nonumber\\
\hat f(\hat\RR', \cL, \cG, \hat\rr',\hat\ell, \hat\g; \hat\CC, \varepsilon, \m)&:=&\hat v(\hat\rr', \GG_1; \hat\CC, \varepsilon)+\varepsilon\m\UU(\cL, \cG, \hat\rr';\m)\nonumber\\
&+&\frac{2\varepsilon^2\hat\RR'\hat\r}{2\mm'}+\frac{\varepsilon^2\hat\r^2}{2\mm'}-
\frac{\varepsilon(\hat\CC-\varepsilon\cG)\GG_1}{\mm'{\hat\rr'}^2}
+
\frac{\varepsilon^2\GG_1^2}{2\mm'{\hat\rr'}^2}
\nonumber\\
&+&\frac{\m\varepsilon^2}{m_0}\hat\RR' {\rm y}_{\cA, 2}+\frac{\m\varepsilon^2}{m_0}\hat\r{\rm y}_{\cA, 2}-\frac{\m\varepsilon}{m_0}\frac{\hat\CC-\varepsilon\cG}{\hat\rr'}{\rm y}_{\cA, 1}+\frac{\m\varepsilon^2}{m_0}\frac{\GG_1}{\hat\rr'}{\rm y}_{\cA, 1}\nonumber\\
\eeqano

\nl We consider the holomorphic extension of $\hat\HH$ on  the domain
$$\cD_{\eta, \varepsilon, \r, s}:=\hat\cB^2_{\eta, \varepsilon}\times {\cal W}_\r\times {\mathbb T}^2_s$$
where ${\cal W}:={\cal W}^\ppd_{\m, {\tt K}}$, with ${\cal W}^\ppd_{\m, {\tt K}}$ as in defined as in Proposition~\ref{cor: action-angle},   $\r$, $s$ are suitably small numbers.
Moreover, letting
$$ \r_-:=\min\{\inf|\hat\CC-\varepsilon\cG|, \ \inf |\cL|,\ \inf \frac{1}{|{\rm y}_{\cA, 1}|},\ \inf \frac{1}{|{\rm y}_{\cA, 2}|},\ \inf \frac{1}{|\hat\r|},\ \r\}$$ 
and assuming that
$$\r_-\le |\hat\CC-\varepsilon\cG|\le \r_+\ , $$
we have let
\beqa{initial domain}\hat\cB^2_{\eta, \varepsilon}:=\big\{(\hat\RR, \hat\rr):\  |\hat\RR|\le \frac{{\mm'}^2{\cal M}}{2\varepsilon\r_+}\eta \ ,\ |\hat\rr'-\hat\rr'_0|\le\frac{\r_-^2}{2{\mm'}^2{{\cal M}'}}\eta\big\}\ .
\eeqa

\nl
Observe that, in the case $\CC=\OO(1)$, hence $\hat\CC=\OO(\varepsilon)$ (see \equ{CC}),  $\r_\pm$ are $\OO(\varepsilon)$ and we are precisely, for the coordinates $\hat\RR'$, $\hat\rr'$, in the range described in comment (i) of the introduction.  The check that  the coordinates $\hat\RR'$, $\hat\rr'$ will remain in their domain for the whole time will be part of the proof.

\nl
We shall prove the following result, which is a more quantitative version of Theorem~A.

 \begin{theorem}\label{theor: Euler}
Fix $\epsilon_0$ small. Let $\kappa$ be an upper bound for the ratio $\frac{|\hat\CC-\varepsilon\cG|}{|\cG|}$.
There exists $\epsilon_*>0$ such that, if

$$\epsilon:=\epsilon_*\max\left\{\frac{\varepsilon}{\eta^2},\ \frac{\eta^3}{\varepsilon},\ \eta,\  \kappa,\ \m,\ \varepsilon\right\}<\epsilon_0$$
the action $\cG$  varies a little in the course of an exponentially long time interval:
$$|\cG(T)-\cG(0)|\le\epsilon_0\r\qquad \forall\ T:\quad |T|\le T_1(\varepsilon, \m, \eta, \kappa; \epsilon_0) 2^{\frac{1}{\epsilon}}$$
where 
$T_1(\varepsilon, \m, \eta, \kappa; \epsilon_0)=T_*\frac{\epsilon_0}{\varepsilon\epsilon(\varepsilon, \m, \eta, \kappa)}$.
\end{theorem}

\nl
We shall need  the following information on the function $
{\tt G}_0(\cL_0, \cG_0, {\rm g}_0, \hat\rr_0')$ in~\equ{lift}.

\nl
 For a given open set $\cA\subset {\mathbb R}^n$ and $\r>0$, we denote as ${\cal A}_{-\r}:=\big\{x\in \cA:\ B^n_{\r}(x)\subset\cA\big\}$.
\begin{lemma}\label{prop small G1} Fix $\d_0$ small. There exists $b_*>0$ such that
\beqa{hatG} \sup_{{\cal W}_{-\d_0}\times {\mathbb T}}|{\tt G}_0-\cG_0|\le b_* \frac{\rr_0'}{\cG_0}\eeqa
\end{lemma}
The proof of Lemma~\ref{prop small G1} is postponed to the end of this section. We now proceed with the

\proof  {\bf of Theorem~\ref{theor: Euler}} We proceed in three steps.

\paragraph{\it a) evaluation of $f$}
Using the definitions above, one sees that the terms $f$ is composed of can be bounded as follows:

\nl
\beqano
&&\dst |\hat v|\le \frac{\eta^3}{\r_-^2}\ ,\quad  \left|\varepsilon\m\UU\right|\le \frac{\varepsilon\m}{\r_-^2}\ ,\quad  \left|\frac{2\varepsilon^2\hat\RR'\hat\r}{2\mm'}\right|\le\frac{\varepsilon\eta}{\r_-^2}\ ,\quad \left |\frac{\varepsilon^2\hat\r^2}{2\mm'}\right|\le\frac{\varepsilon^2}{\r_-^2}\nonumber\\
&&\dst \left|\frac{\varepsilon(\hat\CC-\varepsilon\cG)\GG_1}{\mm'{\hat\rr'}^2}\right|\le \frac{\varepsilon  (\kappa+\m)}{\r_-^2}\ ,\quad \left|\frac{\varepsilon^2\GG_1^2}{2\mm'{\hat\rr'}^2}\right|\le \frac{\varepsilon^2 (\kappa+\m)^2}{\r_-^2}\ ,\quad \left|
 \frac{\m\varepsilon^2}{m_0}\hat\RR' {\rm y}_{\cA, 2}
 \right|\le \frac{\m\varepsilon\eta}{\r_-^2} \nonumber\\
 &&\left|\frac{\m\varepsilon^2}{m_0}\hat\r{\rm y}_{\cA, 2}\right|\le\frac{\m\varepsilon^2}{\r_-^2}\ ,\quad \left|
 \frac{\m\varepsilon}{m_0}\frac{\hat\CC-\varepsilon\cG}{\hat\rr'}{\rm y}_{\cA, 1}
 \right| \le\frac{ \m\varepsilon}{\r_-^2}\ ,\quad\left|
 \frac{\m\varepsilon^2}{m_0}\frac{\GG_1}{\hat\rr'}{\rm y}_{\cA, 1}
 \right|\le \frac{\m\varepsilon^2(\kappa+\m)}{\r_-^2}
 \eeqano
Here,  we have let
$$  \kappa:=\sup\left|\frac{\hat\CC-\varepsilon\cG}{\cG}\right| $$
and we have used, for $|\GG_1|$, the bound
$$|\GG_1|=|{\tt G}_\m-\cG|\le |{\tt G}_0-\cG_0|+|{\tt G}_\m-{\tt G}_0|+|\cG-\cG_0|\le \frac{b_*\hat\rr'}{\cG_0}+\m b_*\cG$$
implied by~\equ{hatG}, for a suitably larger $b_*$. In count of the  previous bounds, we can assert
\beqa{E***}\|f\|\le \EE:=\max\big\{
{\varepsilon\eta}\ ,\ {\varepsilon\kappa}\ ,\ {\varepsilon\m}\ ,\ {\varepsilon^2}\ ,\ {\eta^3}\big\}\r_-^{-2}\eeqa

\paragraph{\it b) rescaling}

\nl
We introduce the transformation
$$\tilde\phi:\quad (\tilde y, \tilde x, \tilde\cL, \tilde\cG, \tilde\l, \tilde{\rm g})\in \tilde\cB^2_{\eta, \varepsilon}\times{\cal W}_\r\times {\mathbb T}^n_s\to (\hat y, \hat x, \cL, \cG, \l, {\rm g})\in  \hat\cB^2_{\eta, \varepsilon}\times{\cal W}_\r\times {\mathbb T}^n_s$$
defined via
\beqa{homothetic}&& \hat\RR'=\frac{{\hat\mm'}^2{\hat{\cal M}'}}{\sqrt\varepsilon(\hat\CC-\varepsilon\tilde\cG)^{3/2}} \tilde y\ ,\qquad \hat\rr'=\hat\rr_0+\frac{\sqrt\varepsilon(\hat\CC-\varepsilon\tilde\cG)^{3/2}}{{\hat\mm'}^2{\hat{\cal M}'}}\tilde x\nonumber\\
&&\cG=\tilde\cG\ ,\quad{\rm g}=\tilde{\rm g}-
\frac{{\hat\mm'}^2{\hat{\cal M}'}}{\sqrt\varepsilon(\hat\CC-\varepsilon\tilde\cG)^{3/2}} \tilde y
\partial_{\tilde\cG}\left(\hat\rr_0+\frac{\sqrt\varepsilon(\hat\CC-\varepsilon\tilde\cG)^{3/2}}{{\hat\mm'}^2{\hat{\cal M}'}}\tilde x\right)\nonumber\\
&&\cL=\tilde\cL\ ,\quad  \l=\tilde\l
\eeqa
The transformation $\tilde\phi$ is canonical, being generated by
$$S(\tilde\cL, \tilde\cG,\hat\RR', \l, {\rm g}, \tilde x)=-\hat\RR'\left(\hat\rr_0+\frac{\sqrt\varepsilon(\hat\CC-\varepsilon\tilde\cG)^{3/2}}{{\hat\mm'}^2{\hat{\cal M}'}}\tilde x\right)+\tilde\cG{\rm g}+\tilde\cL \l$$
By~\equ{initial domain}, the coordinates $(\tilde y, \tilde x)$ can be taken to vary in the set
\beqa{new domain}\tilde\cB^2_{\eta, \varepsilon}:=\Big\{(\tilde y, \tilde x):\  |\tilde y|\le \frac{\r_-^{3/2}}{\r_+}\frac{\eta}{2\sqrt\varepsilon} \ ,\ |\tilde x| \le \frac{\r_-^2}{\r_+^{3/2}}\frac{\eta}{2\sqrt\varepsilon}\Big\}\ ,
\eeqa
while the domain for the coordinates $(\tilde\cL, \tilde\cG, \tilde\l, \tilde{\rm g})$ is left unvaried, ${\cal W}_\r\times {\mathbb T}^n_s$, since the shift in $\tilde{\rm g}$ is real.
The transformation
$\tilde\phi$ yields $\HH$ to
$$\tilde\HH=\HH\circ\tilde\phi=\tilde\hh+ \frac{\tilde\o_0}{2}(\tilde y^2+\tilde x^2)+\tilde f$$
with
$$\tilde\hh(\tilde\cL, \tilde\cG; \hat\CC, \varepsilon):=\hat\hh(\cL,\cG; \hat\CC, \varepsilon)\ ,\quad \tilde\o_0=\varepsilon\frac{{\mm'}^3{{\cal M}'}^2}{(\hat\CC-\varepsilon\cG)^3}\ ,\quad \tilde f:=f\circ \tilde\phi\ .$$

\paragraph{\it c) application of Theorem~\ref{NFC}}

\nl
We  apply Theorem~\ref{NFC}. Indeed, In view of~\equ{new domain}, the Hamiltonian $\tilde\HH$ is real--analytic in $B^2_{\d+\D}\times {\cal W}_\r\times {\mathbb T}^2_s$, with
\beqa{Deltadelta} \d=\a\frac{\eta}{2\sqrt{\varepsilon}} s_0\ ,\quad  \D=\b\frac{\eta}{2\sqrt{\varepsilon}} s_0\eeqa
with some fixed $0<\a<\b<1$ satisfying $\a+\b=1$, and $s_0:=\min\{\frac{\r_-^{3/2}}{\r_+},\ \frac{\r_-^{2}}{\r^{3/2}_+}\}$.
Then we can 
take $n=2$, $ \cI=\cal W$,
\beqa{f}\II=(\tilde\cL, \tilde\cG)\ ,\quad \f=(\tilde\l, \tilde{\rm g})\ ,\quad (y, x)=(\tilde y, \tilde x)\ ,\quad \hh=\tilde\hh\ ,\quad \o_0=\tilde\o_0\ ,\quad f=\tilde f\eeqa
Let us evaluate the constants $a$, $M_0$ $M_1$, $M$, $M_0'$ in
\equ{IN}, in order to check conditions~\equ{assump3} and~\equ{simplify}. We have
\beqano
\tilde\o(\tilde\cL, \tilde\cG)&:=&\partial_{\tilde\cL, \tilde\cG}
\tilde\hh
=\varepsilon\left(\frac{{\mm}^3{{\cal M}}^2}{\cL^3}\ ,\ - \frac{{\mm}^3{{\cal M}}^2}{(\hat\CC-\varepsilon\tilde\cG)^3}\right)\nonumber\\
\tilde\o_1(\tilde\cL, \tilde\cG, \tilde y)&:=&\partial_{\tilde\cL, \tilde\cG}
\left(\tilde\hh+ \frac{\tilde\o_0}{2}\tilde y^2\right)
=\varepsilon\left(\frac{{\mm}^3{{\cal M}}^2}{\cL^3}\ ,\ - \frac{{\mm}^3{{\cal M}}^2}{(\hat\CC-\varepsilon\tilde\cG)^3}+3\varepsilon \frac{{\mm'}^3{{\cal M'}}^2}{(\hat\CC-\varepsilon\cG)^4}\tilde y^2\right)\nonumber\\
\tilde\o_0(\tilde\cL, \tilde\cG)&=&\varepsilon \frac{{\mm'}^3{{\cal M}'}^2}{(\hat\CC-\varepsilon\cG)^3}\nonumber\\
\partial_{\tilde\cL, \tilde\cG}\tilde\o_0(\tilde\cL, \tilde\cG)&=&\left(0,\ 3\varepsilon^2 \frac{{\mm'}^3{{\cal M}'}^2}{(\hat\CC-\varepsilon\cG)^4}\right)
\eeqano
Therefore,
\beqano
&&|\tilde\o_0|=-\varepsilon \frac{{\mm'}^3{{\cal M}'}^2}{|\hat\CC-\varepsilon\cG|^3}\ge \frac{\varepsilon }{\r_+^3}=:a\ ,\quad \|\tilde\o_0\|\le \frac{\varepsilon}{\r_-^3}=:M_0\nonumber\\
&&\left\|
\tilde\o_1
\right\|\le \frac{\varepsilon}{\r_-^3}+\varepsilon^2\frac{\frac{\eta^2}{\varepsilon}}{\r_-^4}=\varepsilon\frac{\r_-+\eta^2}{\r_-^4}=:M_1\ ,\quad \|\tilde\o\|\le \frac{\varepsilon}{\r_-^3}=:M\nonumber\\
&&\|\partial_{\tilde\cL, \tilde\cG}\tilde\o_0\|\le \frac{\varepsilon^2}{\r_-^4}=:M_0'
\eeqano for some $\r_+>1$, with an eventually smaller $\r_-$.
These bounds give

\beqano
&&\frac{M_0'}{a}=\varepsilon\frac{\r_+^3}{\r_-^4}\ ,\quad c=\sqrt{\r_+}\frac{\sqrt\varepsilon}{\eta}\ ,\quad \epsilon=\epsilon_*\max\left\{\frac{\varepsilon}{\eta^2},\ \frac{\eta^3}{\varepsilon},\ \eta,\  \kappa,\ \m,\ \varepsilon\right\}\ ,\quad N=\left[\frac{1}{\epsilon}\right]\nonumber\\
&&\epsilon'=\epsilon'_*\max\left\{\frac{\varepsilon^2}{\eta^4},\ \eta,\ \frac{\varepsilon}{\eta},\  \kappa\frac{\varepsilon}{\eta^2},\ \m\frac{\varepsilon}{\eta^2},\ \frac{\varepsilon^2}{\eta^2}\right\}\ ,\qquad T_1= T_*\epsilon_0\varepsilon^{-1}\epsilon^{-1}
\eeqano
where $\epsilon_*$, $\epsilon'_*$, $T_*$ depend on $\r_+/\r_-$, $\r_+/\r_-$. $\quad \square$

\nl
\proof {\bf of Lemma~\ref{prop small G1}} The thesis is an immediate consequence of  the triangular inequality
$$|\cG_0-\GG|\le |\GG-\sqrt{\EE_0}|+|\cG_0-\sqrt{\EE_0}|$$
the formulae (implied by~\equ{level curves})
\beqa{GG}\GG=\L\sqrt{\frac{\EE_0}{\L^2}-\frac{\d^2}{2}\cos^2{\rm g}-\d\cos{\rm g}\sqrt{\frac{\d^2}{4}\cos^2{\rm g}+1-\frac{\EE_0}{\L^2}}} \eeqa
$$\cG_0=\frac{\L}{2\p}\int_0^{2\p}\sqrt{\frac{\EE_0}{\L^2}-\frac{\d^2}{2}\cos^2{\rm g}'-\d\cos{\rm g}'\sqrt{\frac{\d^2}{4}\cos^2{\rm g}'+1-\frac{\EE_0}{\L^2}}} d{\rm g}'$$
where $\d:=\frac{\mm^2\cM\rr'}{\L^2}$, the Taylor formula 
around $\rr'=0$ and the observation that, for $(\cL_0, \cG_0)\in {\cal W}_{-\d_0}$, the right hand of~\equ{GG} has a positive minimum as ${\rm g}\in {\mathbb T}$. The details are omitted. $\quad\square$
\newpage

\appendix

\section{Two--centre problem and elliptic coordinates}\label{2centres}

In this section we describe the derivation of the formulae~\equ{G1} and~\equ{EEE}.

\nl
As a first step,  we need recall the classical argument, reviewed in~\cite{bekovO78}, which shows the integrability of the Hamiltonian~\equ{2Cold} by separation of variables. 

\nl
After fixing a reference frame with the third axis in the direction of ${\tt v}_0$ and denoting as $(v_1,  v_2,  v_3)$ the coordinates of $ {\tt v}$ with respect to such frame,
one introduces  
the so--called  ``elliptic coordinates''  

 \beqa{lambdabeta} \l=\frac{1}{2}\big(\frac{\rr_+}{\rr_0}+\frac{\rr_-}{\rr_0}\big)\ ,\quad \b=\frac{1}{2}\big(\frac{\rr_+}{\rr_0}-\frac{\rr_-}{\rr_0}\big)\ ,\quad \o:=\arg{(-v_2, v_1)}\eeqa
 where we have let, for short,
 $$\rr_0:=\|{\tt v}_0\|\ ,\quad \rr_\pm:=\|{\tt v}\pm {\tt v}_0\|\ .$$

\nl
Regarding $\rr_0$ as a fixed external parameter and calling $p_\l$, $p_\b$, $p_\o$ the generalized momenta associated to $\l$, $\b$ and $\o$, it turns out that
the Hamiltonian~\equ{2Cold}, written in the coordinates $(p_\l, p_\b, \l, \b)$ is independent of $\o$  and has the expression
\beqa{H2C}\ovl\JJ&=&\frac{1}{\l^2-\b^2}\Big[\frac{p^2_\l(\l^2-1)}{2 \rr_0^2}+\frac{p^2_\beta(1-\beta^2)}{2 \rr_0^2}+\frac{p_\o^2}{2 \rr_0^2}\big(\frac{1}{1-\beta^2}+\frac{1}{\l^2-1}\big)\nonumber\\
&&-\frac{(\mm_++\mm_-)\l}{\rr_0^2}+\frac{(\mm_+-\mm_-)\beta}{\rr_0^2}\Big]\ .
\eeqa
It follows that the ``Hamilton--Jacobi'' equation
$$\ovl\JJ=h$$
separates completely as
\beq{split}{\cF}_\l(p_\l,\l,{p_\o}, \rr_0, h)+{\cF}_\b(p_\beta,\beta,{p_\o},  \rr_0, h)=0\eeq
with

$${\cF}_\l=p^2_\l(\l^2-1)+\frac{p_\o^2}{\l^2-1}-2(\mm_++\mm_-)\l-2 \rr_0^2\l^2h$$
$${\cF}_\b=p^2_\beta(1-\beta^2)+\frac{p_\o^2}{1-\beta^2}+2(\mm_+-\mm_-)\beta+2 \rr_0^2\beta^2h\ .$$

\nl
Taking the derivatives of Equation~\equ{split} with respect to $(p_\l, p_\b, \l, \b)$, one finds that ${\cF}_\l$, ${\cF}_\b$ have to be separately constant with respect to $(p_\l,  \l)$, $(p_\b,  \b)$, respectively. Hence, due to~\equ{split},
there must exist a function $\ovl\EE$, depending on the arguments $({p_\o}, \rr_0, h)$ only,  such that
$${\cF}_\l(p_\l,\l,{p_\o}, \rr_0, h)=-{\cF}_\b(p_\l,\l,{p_\o}, \rr_0, h)=\ovl\EE({p_\o}, \rr_0, h)\quad \forall\ (p_\l, p_\b, \l, \b)$$
We write $\ovl\EE$ as
\beqa{Es}
\ovl\EE&=&\frac{1}{2}({\cF}_\b-{\cF}_\l)\nonumber\\
&=&\frac{p^2_\beta}{2}(1-\beta^2)-\frac{p^2_\l}{2}(\l^2-1)+\frac{p_\o^2}{2}\big(\frac{1}{1-\beta^2}-\frac{1}{\l^2-1}\big)\nonumber\\
&&+\mm_+(\l+\beta)+\mm_-(\l-\beta)+\rr_0^2(\l^2+\beta^2)h\ .
\eeqa

\paragraph{Proof of~\equ{G1}}
 We now check that the function $\ovl\EE$, written in the initial coordinates ${\tt u}$, ${\tt v}$,  coincides with~\equ{G1}.
To this end, we introduce the {\it Delaunay coordinates $\cD_{{\tt v}_0}$, 
relatively to ${\tt v}_0$}.
 Their definition is as follows. If
$${\tt M}:={\tt v}\times {\tt u}\ ,\quad {\tt n}_0:={\tt v}_0\times{\tt M}\ ,\quad {\tt n}:={\tt M}\times{\tt v}$$
and, given three vectors  ${\tt n}_1$, ${\tt n}_2$, ${\tt b}\in {\mathbb R}^3$, with  ${\tt n}_1$, ${\tt n}_2\perp$, ${\tt b}$,
$\a_{{\tt b}}({\tt n}_1,{\tt n}_2)$ denotes the oriented angle defined by the ordered couple $({\tt n}_1,{\tt n}_2)$, relatively to the positive verse established by ${\tt b}$,
Then we define
$$\cD_{{\tt v}_0}:=(\Theta, {\MM}, \RR,\vartheta,  {m}, \rr)$$
via the formulae
\beqa{delaunay}
 \arr{\Theta:=\frac{{\tt M}\cdot {\tt v}_0}{\|{\tt v}_0\|}\\\\
{\MM}:=\|{\tt M}\|\\\\
\RR:=\frac{{\tt u}\cdot {\tt v}}{\|{\tt v}\|}}\qquad 
\arr{
\dst\vartheta:=\a_{{\tt v}_0}({\tt i}, {\tt n}_0)\\\\
\dst {m}:=\a_{{\tt M}}({\tt n}_0, {\tt v})\\\\
\dst  \rr:=\|{\tt v}\|
}
\eeqa
As it is well known, the coordinates $\cD_{{\tt v}_0}$ are homogeneous--canonical (see, e.g.,~\cite{fejoz13b} for a proof):
$${\tt u}\cdot d{\tt v}:=\sum_{i=1}^3 u_i dv_i=\Theta d\vartheta+\MM d m+\RR d\rr$$
The coordinates above are canonical, since they correspond to the well known Deluanay coordinates with respect to a frame 
having the third axis in the direction of the constant vector ${\tt v}_0$ and the first axis in the direction of a fixed ${\tt i}\in {\mathbb R}^3$, ${\tt i}\perp {\tt v}_0$. In the next section we shall define a set of coordinates $\cP$, for a two--particles system, which includes the~\equ{delaunay}'s and simultaneously reduces rotation invariance.

\nl
Note that, since $\Theta$ is a first integral to $\ovl\JJ$,  this Hamiltonian  will depend only on  the four coordinates
$$({\MM}, \RR, {m}, \rr)$$
and on $\rr_0$, $\Theta$ as ``fixed parameters''. Using such coordinates, $\ovl\JJ$ becomes
\beqa{hsDel}\ovl\JJ=\frac{\RR^2}{2}+\frac{{\MM}^2}{2\rr^2}-\frac{\mm_+}{\rr_+}-\frac{\mm_-}{\rr_-}\eeqa
with
\beqa{rpm*}{\rr_\pm:=\sqrt{\rr_0^2\mp2 \rr_0 \rr \sqrt{1-\frac{\Theta^2}{{\MM}^2}}\cos{m} +\rr^2}}\ .\eeqa
Combining this and
~\equ{lambdabeta},
one obtains
\beq{r}{\rr=\rr_0\sqrt{\l^2+\beta^2-1}\qquad {m}=\cos^{-1}\Big(-\frac{\l\beta}{\sqrt{\l^2+\beta^2-1}\sqrt{1-\frac{\Theta^2}{{\MM}^2}}}\Big)}\ .\eeq
The use of the associated generating function
\beqano
S({\MM}, {\Theta}, \l, \b)&=&\RR \rr_0\sqrt{\l^2+\beta^2-1}+\int^{\MM}  \cos^{-1}\left(-\frac{\l\beta}{\sqrt{\l^2+\beta^2-1}\sqrt{1-\frac{\Theta^2}{{{\MM}'}^2}}}\right)d{\MM}'\ .
\eeqano
allows to find the generalized impulses $\ovl p_\l$, $\ovl p_\b$ associated to
$\l$, $\b$ as
$$\arr{
\dst \ovl p_\l=\frac{\rr_0\l \RR}{\sqrt{\l^2+\beta^2-1}}-\frac{\beta\sqrt{(1-\beta^2)(\l^2-1){\MM}^2-(\l^2+\beta^2-1)	\Theta^2}}{(\l^2+\beta^2-1)(\l^2-1)}\\\\
\dst \ovl p_\beta=\frac{\rr_0\beta \RR}{\sqrt{\l^2+\beta^2-1}}+\frac{\l \sqrt{(1-\beta^2)(\l^2-1){\MM}^2-(\l^2+\beta^2-1)	\Theta^2}}{(\l^2+\beta^2-1)(1-\beta^2)}
}
$$
We invert such relations with respect to $\RR$, ${\MM}^2$:
\beqa{RPsi}
\arr{\dst \RR=\frac{\l(\l^2-1)\ovl p_\l+\beta(1-\beta^2)\ovl p_\beta}{\rr_0(\l^2-\beta^2)\sqrt{\l^2+\beta^2-1}}\\\\
\dst {\MM}^2=\frac{(\l \ovl p_\beta-\beta \ovl p_\l)^2(\l^2-1)(1-\beta^2)}{(\l^2-\beta^2)}+\frac{\l^2+\beta^2-1}{(1-\beta^2)(\l^2-1)}\Theta^2}
\eeqa
Using these formulae and the~\equ{r} inside the Hamiltonian~\equ{hsDel}, we find exactly the expression in~\equ{H2C}, with $ p_\l$, $p_\b$, $p_\o$ replaced by $\ovl p_\l$, $\ovl p_\b$, $\Theta$. Therefore, 
the Euler integral will be exactly as in~\equ{Es}, with the same substitutions.
After some 
 elementary computation, we find that the $\ovl\EE$ has, in terms of $\cD_{{\tt v}_0}$, the expression
\beqano
\ovl\EE={\MM}^2+\rr_0^2(1-\frac{\Theta^2}{{\MM}^2})(-\RR\cos{m}+\frac{{\MM}}{\rr}\sin{m})^2- 2 \rr \rr_0\cos{m}\sqrt{1-\frac{\Theta^2}{{\MM}^2}}\big(\frac{\mm_+}{\rr_+}-\frac{\mm_-}{\rr_-}\big)
\eeqano
with $\rr_\pm$ as in~\equ{lambdabeta}. Turning back to the coordinates ${\tt u}$, ${\tt v}$ via~\equ{delaunay}, one sees that $\ovl\EE$ has the expression in~\equ{G1}. The details are omitted.

\paragraph{Proof of~\equ{EEE}}
We finally check that, if the two--centre Hamiltonian is written in the form~\equ{newH2C}, its Euler integral takes the expression in~\equ{EEE} (up to an unessential constant). To this end, we let
$$\widehat{\JJ}(\widehat{\tt y},\widehat{\tt x}, \widehat{\tt x}'):=\frac{1}{\mm}{\JJ}(\mm \widehat{\tt y}, \widehat{\tt x}, \widehat{\tt x}')=\frac{\|  {\widehat{\tt y}}\|^2}{2}-\frac{{\cal M}}{\|{\widehat{\tt x}}\|}-\frac{\m{\cal M}}{\|{\widehat{\tt x}}-{\widehat{\tt x}'}\|}$$
and then we change, canonically,
$$\widehat{\tt x}'=2{\tt v}_0\ ,\quad \widehat{\tt x}={\tt v}_0+{\tt v}\ ,\quad \widehat{\tt y}'=
\frac{1}{2}({\tt u}_0-{\tt u})
\ ,\quad  \widehat{\tt y}={\tt u} $$
(where $\widehat{\tt y}'$, $\widehat{\tt u}_0$ denote the generalized impulses conjugated to $\widehat{\tt x}'$, $\widehat{\tt v}_0$, respectively)
we reach the Hamiltonian $\ovl\JJ$ in~\equ{2Cold}, with $\mm_+={\cal M}$, $\mm_-=\m{\cal M}$.  Turning back with the transformations, one sees that the function $\ovl\EE$ in~\equ{G1} takes the expression
\beqano
\frac{\EE}{\mm}&:=&\frac{1}{{\rm m}}\Big\|\Big({\tt x}-\frac{{\tt x}'}{2}\Big)\times {\tt y}\Big\|^2+\frac{1}{4{\rm m}}({\tt x}'\cdot {\tt y})^2\nonumber\\
&+&\mm{\tt x}'\cdot\Big({\tt x}-\frac{{\tt x}'}{2}\Big)\Big(\frac{{\cal M}}{\|{\tt x}\|}-\frac{\m{\cal M} }{\|{\tt x}'-{\tt x}\|}\Big)
\eeqano
After multiplying  by $\mm$, we rewrite the latter integral as
\beqa{cal N}
\EE=\EE_0+\m\EE_1+\EE_2\eeqa
with 
\beqa{G0G1}\EE_0:=\|{\tt M}\|^2-{\tt x}'\cdot {\tt L}\qquad 
\EE_1:= \mm^2{\cal M}\frac{({\tt x}'-{\tt x})\cdot {\tt x}'}{\|{\tt x}'-{\tt x}\|}\eeqa
where ${\tt M}$, ${\tt L}$
are as  in~\equ{CL},
 and, finally,
$$\EE_2:=\mm\frac{\|{\tt x}'\|^2}{2}{{\JJ}}\ .$$
  Since $\EE_2$ is  itself an integral for ${\JJ}$, we can neglect it and rename 
    \beqa{cal G new}
\EE:=\EE_0+\m\EE_1\eeqa
the Euler integral to ${\JJ}$; just as in~\equ{EEE}. 

 \section{The Implicit Function Theorem}
 Below, for a given positive number $\r$ and a graph
 $$\cF:=\cF (z,A):=\big\{(\a, z(\a):\quad \a\in A)\big\}$$
of a suitable $z:\ A\to {\mathbb R}^n$, with $A\subset {\mathbb R}^p$, we denote as
$$\cF_{\r}:=\big\{(\a,z')\in A\times {\mathbb R}^n:\ |z'-z(\a)|\le {\r}\quad \forall\ \a\in A\big\}\supset \cF\ .$$

\begin{lemma}\label{IFT}
Let $A\subset {\mathbb R}^p$  compact; ${\r_0}>0$; $z_0:\ A\to {\mathbb R}^n$ a continuous function. Let $$\cF_0:=\big\{(\a, z_0(\a):\quad \a\in A)\big\}$$ be the graph of $z_0$ for $\a\in A$.
Let
$\FF :\quad \cF_{0\r_0}\to {\mathbb R}^n$ a continuous function such  that the  matrix
$M(\a):=\partial_z \FF (\a, z_0(\a))$ is continuous and invertible for all $\a\in A$, and let $m$, $\r\le \r_0$ be such that
\beqa{IFT cond}\sup_A\|M^{-1}(\a)\|\le m\ ,\quad 2m\sup_A |\FF(\a, z_0(\a))|\le \r\ ,\quad 2 m\r\sup_{\cF_{0\r}}\|\partial^2_z\FF(\a, z)\|\le 1\ .\eeqa
Then there exist a unique continuous function $z:\ A\to {\mathbb R}^n$ such that
$$z(\a)\in \cF_{0\r}\ \forall\ \a\in A\qquad{\rm and}\qquad \FF(\a, z(\a))~\equiv0\ .$$
If in addition, $\FF\in C^k(\cF_{\r_0})$ with some $k$, then $z\in C^k(A)$. 
\end{lemma}

\section{Basics on the Liouville--Arnold Theorem}\label{Liouville Arnold}

In this section we recall the main content of the Liouville--Arnold theorem, referring to the wide dedicated literature (e.g.~\cite{arnold63a, gallavotti86, zehnder10} and references therein) for proofs and exact statements.

\nl
This milestone result of the 60s, due to V.I. Arnold~\cite{arnold63a}, states that,
given a $n$--degrees of freedom Hamiltonian
$$\FF_1:\quad(p, q)=(p_1, \cdots, p_n, q_1, \cdots, q_n)\in \cM\subset {\mathbb R}^n\times {\mathbb R}^n\to {\mathbb R}$$
equipped $n-1$ independent and Poisson--commuting first integrals  $\FF_2$, $\cdots$, $\FF_n$ and such that
 $\cM$ is an open, connected set of ${\mathbb R}^n$ such that the invariant manifolds 
$$\cM(f):=\big\{(p, q)\in \cM:\quad \FF_i(p, q)=f_i\ ,\ i=1, \cdots, n\big\}\quad \ f=(f_1, \cdots, f_n)\in \cF\subset {\mathbb R}^n$$
are smooth, connected and compact and foliate $\cM$, 
one can find an open and connected set ${\cal I}\subset {\mathbb R}^n$ and a smooth and canonical change coordinates
\beqa{action angle}
\cM&\to&  {\cal I}\times {\mathbb T}^n\nonumber\\
(p, q)&\to& (\II(p, q), \f(p, q))=(\II_1(p, q), \cdots, \II_n(p, q), \f_1(p, q), \cdots, \f_n(p, q))
\eeqa
  such that $\hh:=\FF_1\circ \phi$ is a function of $\II$ only
so the solutions of $\hh$  are linear in the angles:
$$\II(t)=\II(0)\ ,\quad \f(t)=\f(0)+\partial_\II\hh(\II(0)) t\ ,\qquad \forall t\ .$$
The coordinates $(\II, \f)$ are usually called {\it action--angle}.

\nl
The first step to obtain the change~\equ{action angle} is the construction of a (non--canonical)
  diffeomorphism
\beqa{phi***}\phi(f):\qquad \psi\in {\mathbb T}^n\to \big(p(\psi, f), q(\psi, f)\big)\in\cM(f)\qquad \forall\ f\in \cF\eeqa
the existence is proved via abstract arguments of differential topology.

\nl
Next, if 
$${\mathbb T}_k=\{0\}\times \{0\}\times \cdots\times {\mathbb T}\times \cdots\times \{0\}$$
is the $k^{\rm th}$ circle of ${\mathbb T}^n$ obtained letting $\psi_k$ vary in ${\mathbb T}$ and fixing the remaining $\psi_j$ at a fixed value, e.g., 0,
and 
$$\cC_k(f):=\phi({\mathbb T}_k, f)$$
the $k^{\rm th}$ {\it base circle} as the
image of ${\mathbb T}_k$ in $\cM(f)$, one firstly defines
$${\hat\II}_k(f):=\frac{1}{2\p}\oint_{\cC_k(f)} p\cdot dq\ .$$
Under the additional assumption that equations
\beqa{invertibilities}\hat\II(f)=\II\ ,\quad \tilde\II(p, q):=\hat\II(\FF(p, q))=\II\eeqa
can be inverted with respect to $f$, $p$, respectively,
the functions at right hand side in~\equ{action angle} are given by
\beqa{phik}\II(p, q):=\tilde\II(p, q)\ ,\quad \f(p, q):=\hat\f(\II(p, q), q)\ ,\eeqa
where
$$\hat\f_k(\II, q)=\partial_{\II_k}\int^q p(\II, q)\cdot dq$$
with $p(\II, q)$being the inverse of $p\to\tilde\II(p, q)$.
Note that, making use of the canonical changes
$$(p_k, q_k)\to (-q_k, p_k)$$
it is not really needed that the second map in~\equ{invertibilities} is invertible with respect to $p$, but it is sufficient that it can be inverted with respect to one half of its arguments.

\vskip.1in
\noi
{\bf Acknowledgments} I wish to  thank M. Berti,  L. Biasco, A. Celletti, R. de la Llave, A. Delshams, S. Di Ruzza, H. Dullin, A. Giorgilli,  M. Guardia, M. Guzzo, V. Kaloshin and T. Seara for their interest.

\newpage

% \bibliographystyle{plain}
%\bibliography{REFERENCES.bib}

\begin{thebibliography}{10}

\bibitem{arnold63a}
V.~I. Arnol{\cprime}d.
\newblock A theorem of {L}iouville concerning integrable problems of dynamics.
\newblock {\em Sibirsk. Mat. \v Z.}, 4:471--474, 1963.

\bibitem{bekovO78}
A.~A. {Bekov} and T.~B. {Omarov}.
\newblock {Integrable cases of the Hamilton-Jacobi equation and some nonsteady
  problems of celestial mechanics}.
\newblock {\em Soviet Astronomy}, 22:366--370, June 1978.

\bibitem{biscaniI16}
F.~Biscani and D.~Izzo.
\newblock A complete and explicit solution to the three-dimensional problem of
  two fixed centres.
\newblock {\em Monthly Notices of the Royal Astronomical Society},
  455(4):3480--3493, 2016.

\bibitem{chencinerM00}
A.~Chenciner and R.~Montgomery.
\newblock A remarkable periodic solution of the three-body problem in the case
  of equal masses.
\newblock {\em Ann. of Math. (2)}, 152(3):881--901, 2000.

\bibitem{llave01}
R.~de~la Llave.
\newblock A tutorial on {KAM} theory.
\newblock In {\em Smooth ergodic theory and its applications ({S}eattle, {WA},
  1999)}, volume~69 of {\em Proc. Sympos. Pure Math.}, pages 175--292. Amer.
  Math. Soc., Providence, RI, 2001.

\bibitem{fejoz13b}
J.~F{\'e}joz.
\newblock On action-angle coordinates and the {P}oincar\'e coordinates.
\newblock {\em Regul. Chaotic Dyn.}, 18(6):703--718, 2013.

\bibitem{ferrarioT04}
D.~L. Ferrario and S.~Terracini.
\newblock On the existence of collisionless equivariant minimizers for the
  classical {$n$}-body problem.
\newblock {\em Invent. Math.}, 155(2):305--362, 2004.

\bibitem{fortunatiW16}
A.~Fortunati and S.~Wiggins.
\newblock Negligibility of small divisor effects in the normal form theory for
  nearly-integrable {H}amiltonians with decaying non-autonomous perturbations.
\newblock {\em Celestial Mech. Dynam. Astronom.}, 125(2):247--262, 2016.

\bibitem{gallavotti86}
G.~Gallavotti.
\newblock Quasi-integrable mechanical systems.
\newblock In {\em Ph\'enom\`enes critiques, syst\`emes al\'eatoires, th\'eories
  de jauge, {P}art {I}, {II} ({L}es {H}ouches, 1984)}, pages 539--624.
  North-Holland, Amsterdam, 1986.

\bibitem{giorgilliLS09}
A.~Giorgilli, U.~Locatelli, and M.~Sansottera.
\newblock Kolmogorov and {N}ekhoroshev theory for the problem of three bodies.
\newblock {\em Celestial Mech. Dynam. Astronom.}, 104(1-2):159--173, 2009.

\bibitem{guardiaKZ18}
M.~Guardia, V.~Kaloshin, and J.~Zhang.
\newblock Asymptotic density of collision orbits in the restricted circular
  planar 3 body problem.
\newblock {\em arXiv:1805.00800}, 2018.

\bibitem{guzzoCB16}
M.~Guzzo, L.~Chierchia, and G.~Benettin.
\newblock The steep {N}ekhoroshev's theorem.
\newblock {\em Comm. Math. Phys.}, 342(2):569--601, 2016.

\bibitem{jacobi09}
C.~G.~J. Jacobi.
\newblock {\em Jacobi's lectures on dynamics}, volume~51 of {\em Texts and
  Readings in Mathematics}.
\newblock Hindustan Book Agency, New Delhi, revised edition, 2009.
\newblock Delivered at the University of K\"onigsberg in the winter semester
  1842--1843 and according to the notes prepared by C. W. Brockardt, Edited by
  A. Clebsch, Translated from the original German by K. Balagangadharan,
  Translation edited by Biswarup Banerjee.

\bibitem{KustaanheimoS65}
P.~Kustaanheimo and E.~Stiefel.
\newblock Perturbation theory of {K}epler motion based on spinor
  regularization.
\newblock {\em J. Reine Angew. Math.}, 218:204--219, 1965.

\bibitem{milaniG10}
A.~Milani and G.~F. Gronchi.
\newblock {\em Theory of orbit determination}.
\newblock Cambridge University Press, Cambridge, 2010.

\bibitem{nehorosev77}
N.~N. Nehoro{\v{s}}ev.
\newblock An exponential estimate of the time of stability of nearly integrable
  {H}amiltonian systems.
\newblock {\em Uspehi Mat. Nauk}, 32(6(198)):5--66, 287, 1977.

\bibitem{pinzari13}
G.~Pinzari.
\newblock Aspects of the planetary {B}irkhoff normal form.
\newblock {\em Regul. Chaotic Dyn.}, 18(6):860--906, 2013.

\bibitem{pinzari17}
G.~Pinzari.
\newblock An analysis of the sun-earth-asteroid systems based on the two-centre
  problem.
\newblock {\em arXiv:1702.03680}, 2017.

\bibitem{pinzari18}
G.~Pinzari.
\newblock Perihelia reduction and global {K}olmogorov tori in the planetary
  problem.
\newblock {\em Mem. Amer. Math. Soc.}, 255(1218), 2018.

\bibitem{poschel93}
J.~P{\"o}schel.
\newblock Nekhoroshev estimates for quasi-convex {H}amiltonian systems.
\newblock {\em Math. Z.}, 213(2):187--216, 1993.

\bibitem{siegelM71}
C.~L. {S}iegel{,} and J.~K. Moser.
\newblock {\em Lectures on Celestial Mechanics}.
\newblock Springer Verlag, 1995.
\newblock Reprint of the 1971 Edition.

\bibitem{waalkensDR06}
H.~Waalkens, H.~R. Dullin, and P.~H. Richter.
\newblock The problem of two fixed centers: bifurcations, actions, monodromy.
\newblock {\em Phys. D}, 196(3-4):265--310, 2004.

\bibitem{waldvogel76}
J.~Waldvogel.
\newblock The three-body problem near triple collision.
\newblock {\em Celestial Mech.}, 14(3):287--300, 1976.

\bibitem{zehnder10}
E.~Zehnder.
\newblock {\em Lectures on dynamical systems}.
\newblock EMS Textbooks in Mathematics. European Mathematical Society (EMS),
  Z\"urich, 2010.
\newblock Hamiltonian vector fields and symplectic capacities.

\end{thebibliography}
\addcontentsline{toc}{section}{References}
%}
\def\cprime{$'$} \def\cprime{$'$}

\end{document}